\documentclass{article}

\usepackage{arxiv}
\usepackage[utf8]{inputenc} 
\usepackage[T1]{fontenc}    
\usepackage{nicefrac}       
\usepackage{microtype}      
\usepackage{lipsum}
\usepackage[overload]{textcase}
\usepackage{amssymb}
\usepackage{graphicx}
\usepackage{upgreek}
\usepackage{amsmath}
\usepackage{mathtools}
\usepackage[boxruled,algoruled]{algorithm2e}
\usepackage{float}
\usepackage{bigints}
\usepackage{color}
\usepackage{framed}
\usepackage[htt]{hyphenat}
\usepackage{url}
\usepackage{makecell}
\usepackage{tabularx}
\usepackage{xr-hyper}

\newcommand{\adorym}{Adorym}
\newcommand{\autograd}{\textit{Autograd}}
\newcommand{\torch}{\textit{PyTorch}}
\newcommand{\scipy}{SciPy}
\newcommand{\tensorflow}{\textit{TensorFlow}}
\newcommand{\modfm}{\texttt{ForwardModel}}

\newcommand{\modwpr}{\texttt{wrapper}}
\newcommand{\modprop}{\texttt{propagate}}
\newcommand{\modreg}{\texttt{Regularizer}}

\newcommand{\micron}{$\upmu$m}

\newcommand{\vecxx}{\boldsymbol{x}}
\newcommand{\vecr}{\boldsymbol{r}}
\newcommand{\vecrxy}{\boldsymbol{r}_{x,y}}

\newcommand{\vectheta}{\boldsymbol{\theta}}
\newcommand{\vecy}{\boldsymbol{y}}
\newcommand{\veco}{\boldsymbol{o}}
\newcommand{\vecp}{\boldsymbol{p}}
\newcommand{\vecdr}{\Delta\boldsymbol{r}}
\newcommand{\vecymeas}{\boldsymbol{y}_{\text{meas}}}

\newcommand{\mataa}{\boldsymbol{A}}

\newcommand{\ipredm}{I_{\text{pred},m}}
\newcommand{\imeasm}{I_{\text{meas},m}}

\newcommand{\norm}[1]{\left\lVert#1\right\rVert}

\newcommand{\eg}{\textit{e.g.}}
\newcommand{\degree}{^\circ}
\newcommand{\fny}{f_\text{Ny}}
\newcommand{\daffine}{d_{\text{affine}}}
\newcommand{\jacob}[2]{\frac{\partial #1}{\partial #2}}

\graphicspath{{figures/}}

\title{Adorym: A multi-platform generic x-ray image reconstruction framework based on automatic differentiation}

\author{
 Ming Du \\
  Advanced Photon Source \\
  Argonne National Laboratory \\
  Lemont, Illinois 60439, USA \\
  mingdu@anl.gov \\
   \And
 Saugat Kandel \\
  Applied Physics Program \\
  Northwestern University \\
  Evanston, Illinois 60208, USA \\
  \And
 Junjing Deng \\
  Advanced Photon Source \\
  Argonne National Laboratory \\
  Lemont, Illinois 60439, USA \\
  \And
 Xiaojing Huang \\
  National Synchrotron Light Source II \\
  Brookhaven National Laboratory \\
  Upton, New York 11973, USA \\
  \And
 Arnaud Demortiere \\
  Laboratoire de R\'{e}activit\'{e} et Chimie des Solides (LRCS), CNRS UMR 7314 \\
  Universit\'{e} de Picardie Jules Verne, Hub de l'Energie \\
  80039 Amiens Cedex, France \\
  \And
 Tuan Tu Nguyen \\
  Laboratoire de R\'{e}activit\'{e} et Chimie des Solides (LRCS), CNRS UMR 7314 \\
  Universit\'{e} de Picardie Jules Verne, Hub de l'Energie \\
  80039 Amiens Cedex, France \\
  \And
 Remi Tucoulou \\
  European Synchrotron Radiation Facility \\
  38000 Grenoble, France \\
  \And
 Vincent De Andrade \\
  Advanced Photon Source \\
  Argonne National Laboratory \\
  Lemont, Illinois 60439, USA \\
  \And
 Qiaoling Jin \\
  \{Department of Physics \&{} Astronomy, Chemistry of Life Processes Institute\} \\
  Northwestern University \\
  Evanston, Illinois 60208, USA \\
  \And
 Chris Jacobsen \\
  Advanced Photon Source \\
  Argonne National Laboratory, Lemont, Illinois 60439, USA \\
  \{Department of Physics \&{} Astronomy, Chemistry of Life Processes Institute\} \\
  Northwestern University, Evanston, Illinois 60208, USA \\
  cjacobsen@anl.gov
}

\begin{document}

\maketitle

\begin{abstract}
We describe and demonstrate an optimization-based x-ray image
reconstruction framework called \adorym{}. Our framework provides a generic
forward model, allowing one code framework to be used for a wide range
of imaging methods ranging from near-field holography to and fly-scan
ptychographic tomography.  By using automatic differentiation for
optimization, \adorym{} has the flexibility to refine experimental
parameters including probe positions, multiple hologram alignment, and
object tilts.  It is written with strong support for parallel
processing, allowing large datasets to be processed on high-performance
computing systems.  We demonstrate its use on several experimental
datasets to show improved image quality through parameter refinement.
\end{abstract}

\section{Introduction}
\label{sec:intro}

Most image reconstruction problems can be categorized as inverse problem solving.  
One begins with the assumption that a measurable
set of data $\vecy$ arises from a forward model $F(\vecxx, \vectheta)$,
which in turn depends on an object function $\vecxx$ and
parameters $\vectheta$, giving
\begin{equation}
  \vecy = F(\vecxx, \vectheta).
  \label{eqn:forward_model}
\end{equation}
With experimental data $\vecymeas$ and additive noise
  $\boldsymbol{\epsilon}$, imaging experiments build a relation of
\begin{equation}
  \vecymeas = F(\vecxx, \vectheta) + \boldsymbol{\epsilon}.
  \label{eqn:problem_equation}
\end{equation}
When $F$ is a non-linear function of $\vecxx$, or when the problem size
is very large, a direct solution of Eq.~\ref{eqn:problem_equation} is either intractable
or non-existent at all. In other scenarios where a computationally feasible direct
solution does exist, the quality of the solution can degrade when the known information
is noisy or incomplete. A common example is standard
filtered-backprojection tomography, which is known to produce significant artifacts when
the projection images are sparse in viewing angles \cite{kak_1988}.
These issues motivate the use of iterative methods in solving Eq.~\ref{eqn:problem_equation},
where $\vecxx$ is gradually adjusted to find a
minimum of a loss function $L$ that is often formulated as
\begin{equation}
  L = \norm{\vecymeas - \vecy}^2=\norm{\vecymeas - F(\vecxx, \vectheta)}^{2},
  \label{eqn:loss_function}
\end{equation}
where the $\norm{\cdot}^2$ operation computes the Euclidean distance between
$\vecymeas$ and $F(\vecxx)$. Loss function $L$ formulations other than
Eq.~\ref{eqn:loss_function} are also frequently used \cite{godard_optexp_2012},
but they all include a metric measuring the mismatch between $\vecymeas$ and $F(\vecxx)$.
The object function $\vecxx$ can be a collection of
coefficients for a certain basis set (such as for Zernike polynomials
\cite{zingarelli_ao_2013}), which might yield a relatively small
number of unknowns with easy solution.  However, letting $\vecxx$ be a
complex 2D pixel or 3D voxel array provides a more general description
for complicated objects. Furthermore, the parameter $\vectheta$ may
include a complex
finite-sized illumination wavefield (a probe function) with a variety
of positions or angles that sample the object array, and a variety of
propagation distances leading to a set of intensity measurements
$\vecymeas$. In some cases, $\vectheta$ is also unknown and need to be
solved along with $\vecxx$. For example, in more complicated imaging methods such as far-field
ptychography \cite{hoppe_aca1_1969,rodenburg_apl_2004} and its
near-field counterpart \cite{stockmar_scirep_2013}, the probe function
might itself be finite sized but unknown, but it can be recovered
along with the object as part of recovering $\vecxx$.  In other
methods such as holotomography \cite{cloetens_apl_1999}, at each
object rotational angle one might
have a set of intensity measurements $\vecymeas$ acquired at different
Fresnel propagation distances from the object $\vecxx$, and each
measurement might have experimental errors in position or rotation,
collectively symbolized by $\vectheta$.

The fact that a wide range of imaging problems can be framed in terms
of minimizing the loss function $L$ of Eq.~\ref{eqn:loss_function}
suggests that object reconstruction algorithms should be generic to
some degree, or at least highly modular.  One example of this can be
found in multislice ptychography \cite{maiden_josaa_2012}, which can
be thought of as combining the probe-scanning of ptychography with
multiple Fresnel propagation distances as in holotomography.  There
has also been an evolution in approaches to ptychographic tomography.
It was first demonstrated using 2D ptychography to recover a complex
exit wave at each rotation angle, followed by phase unwrapping,
followed by standard tomographic reconstruction of these set of
projection images \cite{dierolf_nature_2010}.  An extension of this
approach to objects that are thick enough that propagation effects
arise was done by using multislice ptychography reconstruction to
generate a synthesized projection, followed by phase unwrapping and
standard tomographic reconstruction \cite{li_scirep_2018}.  Since
ptychographic image reconstruction \cite{rodenburg_apl_2004} by itself
can be treated as a nonlinear optimization problem
\cite{guizar_optexp_2008}, an alternative approach is to do the entire
ptychographic tomography reconstruction as a nonlinear optimization
problem.  One can then be flexible on the ordering of scanned probe
positions versus rotations \cite{gursoy_optlett_2017}, and one can
incorporate multislice propagation effects
\cite{gilles_optica_2018,du_sciadv_2020}.  In this case, one uses the
full set of intensities recorded at all probe positions and object
rotations as $\vecymeas$, and recovers the 3D object as well as the
probe function using the loss function minimization approach of
Eq.~\ref{eqn:loss_function}.  One can also include a noise model in
the loss function, which might be the Gaussian approximation to the
Poisson distribution at higher photon count, or a Poisson model at
lower photon count \cite{godard_optexp_2012}.  The point of these
examples is that a nonlinear optimization approach allows one to
change the forward model $F$ and the loss function $L$ to match the
conditions of the experiment, and use a sufficiently flexible
optimization approach to minimize $L$ and thus find a solution for
$\vecxx$.

The loss function $L$ of Eq.~\ref{eqn:loss_function} is a scalar, but
it depends on a large number of unknowns in $\vecxx$, and furthermore
the forward model $F(\vecxx)$ may by itself be nonlinear.  In
addition, the loss function $L$ can also include many regularizers $R$
with weighting $\lambda$ \cite{neumaier_siamr_1998} to incorporate
known or desired properties of the image such as sparsity
\cite{tibshirani_jrssb_1996}.  Minimizing a multivariate scalar
quantity can be done using gradient-based methods, which requires one to
calculate the partial derivative of $L$ with regards to each element of
$\vecxx$ to give $\jacob{L}{x_i}$.

\begin{figure}
  \centering
  \includegraphics[width=0.95\textwidth]{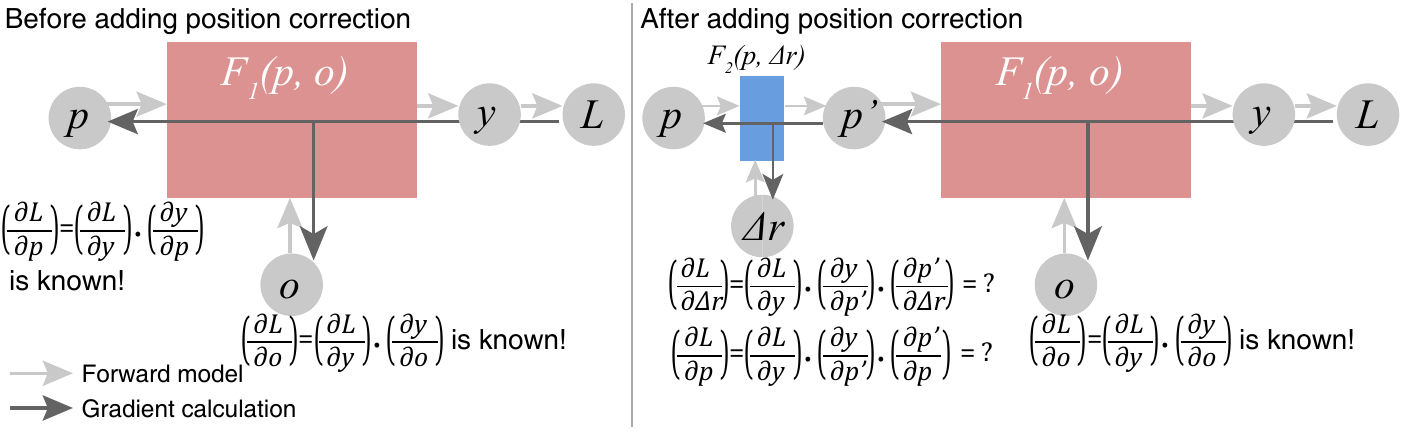}
  \caption{When modifying an existing forward model, the gradient of the new
    loss function often needs to be re-derived by going into the depth
    of the model. }
  \label{fig:gradient_break}
\end{figure}

For an example, we can look at the gradient-based minimization strategy
for the ptychography application (the PIE algorithm \cite{rodenburg_apl_2004} is one such procedure).
PIE can be shown
\cite{godard_optexp_2012} to be the equivalent of gradient descent
minimization of a loss function
$L = \norm{F_1(\veco, \vecp) - \vecymeas}^2$.  This loss function is the $l_2$-norm of the
difference between $F_1(\vecxx)$ and $\vecymeas$.  In this case $F_1$ is the
ptychographic forward model as a function of the object $\veco$ and
probe $\vecp$, with the forward model accounting for object sampling, probe
modulation, and far-field diffraction. The structure of the problem is
illustrated schematically at left in
Fig.~\ref{fig:gradient_break},
where the gradient of $L$ with regards to both $\veco$ and $\vecp$ (denoted as $\jacob{L}{\veco}$ and $\jacob{L}{\vecp}$)
is used to
optimize the guesses of both the object and the probe.
Now, let us consider optimization of probe positions (so as to account
for differences between intended and actual positions) along with reconstructing
the object \cite{guizar_optexp_2008,maiden_ultramic_2012,zhang_optexp_2013,tripathi_optexp_2014}. One way to incorporate probe
position corrections $\vecdr$ into the forward model is to
use the Fourier shift theorem \cite{lim_1990},
an operation denoted by $F_2$. The new forward model is then constructed
by placing $F_2$ before $F_1$ -- that is, first apply Fourier shift to get the shifted
probe $\vecp'$ at the current scan position, then perform object modulation and far-field
propagation. This is shown schematically at right
in Fig.~\ref{fig:gradient_break}. Modifying the forward model can be
done trivially by simply inserting the Fourier shift code at the right location,
but we would also have to re-derive the gradients in order to solve the unknowns.
Since $\veco$ enters the forward model
after $F_2$, we still know the gradient $\jacob{L}{\veco}$. However, our previous knowledge
of $\jacob{L}{\vecp}$ no longer holds, since we now must differentiate through
$F_2$ to get the new gradient with regards to $\vecp$. In other words, the updated
$\jacob{L}{\vecp}$ now requires additional derivatives $\jacob{y}{\vecp'}$
and $\jacob{\vecp'}{\vecp}$ (Fig.~\ref{fig:gradient_break}),
which are not known before.
Moreover, the gradient with
regards to $\vecdr$ should also be derived in order to optimize $\vecdr$.
Modifying the gradients requires much more effort than
modifying the forward model! The former can be tedious,
and prone to errors, and often needs to be redone when changes to the forward
model are made. The disproportional difficulty of re-deriving the gradients
introduces heavy burdens to algorithm development tasks beyond the parameter refinement
example described above -- for example, testing out new noise models,
developing more complicated physical models (\eg{}, one that accounts for multiple scattering),
and extending existing algorithms from one imaging setup to another.

These complexities have inspired the use of an alternative approach
for image reconstruction problems: the use of automatic
differentiation (AD) \cite{griewank_2008} for loss function
minimization.  Automatic differentiation involves the estimation of
partial derivatives of mathematical operations as expressed in
computer code, by storing intermediate results from small variations
of the input vector $\vecxx$ on a forward pass and using these in a
backward pass to accumulate the final derivative. The use of AD was
suggested for iterative phase retrieval at a time before easy-to-use
AD tools were available \cite{jurling_josaa_2014}.  However, even in
the short time since that suggestion, the explosion in machine
learning (where AD is often used to train a neural network
\cite{goodfellow_2017}) has led to the development of powerful AD
toolkits, which have then been exploited for parallelized
ptychographic image reconstruction \cite{nashed_procedia_2017}.
Because one does not have to manually calculate a new set of
derivatives as the forward problem is modified, AD has subsequently
been used for image reconstruction in near-field and Bragg
ptychography \cite{kandel_optexp_2019}, and for comparing near-field
and far-field ptychography with near-field holography with both
Gaussian and Poisson noise models \cite{du_jac_2020}.  Automatic
differentiation has also been used to address 3D image reconstruction
beyond the depth-of-focus limit by using multislice propagation for
the forward model \cite{du_sciadv_2020}. These successes have
motivated our development of the software framework we report here,
whose primary aim is to extend the application of AD beyond a specific
imaging modality.

In designing an AD-based framework for image reconstruction, we have
attempted to follow these guidelines:
\begin{itemize}

\item The framework should be generic and able to work with a variety
  of imaging techniques by assuming the general abstract model of
  image reconstruction. That is, the essential components of the
  framework should include the object function, the probe, the forward
  model, the loss, the optimizer, and the AD-based differentiator, but
  each of them should be subject to as few assumptions as possible,
  and whenever an assumption is necessary, it should be preferably a
  general one. For example, the object function is assumed to be 3D,
  which is downward-compatible with 2D image reconstruction
  problems. Another example is in regard to the forward model: one of
  the several forward model classes provided in the framework is a
  ptychotomography model assuming multiple viewing angles and multiple
  diffraction patterns per angle. By setting the number of diffraction
  patterns per angle to 1 and using a larger probe size, it can also
  work for full-field holography.

\item The framework should be able to not only optimize the object
  function, but also refine experimental parameters such as probe
  positions or projection alignments in order to address practical
  issues one would encounter in experiments.

\item The forward model part should have a plug-and-play
  characteristic, which would allow users to conveniently define new
  forward models to work with the framework, or to modify the existing
  ones. We will show in Section \ref{sec:forward_model} that we
  achieved this by packaging each forward model as an individual
  Python class that contains the prediction function and the loss
  function. As long as new variants of the class are created with both
  components following the input/output requirements, users can get it
  to work with AD immediately after telling the program to calculate
  the loss using that forward model. The same plug-and-play readiness
  also applies to refinable parameters and optimizers.

\item The framework should not be bound to a certain AD library. There
  are a variety of AD packages available, and each of them are
  uniquely advantageous in some aspects. We have thus created a
  unified frontend in our framework that wraps two AD libraries,
  namely \torch{} and \autograd{}, and used the application programming
  interfaces (APIs) from that frontend to build all forward models. In
  this way, the user can switch between both AD backends by simply
  changing an option. The frontend may also be expanded to other AD
  backends as long as they are added to the frontend following the
  stipulated input/output of each function.

\item The framework should be able to run on both single workstations
  and high-performance computers (HPCs), which allows one to freely
  choose the right platform and optimally balance convenience and scalability.
  This means that it should support parallelized processing
  using a widely available protocol. As will be introduced in Section
  \ref{sec:distribution_modes}, we used the MPI protocol
  \cite{mpi_31_standard} in our case. We have also implemented several
  strategies for parallelization, ranging from data parallelism
  \cite{xing_engineering_2016} to more memory-efficient schemes based
  on either MPI communication or parallel HDF5 \cite{hdf5}.

\item The framework also needs to be implemented in a language that
  has a large community of scientific computation users, and has the
  API support of most popular AD libraries. The language itself should
  also be portable, so that the program can be run on a wide range of
  platforms. As such, we choose to implement our framework in Python.

\end{itemize}
Our design following these criteria leads to \adorym{}, which stands
for \textbf{A}utomatic \textbf{D}ifferentiation-based \textbf{O}bject
\textbf{R}etrieval with D\textbf{y}namical \textbf{M}odeling. The same
name appeared in our early publication on the algorithm for
reconstructing objects beyond the depth-of-focus limit
\cite{du_sciadv_2020}, where the notion of ``dynamical modeling''
means both that the forward model accounts for dynamical scattering in
thick samples using multislice propagation, and that the forward model
can be dynamically adjusted for various scenarios thanks to AD. The
source code of \adorym{} is openly available on our GitHub repository
(\url{https://github.com/mdw771/adorym}). We have also prepared a detailed
documentation (\url{https://adorym.readthedocs.io/}). The rest of this paper will
describe the overall architecture of \adorym{} and each component of
it. We will then show reconstruction results of both simulated and
experimental, and demonstrate \adorym{}'s performance for varying
experimental types and imaging techniques. 

\section{Methods and theories}
\label{sec:methods}

\begin{figure}
  \centerline{\includegraphics[width=0.98\textwidth]{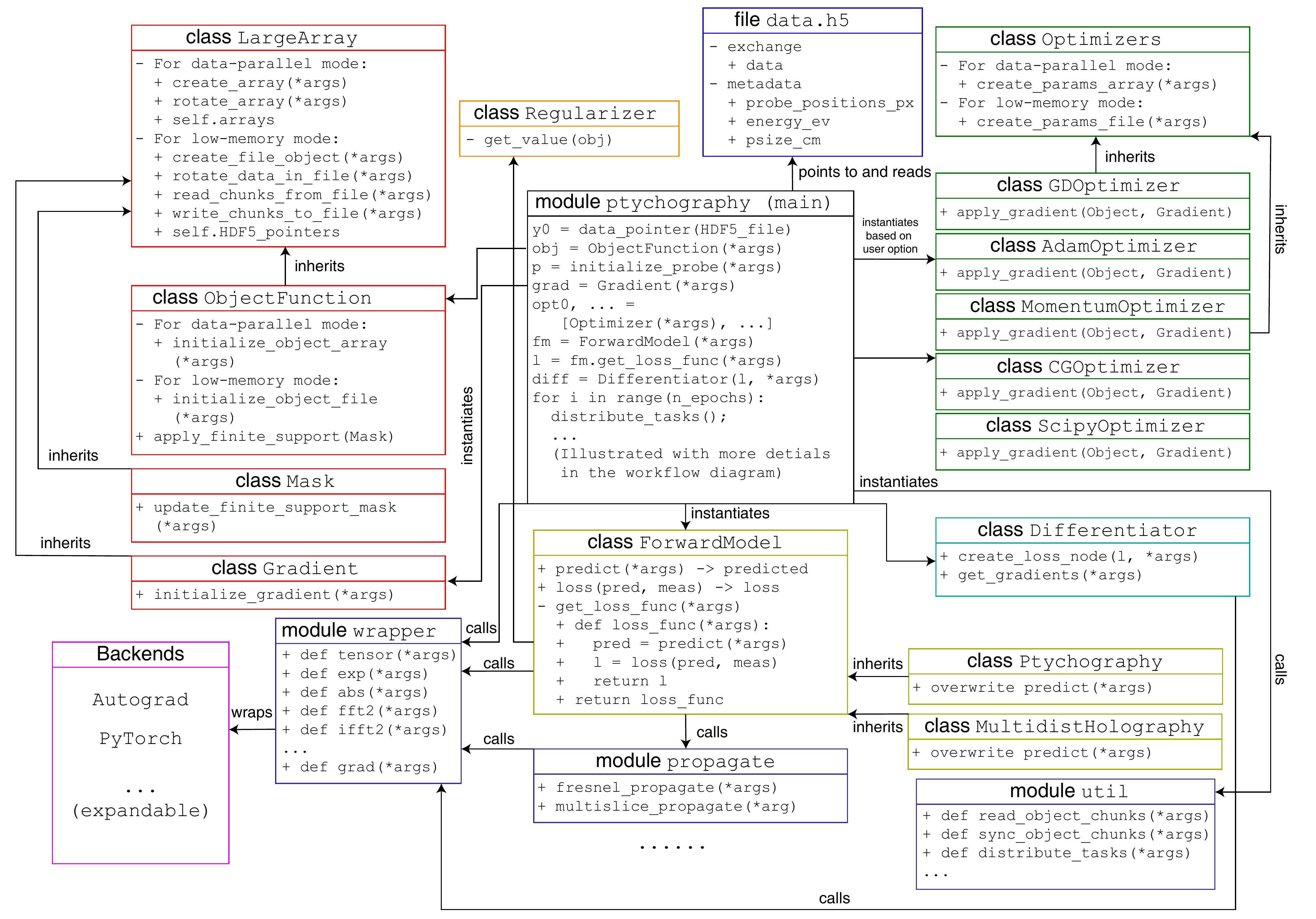}}
  \caption{Architecture of \adorym{}, listing the relationships between
    all modules, classes, and child classes.}
  \label{fig:block_chart}
\end{figure}

With the aim of maximizing the applicability of \adorym{} to a wide
range of x-ray imaging techniques, we adhered to modular and
object-oriented programming principles when designing it.
In this section, we briefly introduce the most crucial modules
and infrastructures of \adorym{} -- namely, its data format, its
forward models and loss functions, its AD backends, and its parallelization
scheme. These parts are directly relevant to the realization of \adorym{}'s
versatility, flexibility, and scalability.
While we will limit our narrative to these
key features of \adorym{} in the main text, readers can refer to the
supplementary materials for further detail.
A full picture illustrating its architecture is shown in
Fig.~\ref{fig:block_chart}.

\subsection{Data format and array structures}
\label{sec:data_format}

\adorym{} uses HDF5 files \cite{hdf5} for input and output, where
acquired images are saved as a 4D dataset with shape
\texttt{[num\_angles, num\_tiles, len\_detector\_y,
  len\_detector\_x]}.  This is a general data format that is
compatible towards more specific imaging types: for example, 2D
ptychography has \texttt{num\_angles = 1} and \texttt{num\_tiles > 1},
while full-field tomography has \texttt{num\_tiles = 1} and
\texttt{num\_angles > 1}. The last two dimensions specify a 2D image
which is referred to as a ``tile.''  For ptychography, a tile is just a
single diffraction pattern.  When dealing with full-field data from
large detector arrays, \adorym{} provides a script to divide each
image into several subblocks or tiles, so that the divided image data
can be treated in a way just like ptychography, where only a small
number of these tiles are processed each time. When
  applied to wavefield propagation, this is known as a ``tiling-based''
  approach; it allows single workstations to work with very large
  arrays \cite{blinder_optexp_2019}, or parallel computation on large
  arrays when using HPCs \cite{ali_optexp_2020}.

\begin{figure}
  \centerline{\includegraphics[width=0.6\textwidth]{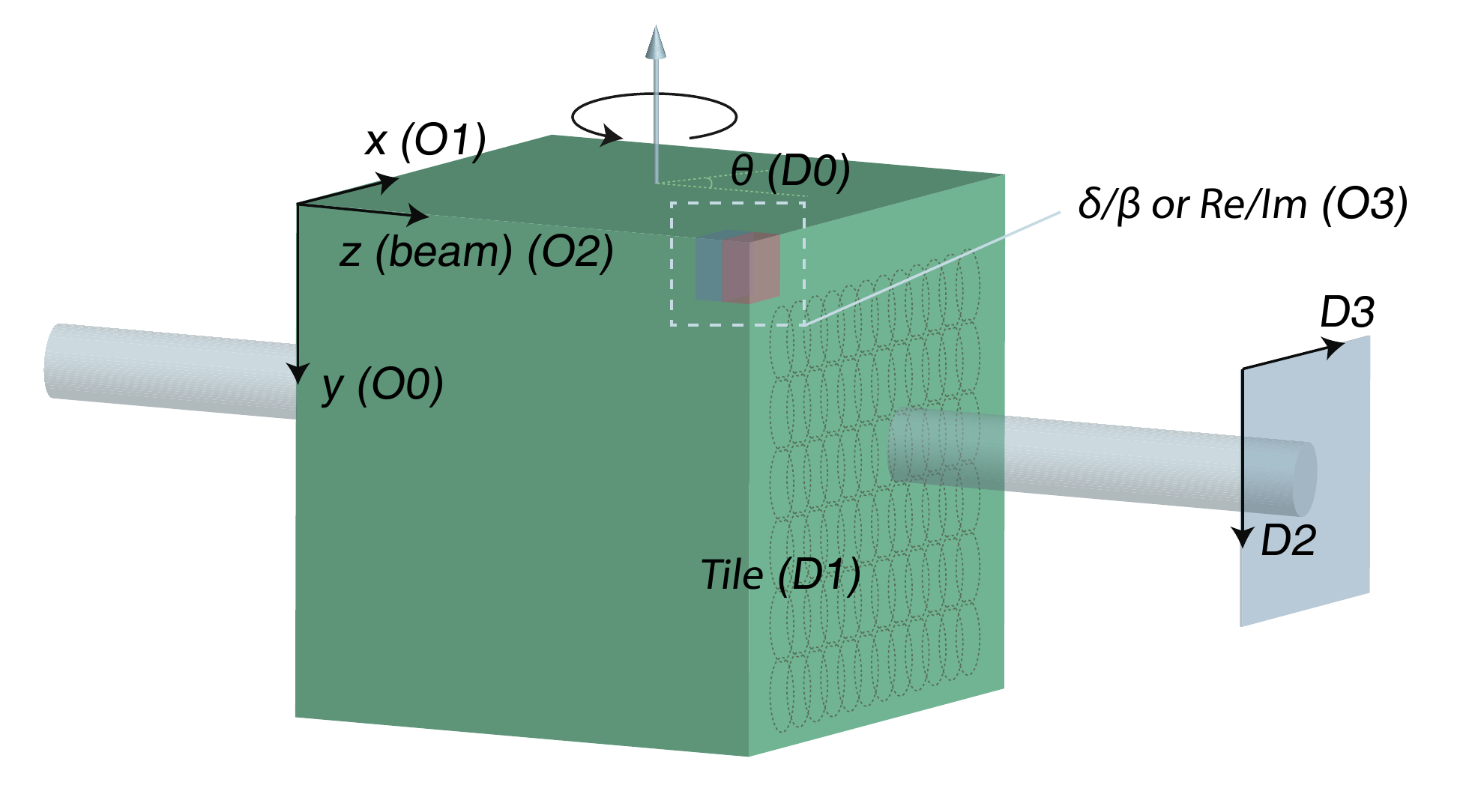}}
  \caption{Representation of experimental coordinates in \adorym{}'s
  readable dataset ($D$) and object function array ($O$). Directions
  and quantities are labeled with the index of dimension in the corresponding
  array; for example, $O2$ means that the associated object axis is
  stored as the 2nd dimension of the object array. }
  \label{fig:geo}
\end{figure}

During reconstruction, the object function being solved is stored as a
4D array, with the first 3 dimensions representing the spatial axes.
Depending on user setting, the last dimension represents either the $\delta$/$\beta$-parts of the
object's complex refractive indices, or the real/imaginary parts of the object's
multiplicative modulation function (\emph{i.e.}, for an incident wavefield
$\psi(\vecrxy)$ on a 2D plane $xy$, the wavefield modulated by object function
$O(\vecrxy)$ is given by $\psi(\vecrxy)\cdot \left(\Re[O(\vecrxy)] + i\Im[O(\vecrxy)]\right)$).
While the real/imaginary representation is more commonly used in
coherent diffraction imaging with non-uniform illumination such in ptychography
\cite{rodenburg_apl_2004}, the $\delta$/$\beta$ representation can potentially
remove the need for phase wrapping when a proper regularizer is used
(this is discussed in Section S1.2 of Supplementary
Material). The array geometry of the raw data file and the object function is shown in
Fig.~\ref{fig:geo}.

\subsection{Forward model and refinable parameters}
\label{sec:forward_model}

\adorym{} comes with a \modprop{} module for simulating wave
propagation. Near-field and far-field propagation can be done using
the built-in Fresnel propagation and Fraunhofer
functions. Furthermore, \adorym{} can also model diffraction in the
object using the multislice method
\cite{cowley_actacryst_1957a,ishizuka_actacryst_1977}, so that
multiple scattering in thick samples beyond the depth of focus can be
accounted for in object reconstruction \cite{du_sciadv_2020}. In
\adorym{}, two variants of the multislice algorithm are implemented:
the first assumes a constant slice spacing, so that the Fresnel
convolutional kernel can be kept constant.  This implementation is
intended for performing joint ptychotomography reconstruction for
thick samples with isotropic voxel size.  The other variant allows the
slices to be separated by unequal spacing, which is more suitable for
single-angle multislice ptychography
\cite{maiden_josaa_2012,suzuki_prl_2014,tsai_optexp_2016b}. For the
latter, \adorym{} allows the slice spacings to be refined in the AD
framework.  Moreover, \adorym{} allows one to use and reconstruct
multiple mutually incoherent probe modes, which can be used to account
for partial coherence \cite{thibault_nature_2013} or continuous motion
scanning \cite{pelz_apl_2014,deng_optexp_2015,huang_scirep_2015}.

\adorym{} also allows other experimental parameters to be refined along with
the minimization of the loss function, as long as these parameters can be
incorporated in the forward model as differentiable functions. An example
is probe position refinement in ptychography, as shown in Algorithm
\ref{algo:example_ptycho} and demonstrated in Sec.~\ref{sec:2idd}.
A full list of refinable parameters currently provided
by \adorym{} can be found in Section S1.D.2 of the supplementary material.

\begin{algorithm}
	\SetAlgoLined\DontPrintSemicolon
	\SetKwFunction{TakeObjectChunk}{TakeObjectChunk}
	\SetKwFunction{MultislicePropagate}{MultislicePropagate}
	\SetKwFunction{FourierShift}{FourierShift}
	\SetKwFunction{FarFieldPropagate}{FarFieldPropagate}
	\SetKwFunction{FourierShift}{FourierShift}
	\SetKwFunction{Rotate}{Rotate}
	\SetKwFunction{append}{append}
	\SetKwInput{Input}{Input}
	\SetKwBlock{Initialization}{Initialization}{end}
	\SetKwBlock{Main}{Procedure}{end}
	\SetKwInput{Output}{Output}
	\Input{
	\par
	detector size [$l_y$, $l_x$], object size [$L_y$, $L_x$, $L_z$], full object function $O$, a list of probe modes $\boldsymbol{\Psi}$, probe positions $\boldsymbol{R}$, probe position corrections $\boldsymbol{\Delta R}$, wavelength $\lambda$, pixel size $\delta$, current rotation angle $\theta$
    }
    \Initialization{
    \par
	$\boldsymbol{o} \leftarrow []$ \tcp*{List of object chunks each in shape of $[l_y, l_x, L_z]$}
	$\boldsymbol{\Psi_{\text{detector}}} \leftarrow []$ \tcp*{List of detector-plane waves}
	$\boldsymbol{\Psi_{\text{shifted}}} \leftarrow []$ \tcp*{Shifted probe functions}
    }
    \Main{
    $O' \leftarrow \Rotate{O}$ \;

	\For {$\vecr$ \textup{in} $\boldsymbol{R}$}{
		\tcc{Get local object chunks corresponding to diffraction patterns (tiles).}
		$\boldsymbol{o}$.\append(\TakeObjectChunk{$O'$, $\vecr$})\;
		\tcc{Shift probes according to the list of position corrections.}
		$\boldsymbol{\Psi_{\text{shifted}}}$.\append(\FourierShift{$\boldsymbol{\Psi}$, $\boldsymbol{\Delta R}$})
		}
	
	\tcc{Do propagation for all probe modes.}
	\For {$\psi_i$ \textup{in} $\boldsymbol{\Psi_{\text{shifted}}}$}{
	$\psi_i \leftarrow$ \MultislicePropagate{$\psi_i$, $\boldsymbol{o}$, $\lambda$, $\delta$} \;
  	$\psi_i \leftarrow$ \FarFieldPropagate{$\psi_i$} \;
  	$\boldsymbol{\Psi_{\text{detector}}}$.\append{$\psi_i$}
     }
    }
	\Output{
	$\boldsymbol{\Psi_{\text{detector}}}$
    }
	
    \caption{An example \texttt{predict} function in the forward model
      of far-field ptychotomography.  This is for data parallelism
      mode, with probe position refinement and multiple probe modes. }
	\label{algo:example_ptycho}
\end{algorithm}

The fact that \adorym{} assumes a 4D data format, and a 3D object function
with 2 channels, means that it can use a ptychotomographic forward model
to work with many imaging techniques. Holography data, for
instance, is interpreted by \adorym{} as a special type of
ptychography with 1 tile per angle, with detector size equal to
the $y$- and $x$-dimensions of the object array, and with a near-field
propagation from the sample plane to the detector plane. However,
this does not prevent us from creating forward models dedicated for
more specific imaging techniques. In fact, since all forward models
are packaged as individual classes, users will find it easy to define
new forward models, and then use them in the reconstruction routine.

\subsection{Loss functions}

For the final loss function, \adorym{} has two built-in types
of data mismatch term.  The first is the least-square error (LSQ), expressed as
\begin{equation}
	\mathcal{D}_{\text{LSQ}} = \frac{1}{N_d}\sum_{m}^{N_d}\norm{\sqrt{\ipredm} - \sqrt{\imeasm}}^2
	\label{eqn:lsq_loss}
\end{equation}
where $N_d$ is the number of detector (or tile) pixels.  The second is
the Poisson maximum likelihood error, expressed as
\cite{godard_optexp_2012}:
\begin{equation}
	\mathcal{D}_{\text{Poisson}} = \frac{1}{N_d}\sum_{m}^{N_d}\ipredm - \imeasm\log(\ipredm).
	\label{eqn:poisson_loss}
\end{equation}
While both loss functions have their own merits in terms of numerical robustness
and noise resistance \cite{godard_optexp_2012,du_jac_2020}, we provide
both of them for users' choice.
Furthermore, \adorym{} also allows users to conveniently add new loss function
types (for example, mixed Gaussian-Poisson).

\adorym{} allows users to provide a
finite support constraint to supply prior knowledge about the sample.
This is commonly used in
coherent diffraction imaging \cite{fienup_optlett_1978,miao_nature_1999,marchesini_prb_2003}. Furthermore,
\adorym{} comes with \modreg{} classes which contains several types of
Tikhonov regularizers \cite{tikhonov_dansssr_1963,mccann_arxiv_2019}.
These include the $l_1$-norm \cite{tibshirani_jrssb_1996},
the reweighted $l_1$-norm \cite{candes_jfaa_2008}, and total variation
\cite{grasmair_amo_2010}. A detailed description of these regularizers
can be found in the supplementary materials.
Additional regularizers can be added by creating new
  child classes of \modreg{}.

\subsection{AD engines and optimizers}
\label{sec:module_wrapper}

The automatic differentiation (AD) engine is the cornerstone of \adorym{}: it provides the
functionality to calculate the gradients of the loss function with
regards to the object, the probe, and other
parameters. 
\adorym{} uses two AD backends: \autograd{} \cite{autograd_docs}, and
\torch{} \cite{pytorch}. \autograd{} builds its data type and functions on the
basis of the popular scientific computation package NumPy \cite{numpy}
and \scipy{} \cite{scipy}; these in turn can make use of hardware-tuned
libraries such as the Intel Math Kernel library
\cite{intel_mkl}. However, \autograd{} does not have built-in support
for graphical processing units (GPUs). This shortcoming is addressed in
\torch{}, which also has a larger user community. Both libraries are dynamic graph tools that
allow the computational graph (representing the forward model) to be altered at runtime \cite{looks_arxiv_2017}.
This allows for more flexible workflow control, and easier debugging.
In order for users to easily switch between both backends, we
have built a common front end for them in the \modwpr{} module. This
module provides a unified set of APIs to create optimizable or
constant variables (arrays in \autograd{}, or tensors in \torch{}), call
functions, and compute gradients.

Optimizers define how gradients should be used to update the object
function or other optimizable quantities. \adorym{} currently provides
4 built-in optimizers: gradient descent (GD), momentum gradient
descent \cite{qian_nn_1999}, adaptive momentum estimation (Adam)
\cite{kingma_arxiv_2014}, and conjugate gradient
(CG) \cite{polyak_ucmmp_1969}. Moreover, \adorym{} also has a \texttt{ScipyOptimizer} class
that wraps the \texttt{optimize.minimize} module of the \scipy{}
library \cite{scipy}, so that users may access more optimization
algorithms coming with \scipy{} when using the \autograd{} backend on
CPU. More details about these optimizers can be found in the supplementary
materials.

\subsection{Parallelization modes}
\label{sec:distribution_modes}

\begin{figure}[H]
  \centering \includegraphics[width=0.98\textwidth]{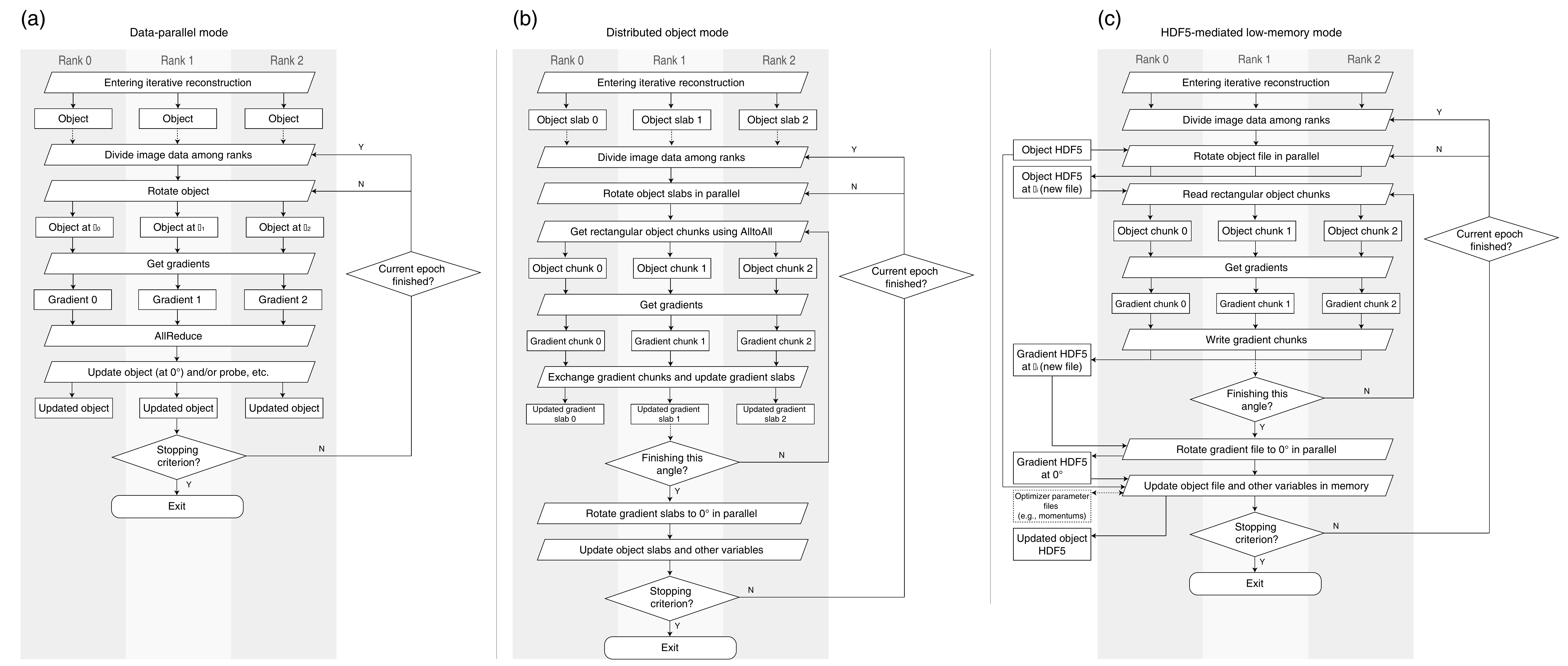}
  \caption{Workflow diagram of \adorym{} in DP mode, DO mode, and H5
    mode.}
  \label{fig:workflow}
\end{figure}

\adorym{} supports parallelized processing based on the message
passing interface (MPI) \cite{mpi_31_standard}, which allows the
software to spawn multiple processes using different GPUs, or many
nodes on an HPC.  Parallel processing based on MPI can be implemented
in several different ways, each of which differs from others in terms
of computational overhead and memory consumption. On the most basic
level, \adorym{} allows users to run in its data parallelism (DP)
mode, where each MPI rank saves a full copy of the object function and all other parameters
\cite{xing_engineering_2016}.  Before updating the parameters, the
gradients are averaged over all ranks.  While the DP mode is low in
overhead, it has high memory consumption.  This in turn limits the
ability to reconstruct objects when one has limited memory resources.
Thus, \adorym{} also provides another two modes of parallelization,
namely the distributed object (DO) mode, and the HDF5-file-mediated
low-memory (H5) mode.

\subsubsection{Distributed object (DO) mode}
\label{sec:do_mode}

\begin{figure}[H]
  \centering \includegraphics[width=0.9\textwidth]{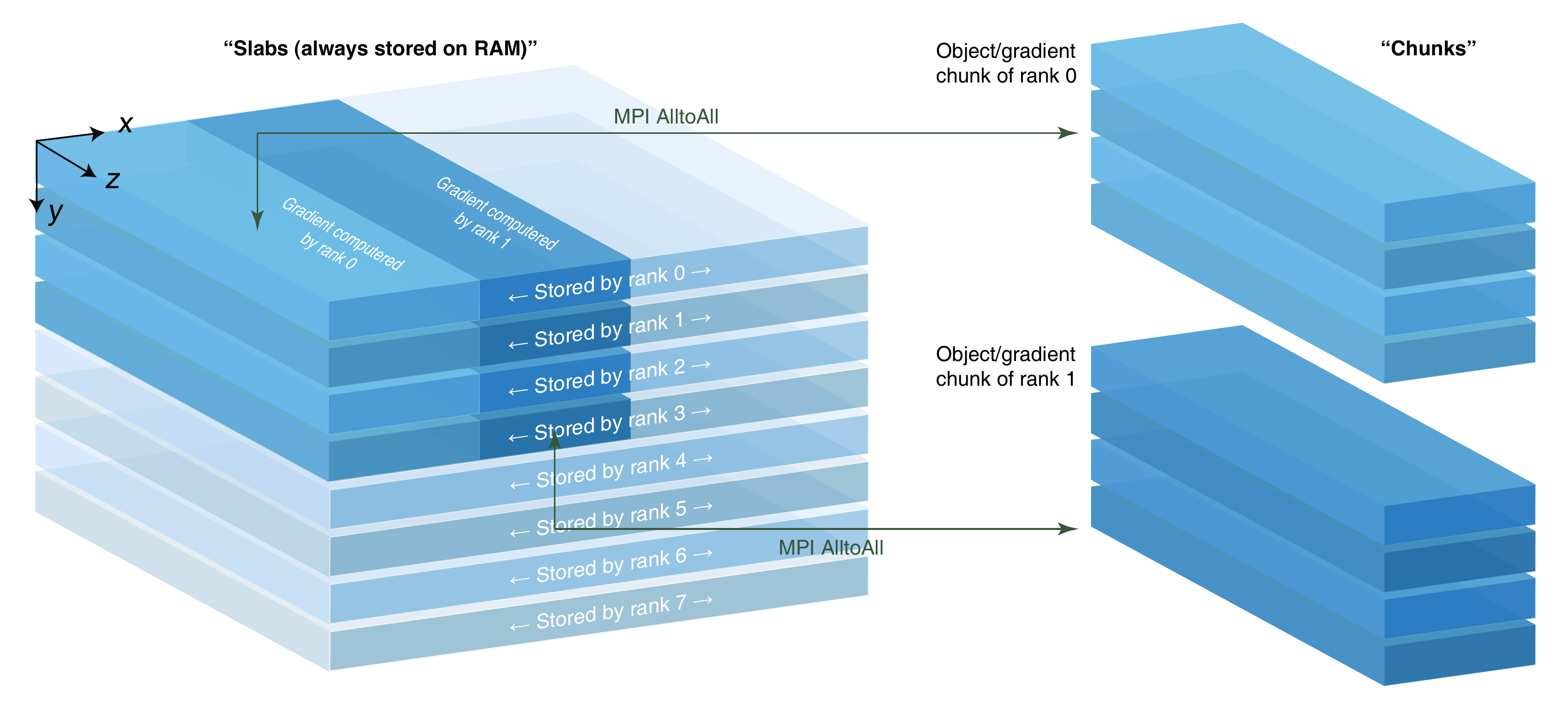}
  \caption{Illustration of the distributed object (DO) scheme. With
    multiple MPI ranks, the object is divided into a vertical stack of
    slabs, which are kept separately by all ranks. When a rank
    processes a diffraction image, it gets data from ranks that
    possess the parts of the object function that it needs, after
    which it assembles these partial slabs into the object chunk
    corresponding to the beam path of the diffraction image that it is
    processing.  After gradient of the chunk is calculated, it is
    scattered back into the same positions of the gradient slabs kept
    by relevant ranks. Object update is done by each rank individually
    using the gathered gradient array.}
  \label{fig:distributed_object}
\end{figure}

In distributed object (DO) mode,
only one object function is jointly kept by all ranks. This is done by
seperating the object along the vertical axis into several slabs, and
letting each rank store one of them in its available RAM as 32-bit
floating point numbers.  In this way, standard tomographic rotation
can be done independently by all ranks in parallel. The stored object
slab does not enter the device memory as a whole when GPU acceleration
is enabled; instead, only partial chunks of the object that are relevant
to the tiles being processed (\emph{i.e.}, on the beam paths of these tiles)
are extracted from the object slabs
in the CPU, and only these partial chunks are sent to GPU memory for
gradient computation. Since the shape of a slab is usually
different from an object chunk (whose $x$-$y$ cross
sections should match the size of a tile), we use MPI's AlltoAll
communication \cite{mpi_31_standard} for a rank to gather the voxels it needs to assemble an
object chunk from other ranks.  This is illustrated in
Fig.~\ref{fig:distributed_object}, where we assume the MPI command
spawns 8 ranks, and the object function is evenly distributed among
these ranks. For simplicity, we assume a batch size of 1, meaning a
rank processes only 1 tile at a time. Before entering the
differentiation loop, the object slab kept by each rank is duplicated
as a new array, which is then rotated to the currently processed
viewing angle. In our demonstrative scenario, rank 0 processes the top
left tile in the first iteration, and the vertical size of the tile
spans four object slabs. As ranks 0--3 have knowledge about rank 0's
job assignment, they extract the part needed by rank 0 from their own
slabs, and send them to rank 0. Rank 0 then
assembles the object chunk from the partial slabs by concatenating
them along $y$ in order. Since rank 0 itself contains the object slab
needed by other ranks, it also needs to send information to them.
The collective send/receive can be done using MPI's AlltoAll communication
in a single step. The ranks
then send the assembled object chunks to their assigned GPUs (if available) for
gradient computation, yielding the gradient chunks. These gradient chunks are
scattered back to the ``storage spaces'' of relevant ranks on RAM -- but this time to
``gradient slab'' arrays that are one-to-one related to the object slabs.
After all tiles on this certain
viewing angle are processed, the gradient slabs kept by all ranks are
reverted back to 0$\degree{}$. Following that, the object slab is
updated by the optimizer independently in each rank. When
reconstruction finishes, the object slabs stored by all ranks are
dumped to a RAM-buffered hard drive, so that they can be stacked to
form the full object. The optimization workflow of the DO mode is
shown in Fig.~\ref{fig:workflow}(b).

While the DO mode has the advantage of
requiring less memory per rank, a limitation is that
tilt refinement about the $x$- and $z$-axis is hard to implement.
That is because tilting about these axes requires a rank to get voxel values from
other slabs, which not only induces excessive MPI communication, but
also requires AD to differentiate through MPI operations. The latter is
not impossible, but existing AD packages may need to be modified in
order to add that feature. Another possible approach among the slabs
kept by different MPI ranks is to introduce overlapping regions along
$y$; in this case, tilting about $x$ and $z$ can be done individually
by each rank, though the degree of tilt is limited by the length of
overlap.  This is not yet implemented in \adorym{}, but could be added
in the future.

\subsubsection{HDF5-file-mediated low-memory mode (H5)}

When running on an HPC with hundreds or thousands of computational nodes,
the DO mode can significantly reduce the memory needed by each node to store the object
function if one uses many nodes (and only a few ranks per node)
to distribute the object. If one is instead using a single workstation or laptop, the total volume
of RAM is fixed, and very large-scale problems are still difficult to solve
even if one uses DO. For such cases, \adorym{} comes with an alternative
option of storing the full object function in a parallel HDF5 file
\cite{hdf5} on the hard drive that is accessible to all MPI ranks.
This is referred to as the H5 mode of data storage.
In the H5 mode, each rank reads or writes only partial chunks of the object or
gradient that are relevant to the current batch of raw data it is processing,
similar to the DO mode except that the MPI communications are replaced with
hard drive read/write instructions.
While the H5 mode
allows the reconstruction of large objects on limited memory machines,
I/O with a hard drive (even with a solid state drive) is slower than
memory access.  Additionally, writing into an HDF5 file with multiple ranks
is subject to contention, which may be mitigated if a parallel file
system with multiple object storage targets (OSTs) is available
\cite{behzad_2013}.  This is more likely to be available at an HPC
facility than on a smaller, locally managed system, and precise
adjustment of striping size and HDF5 chunking are needed to optimize
OST performance \cite{howison_hdf}.
Despite these challenges, the HDF5 mode is valuable because it
enables reconstructions of large objects and/or complex forward models
on limited memory machines. It also provides future-proofing because,
as fourth-generation synchrotron facilities deliver higher brightness and
enable one to image very thick samples \cite{eriksson_jsr_2014}, we may eventually encounter
extra-large objects which might be
difficult to reconstruct even in existing HPC machines.

A comparison of the three parallelization modes is shown in
Fig.~\ref{fig:workflow}.

\section{Results}
\label{sec:results}

Having described the key elements of \adorym{}, we now demonstrate its
use for reconstructing several simulated and experimental datasets.

\subsection{Computing platforms}
\label{sec:platforms}

This section involves several different computing platforms, on which \adorym{}
or other referenced packages were run:
\begin{itemize}

\item The machine ``Godzilla'' is a HP Z8 G4
  workstation with two Intel Xeon Silver 4108 CPUs (8 cores each), 768
  GB DDR4 RAM, one NVIDIA P4000 GPU (8 GB), a 256 GB solid
  state drive for the Red Hat Enterprise Linux 7 operating system and
  swap files, and a 18-TB, 3-drive RAID 5 disk array.

\item The machine ``Blues'' is an HPC system at Argonne's Laboratory
  Computing Resource Center (LCRC).  It includes 300 nodes equipped
  with 16-core Intel Xeon E5-2670 CPUs (Sandy Bridge architecture),
  and 64 GB of memory per node.

\item The machine ``Cooley'' is an HPC system at the Argonne Leadership
  Computing Facility (ALCF). It has 126 compute nodes with Intel Haswell
  architecture. Each node has two 6-core, 2.4-GHz Intel E5–2620 CPUs,
  and NVIDIA Tesla K80 dual GPUs. 

\item The machine ``Theta'' is a HPC system at the Argonne Leadership
  Computing Facility (ALCF).  This Cray XC40 system uses the Aries
  Dragonfly interconnect for inter-node communication
  \cite{parker_2017}. Each node of Theta possesses a 64-core Intel
  Xeon Phi processor (Intel KNL 7230) and 192 GB DDR4 RAM. An
  additional 16 GB multi-channel dynamic RAM (MCDRAM) is also
  available with each node; this is used as the last-level cache
  between L2 and DDR4 memory in our case. Up to 4392
  nodes are available, though we used no more than
  256 in the work reported here.  The storage system
  of Theta uses the Lustre parallel file system, which allows data to
  be striped among 56 of its object storage targets (OSTs).  This
  system was used to analyze the charcoal dataset shown in
  Fig.~\ref{fig:charcoal}.

\end{itemize}
We indicate the platform used for each demonstration of \adorym{}
below, and show compute time metrics in Table \ref{tab:perf}.

\subsection{Multi-distance holography for heterogeneous samples: a simulation case study}
\label{sec:cameraman_mdh}


\begin{figure}[H]
  \centering
  \includegraphics[width=0.98\textwidth]{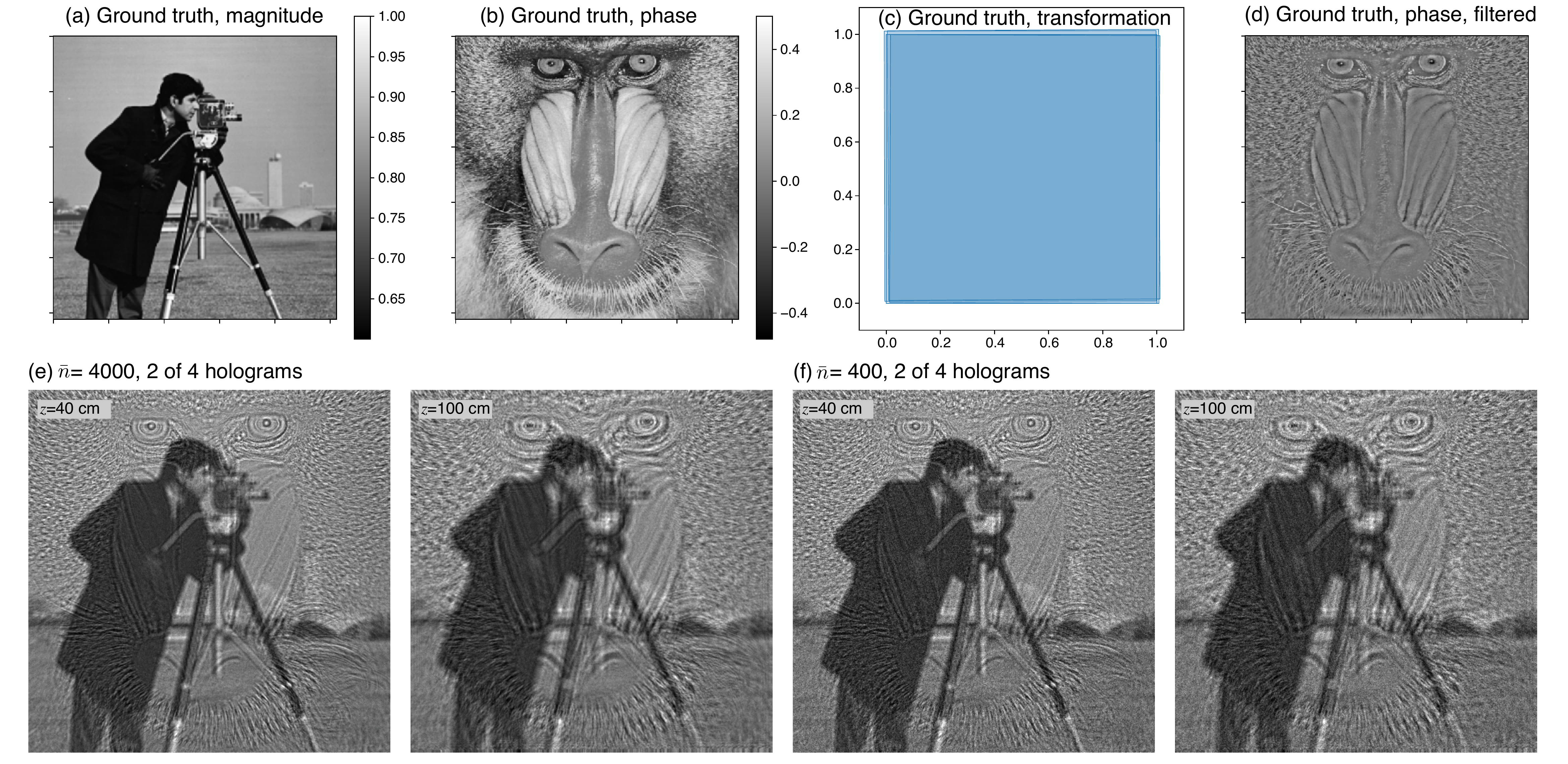}
  \caption{Ground truth images used in the simulation. Images (a) and
    (b) show the magnitude and phase of the simulated object, which
    are uncorrelated.  In (c) we show the artificial affine
    errors added to the holograms. Transformation
    matrices used for all 4 holograms at different defocusing
    distances (where the one for the first hologram is an identity
    transformation matrix) are applied to the vertices of 4
    $1\times 1$ squares, and the distorted squares are shown in the
    figure. In (d) we show a high-pass filtered version of (b), which
    is used for calculating the SSIM of reconstructed phase maps.  The
    first (40 cm distance) and last (100 cm) of four recorded
    holograms are shown in (e) with $\bar{n}=4000$ photons per pixel
    incident, and in (f) with $\bar{n}=400$ photons per pixel
    incident. }
\label{fig:cameraman_mdh_raw_affine}
\end{figure}

Multi-distance holography (MDH)
\cite{cloetens_apl_1999,krenkel_actacrysta_2017} is an in-line
coherent imaging technique, where holograms are collected at multiple
distances to provide diversity for robust phase retrieval.  These
distances should be chosen to minimize the overlap of spatial
frequencies for which the contrast transfer function (CTF) from
Fresnel propagation crosses zero \cite{zabler_rsi_2005}.  The recorded
data are non-linearly related to both the magnitude and phase of the
object, but a linear approximation can be taken if one assumes that
the object is pure-phase \cite{cloetens_apl_1999} or weak in
absorption, and has uniform composition so that
$\delta(\vecr) = (1 / \kappa)\beta(\vecr)$ \cite{turner_optexp_2004}.
In this case the linearized CTF can be easily inverted, and the phase
of the object can be found directly from a combination of the
back-propagated holograms \cite{cloetens_apl_1999}.  Even when these
conditions are not fully satisfied, the multiple hologram distances
provide phase diversity, allowing object reconstruction using an
iteratively regularized Gauss–Newton (IRGN) method with the addition
of either a finite-support constraint \cite{maretzke_optexp_2016} or
the pure-phase object approximation \cite{krenkel_actacrysta_2017}.

In order to test the ability of \adorym{} to reconstruct
multi-distance holography data, we created a $512\times 512$ 2D
simulated object shown in Fig.~\ref{fig:cameraman_mdh_raw_affine}.
The well-known ``cameraman'' image was used as the magnitude of the
object modulation function, in which the magnitude ranges from 0.6 to
1.0. The ``mandrill'' image was used for the phase, with
a range of $-0.5$ to $+0.5$ radians.  
Because an optimization approach allows one
  to use an accurate forward model, there is also no need to require
  compositional uniformity or weak object conditions.  
By using separate images for
magnitude and phase, we effectively gave individual pixels wildly
differing ratios of $\delta/\beta$ in the x-ray refractive index of
$n=1-\delta-i\beta$.  Thus the simulated object satisfied neither the
weak absorption nor the uniform composition approximations.  As a
simplification, we assumed that the object and detector had the same
1~\micron{} pixel size (a simplification from the common practice of
using geometric magnification from a smaller source), which means a
hologram recorded using 17.5 keV X rays would have a
depth of focus of about 7.3~cm.  We then simulated the recording of
holograms at $z=40$, 60, 80, and 100~cm distance from the object, so
these holograms are noticeably different from each other as shown in
Fig.~\ref{fig:cameraman_mdh_raw_affine}.

\begin{figure}
  \centering
  \includegraphics[width=0.85\textwidth]{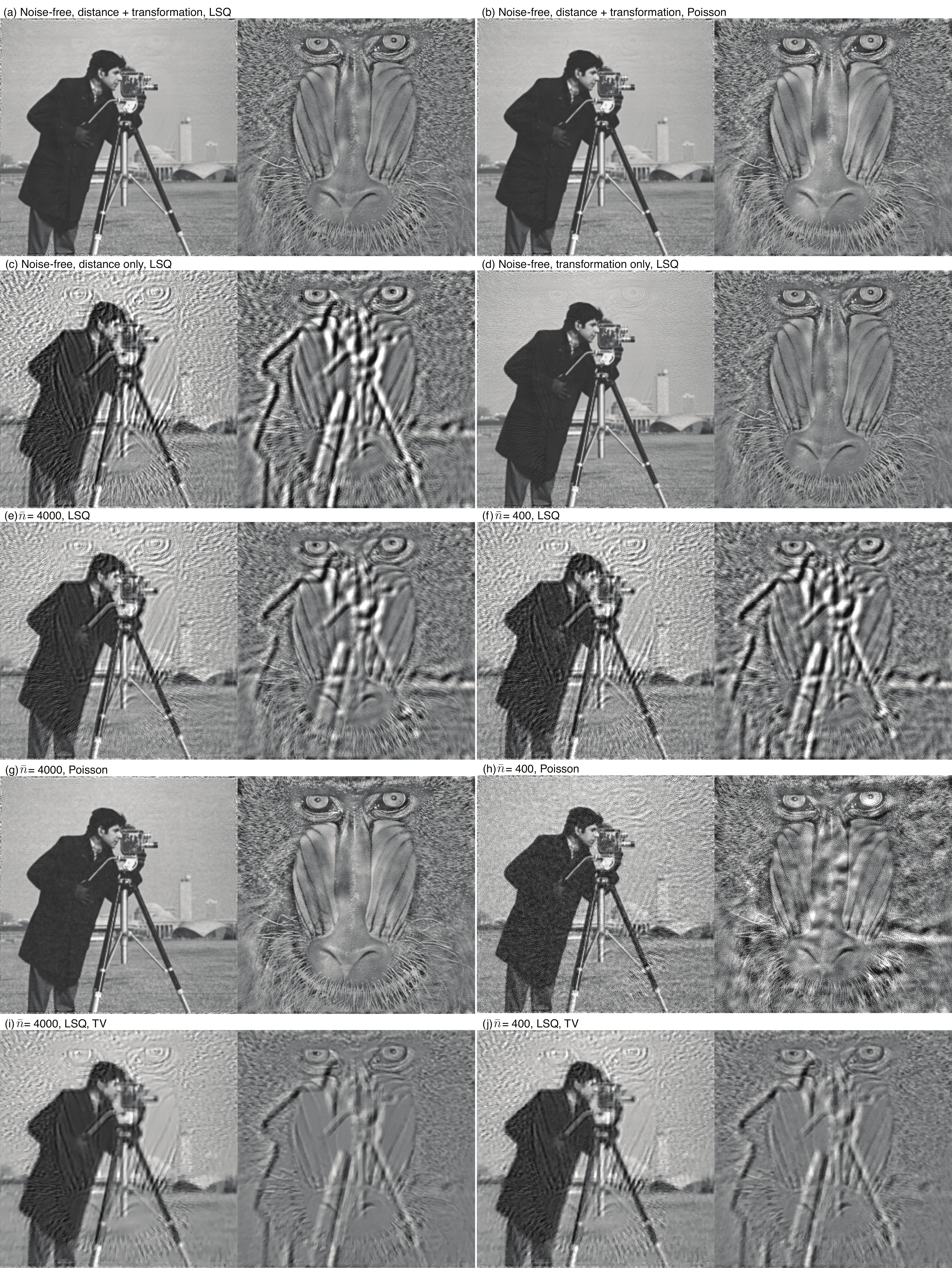}
  \caption{Multi-distance holography test using the ``cameraman''
    image for magnitude and the ``mandrill'' image for phase.  Images
    (a-d) show \adorym{} reconstructions from noise-free holograms,
    either with both affine transformation and distance refinement
    enabled, or with just one of them.  In the former case, both LSQ
    (Eq.~\ref{eqn:lsq_loss}) and Poisson (Eq.~\ref{eqn:poisson_loss})
    loss function results are shown. We then show magnitude and phase
    reconstructions that include both distance and transformation
    refinement with limited per-pixel photons of $\bar{n}=4000$ using
    LSQ (e) and Poisson (g) noise models, and for $\bar{n}=400$ using
    LSQ (f) and Poisson (h) noise models -- in both cases, use of the
    Poisson noise model reduces crosstalk between the magnitude
    ``cameraman'' and phase ``mandrill'' images.  If we add a total
    variation (TV) regularizer to the reconstruction, the image
    quality is degraded (with significant crosstalk between magnitude
    and phase images) because both the ``cameraman'' and ``mandrill''
    images contain high spatial frequency detail.  This is shown in
    (i) for $\bar{n}=4000$ and in (j) for $\bar{n}=400$ with the LSQ
    noise model.}
  \label{fig:cameraman_mdh_recons}
\end{figure}

The intensity of an in-line hologram is unable to record the
zero-spatial-frequency phase of an object; in fact, phase contrast is
only transferred at spatial frequencies that appear outside the first
Fresnel zone of radius $r_{1}=\sqrt{\lambda z}$ in the hologram.
Therefore we show in Fig.~\ref{fig:cameraman_mdh_raw_affine}(d) a
version of the ``mandrill'' phase object which has been highpass
filtered to include only those spatial frequencies above
$1/4\sqrt{\lambda \bar{z}}$, with $\bar{z}=70$ cm representing the
mean hologram recording distance (and with the filter cutoff smoothed
using a Gaussian with $\sigma=2.5$ pixels). Combined with the
``cameraman'' magnitude image, this highpass-filtered phase image
serves as a modified ground-truth object when evaluating multi-distance
hologram reconstructions.  We then incorporated several types of
measurement error into our simulated hologram recordings:
\begin{enumerate}

\item When recording holograms at multiple distances, it is possible
  for the translation stage of the detector system (typically a
  scintillator screen followed by a microscope objective and a visible
  light camera) to introduce slight translation and orientation errors
  between the recordings, and a scaling error can arise from imperfect
  knowledge of the source-to-object and object-to-detector distances
  used to provide geometric magnification.  These misalignment factors
  can be collectively described and refined by \adorym{} as an affine
  transformation matrix.  We therefore
  applied random affine transformations to the second, third, and
  fourth holograms, resulting in misalignment in translation, tilt,
  and non-uniform scaling.  Fig.~\ref{fig:cameraman_mdh_raw_affine}(c)
  shows the distortion that would result if the same affine
  transformations were applied to full $512\times 512$ \micron{}
  hologram recordings in Cartesian coordinates.  To measure the error
  in the recovered inverse affine transformation matrix $\mataa_{r}$
  versus the actual affine matrix $\mataa_{0}$, we used a vector
  $\vecr_{0}=(1,1)$ in a
  Euclidian distance metric $\daffine$ of
  \begin{equation}
    \daffine = \frac{|\mataa_{r}\mataa_{0}\vecr_{0}
      - \vecr_{0}|}{|\vecr_0|}
    \label{eqn:daffine}
  \end{equation}
  which has a value of zero if the recovered affine matrix matches the
  actual matrix.

\item In order to test refinement of propagation distances, we gave
  \adorym{} deliberately erroneous starting values of $z=38$, 58, 78
  and 98~cm for the hologram recording distances, rather than the
  actual distances of $z=40$, 60, 80, and 100~cm.

\item When indicated, we also added Poisson-distributed noise
  corresponding to $\bar{n}=4000$ or $\bar{n}=400$ incident photons
  per pixel, so that each of the four holograms had an average of
  either 1000 or 100 photons per pixel before accounting for
  absorption in the object.

\end{enumerate}
These errors were incorporated into the set of simulated holograms,
with two of the four holograms shown in
Figs.~\ref{fig:cameraman_mdh_raw_affine}(e) and (f). We then set the
object size to $512\times 512\times 1$, and ran 1000 epochs of
reconstruction using \torch{} as the GPU-enabled backend on our workstation ``Godzilla.''
This led to the results shown in Fig.~\ref{fig:cameraman_mdh_recons}.
With both projection affine transformation refinement and propagation
distance refinement turned on with the LSQ loss function
(Eq.~\ref{eqn:lsq_loss}), a clear and sharp magnitude image was
obtained as shown in Fig.~\ref{fig:cameraman_mdh_recons}(a).  The
propagation distances were refined to $z=39.7$, 59.7, 79.9, and
100.1~cm, which are close to the true values of $z=40$, 60, 80, and
100~cm, especially considering the 7.3~cm depth of focus.  When we instead used the Poisson loss function
(Eq.~\ref{eqn:poisson_loss}), we obtained the reconstructed image of
Fig.~\ref{fig:cameraman_mdh_recons}(b), which is as good as the
LSQ result; in addition, the refined hologram
distances of $z=39.9$, 59.9, 80.0, and 100.0~cm were also very accurate.
When we turned off affine transformation refinement while using LSQ,
the reconstruction was significantly degraded, showing both
significant magnitude/phase crosstalk and also fringe artifacts as
shown in Fig.~\ref{fig:cameraman_mdh_recons}(c).  When we turned off
distance refinement but kept transformation refinement, the
reconstructed images showed higher fidelity than in the
distance-refinement-only case, as shown in
Fig.~\ref{fig:cameraman_mdh_recons}(d).  This indicates that affine
transformation refinement plays a crucial role in this example.

\begin{figure}
  \centerline{\includegraphics[width=0.65\textwidth]{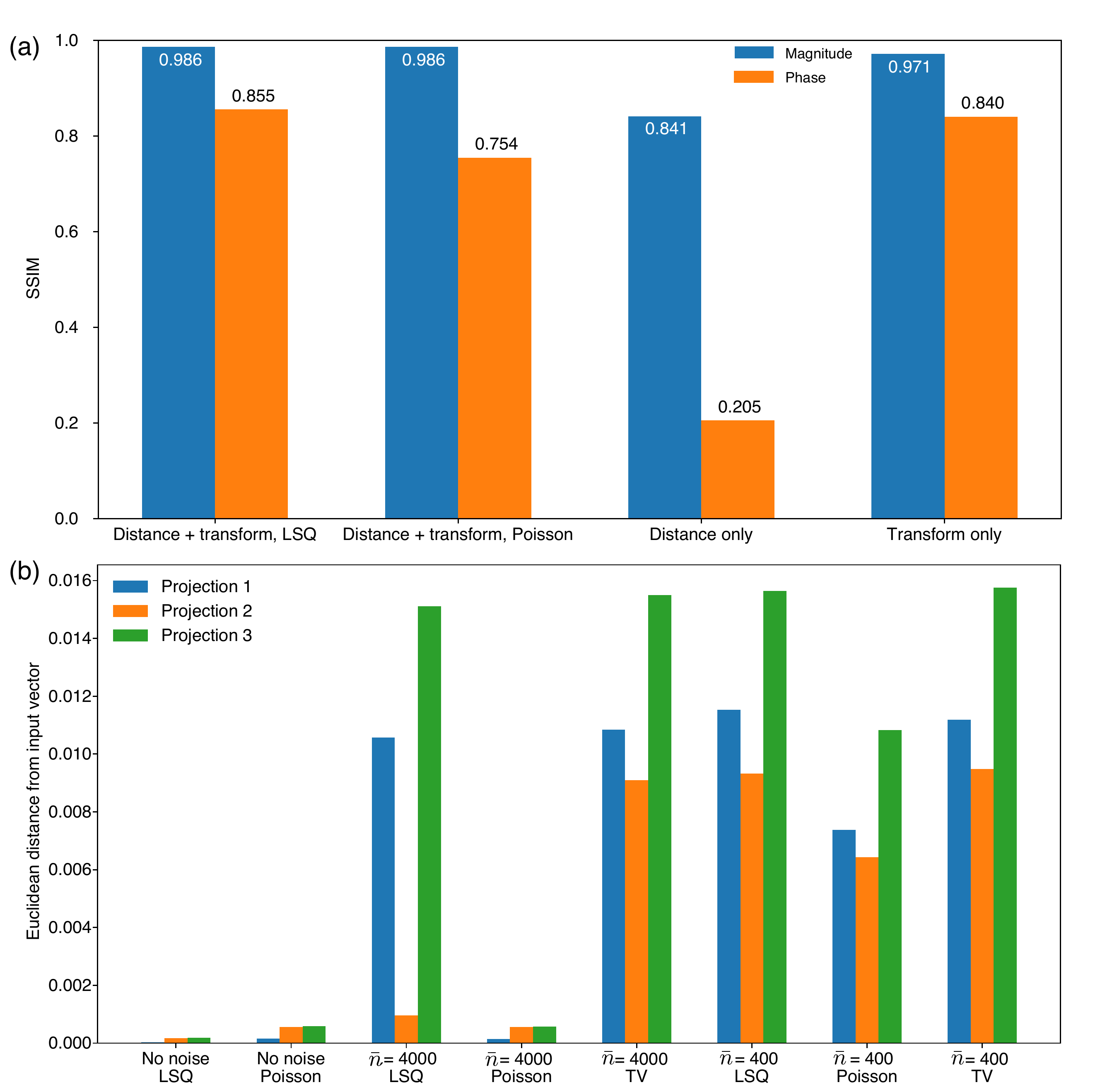}}
  \caption{(a) Structural similarity indices (SSIMs) between ground
    truth and various reconstructed images from noiseless data. The reconstructed phase
    contrast images were compared against the highpass filtered phase
    image shown in Fig.~\ref{fig:cameraman_mdh_raw_affine}(d).  (b)
    Euclidian distance error $\daffine$ (Eq.~\ref{eqn:daffine}) for
    the recovered affine transformation matrix $\mataa_{r}$, under
    different choices for loss function and regularizer, and different
    fluences $\bar{n}$.}
\label{fig:cameraman_mdh_error}
\end{figure}

In order to quantitatively compare these outcomes, we calculated the
structural similarity indices (SSIMs) of the various reconstructed
magnitude and phase images in comparison with the modified ground
truth object as described above. Readers could consult the cited paper, or
the supplementary material, for the detailed definition. We used the magnitude
image of Fig.~\ref{fig:cameraman_mdh_raw_affine}(a) and the highpass
filtered phase image of Fig.~\ref{fig:cameraman_mdh_raw_affine}(d) as
ground truth images, with the phase image normalized using its mean
$\bar{\varphi}$ and standard deviation $\sigma_{\varphi}$ as
$\varphi^{\prime}=(\varphi-\bar{\varphi})/\sigma_{\varphi}$ so as to
eliminate phase offsets and scalings from the SSIM calculation.  The
results of SSIM comparisons of these ground truth images with various
reconstructions are presented in
Fig.~\ref{fig:cameraman_mdh_error}(a), showing significant improvement
with affine transformation refinement turned on. The influence of
distance error, on the other hand, is relatively small.

We now consider a more realistic situation where the measured images
are subject to photon noise. For this, the holograms were created with
Poisson noise assuming two incident photon fluence levels:
$\bar{n}=4000$ photons per pixel total over the 4 holograms, and
$\bar{n}=400$; examples of 2 of the 4 holograms at each fluence are
shown in Figs.~\ref{fig:cameraman_mdh_raw_affine}(e) and (f). For each
noise level, we repeated the reconstruction using both LSQ and Poisson
loss functions, with both transformation and distance refinement
turned on. In addition to that, we also include in our test scenarios
the use of LSQ along with a total variation (TV) regularization
described in Table S1 of the supplementary material (TV regularization
is often used for noise suppression by guiding the reconstruction
algorithm towards a spatially smoother solution
\cite{grasmair_amo_2010,kamilov_ieeetci_2016}). The TV regularizer
weighting hyperparameter was set to $\gamma=0.01$, since higher
weightings resulted in ``blocky'' looking reconstructions with
significant loss of high-frequency information. From the results shown
in Fig.~\ref{fig:cameraman_mdh_recons}, we can see that the Poisson
loss function leads to the best visual appearance for both fluence levels.
The addition of the TV term
reduces the contrast of crosstalk artifacts between phase and magnitude, but it also affects
high-spatial-frequency features.

For distance refinement under noisy imaging conditions, LSQ+TV yielded
the worst performance: the reconstruction from dataset with
$\bar{n}=4000$ total photons/pixel ended up with refined distances of $z=37.8$, 45.2,
107.1, and 116.1~cm.  The average distance error of LSQ+TV at
$\bar{n}=400$ is about 20\%{} lower than that
with $\bar{n}=4000$, which is counterintuitive.  Because distance
refinement is highly dependent on the gradient yielded from the
mismatch of Fresnel diffraction fringes, it may be incompatible with
the preference for smoothing provided by TV.  Use of LSQ and Poisson
error functions without TV regularization work well: for $\bar{n}=400$ photons/pixel total,
LSQ yielded $z=38.8$, 56.1, 79.0, and 101.8~cm, while Poisson gave
$z=39.2$, 59.2, 79.1, and 99.1~cm. With $\bar{n}=4000$ photons/pixel,
distance refinement using Poisson is almost as accurate as the noise
free case.

Now we compare the results of affine transformation refinement for both
noise-free and noisy cases in Fig.~\ref{fig:cameraman_mdh_error}(b).
For noise-free data, the
transformation refinement results are almost the exact inverse of
the original distorting matrix, giving a normalized Euclidean error
(Eq.~\ref{eqn:daffine}) of $\daffine\simeq 10^{-3}$ which, for an
image less than 512 pixels across, corresponds to a residual error of
less than one pixel. When noise is present, using affine refinement
together with Poisson loss function leads to significant improvements.

\subsection{Experimental multidistance holography with material homogeneity constraint}

\begin{figure}[t!]
  \centering \includegraphics[width=0.9\textwidth]{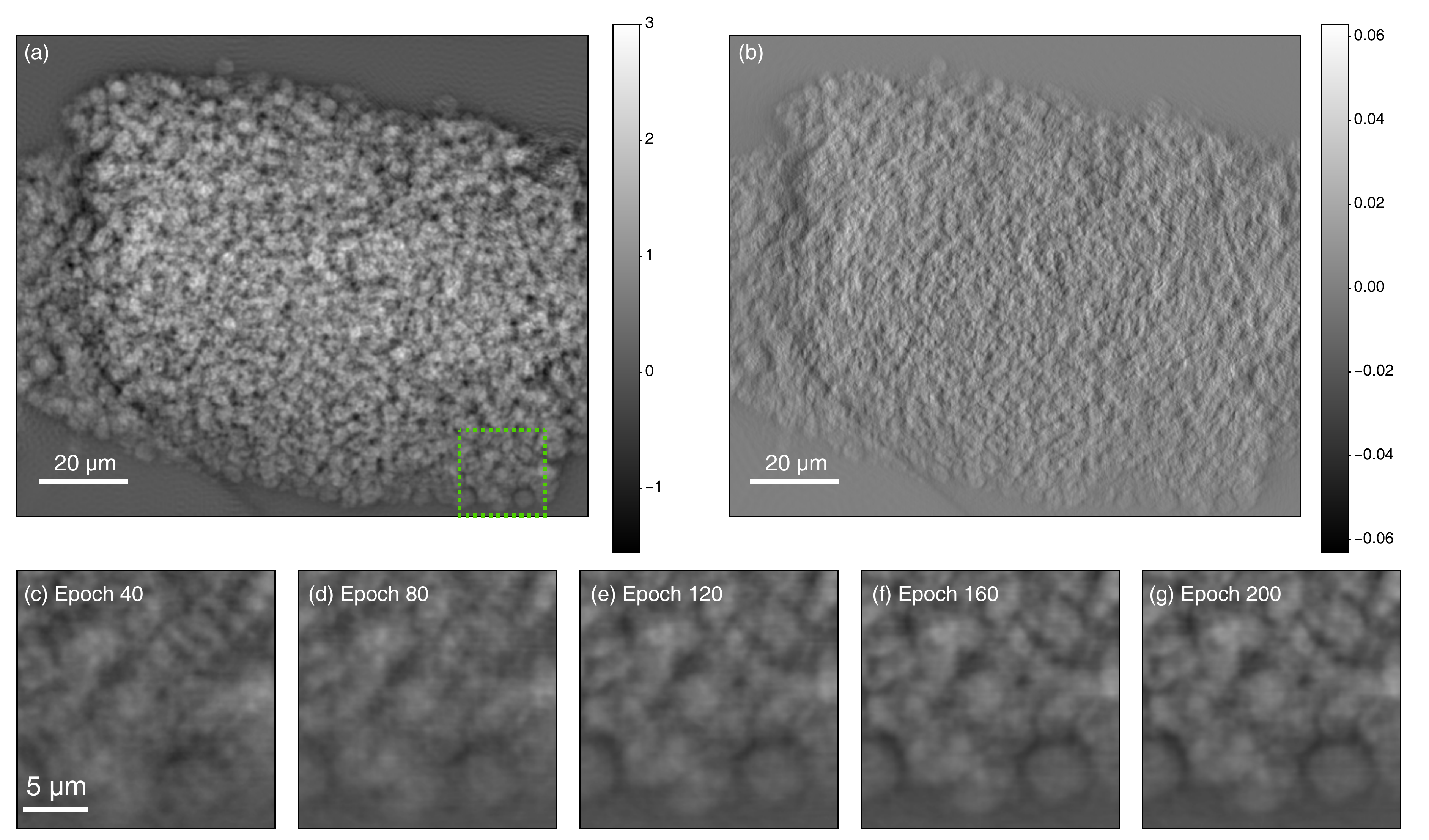}
  \caption{Phase retrieval and tomographic reconstruction results of
    multi-distance holography (MDH) data of a Li-O$_{2}$ battery
    electrode.  (a) Phase contrast projection image 
    retrieved with \adorym{} using 3 holograms at slightly
    different distances.  (b) A reference reconstruction obtained with
    the contrast transfer function (CTF) algorithm, using the same
    data as (a) plus an additional hologram at another distance.
    (c-g) Evolution of the reconstruction for a region indicated by
    the green dashed box in (a) from epoch 40 to 200, sampled every 40
    epochs. Misalignment of holograms produces errors that are more
    severe away from the image center, so that the patch at epoch 40
    exhibits obvious feature duplication.  With transformation
    refinement, the feature duplication gradually diminishes
    throughout the reconstruction process.
    \cite{su_acsaem_2020}.}
\label{fig:esrf_mdh}
\end{figure}

The numerical experiment shown in Sec.~\ref{sec:cameraman_mdh} assumed
an extreme case where the magnitude and phase of the object modulation
function are completely uncorrelated. In reality, it is more often the
case that these two parts show significant spatial correlation, and
this can be exploited as an additional constraint. As noted above, if
one can make the heterogeneous material assumption of
$\delta(\vecr) = (1 / \kappa)\beta(\vecr)$, then both the contrast
transfer function (CTF) approach as well as a transport of intensity
equation (TIE) based approach can be used in a linear approximation to
relax the requirement on the diversity of Fresnel recording distances
\cite{turner_optexp_2004}.  This assumption of fixed $\kappa$ can also
be incorporated into the Fresnel diffraction model of \adorym{}, so it is
our forward model of choice for multi-distance holography (MDH).

We used Adorym to reconstruct a MDH dataset of a battery electrode
that contained porous lithium peroxide (Li$_2$O$_2$) formed as the
product of a cathodic reaction. The sample was prepared following the
procedure used for a transmission x-ray microscope study of the same
type of materials \cite{su_acsaem_2020}. The MDH data were collected
at the ID16B beamline of the European Synchrotron Radiation Facility
(ESRF) using a beam energy of 29.6 keV and a point-projection
microscope setup. The source-to-detector distance was fixed at about
75 cm, and holograms were collected at four sample-to-detector
distances so as to minimize zero-crossings in the CTF phase retrieval
equation \cite{zabler_rsi_2005}.  We used only three of the four
holograms for our Adorym reconstruction in order to challenge it with
reduced information redundancy. The selected sample-to-detector
distances are $z_{\text{sd}}=69.61$, 69.51, and 69.11~cm, which convert to
effective distances of $z_{\text{eff}}=5.366$, 5.450, and 5.785~cm if a plane-wave
illumination were to have been used based on the Fresnel scaling
theorem \cite{paganin_2006}.
The holograms were rescaled so that they have the same pixel size of 50 nm.
Before
reconstruction, the holograms were coarsely aligned using phase
correlation. Reconstruction of the phase image was
performed first on the workstation ``Godzilla'' which was run for 200
epochs with $\kappa$ initialized to be 0.01.  After this,
refinement was enabled for $\kappa$, propagation distance, and
hologram affine transformation. The resulting phase contrast image is
shown in Fig.~\ref{fig:esrf_mdh}(a), and performance metrics are shown
in Table \ref{tab:perf}.  For comparison purposes, we also obtained a
phase contrast image using the CTF algorithm in the weak phase and
weak absorption assumption, and where the
pre-reconstruction alignment was done separately for both translation
and tilt; this image is shown in Fig.~\ref{fig:esrf_mdh}(b).  In
comparing these two images, the \adorym{} reconstruction of
Fig.~\ref{fig:esrf_mdh}(a) allows one to
distinguish many individual particles even at the center of the
deposition cluster where the particles are most densely stacked. In
addition, because the raw holograms are slightly misaligned in tilt
angle while the simple phase correlation method we used only fixes
translational misalignment, our input images to \adorym{} were well
aligned at the center but not near the edges. The affine
transformation refinement feature of \adorym{} was able to effectively
fix the misalignment issue: in the final result of
Fig.~\ref{fig:esrf_mdh}(a), we do not see any duplicated ``ghost
features'' that would typically come from the misalignment of multiple
transmission images taken from the same viewing angle. Rather, the
reconstructed phase contrast image shows good contrast, and standalone particles at
the boundary of the deposited Li$_2$O$_2$ cluster appear to be well
defined. Minor fringe artifacts remain visible in the empty region,
which can be attributed to the slight deviation of the sample from the
homogeneous object assumption. By comparing the intermediate phase map
at different stages of the reconstruction process, we do observe the
effect of misalignment far away from the image center.
Fig.~\ref{fig:esrf_mdh}(b-f) tracks the evolution of a local region
near the lower right corner of the phase map from epoch 40 to 200; the
location of the region in the whole image is indicated by the green
dashed box in Fig.~\ref{fig:esrf_mdh}(a). At epoch 40, this part of
the phase map is severely impacted by feature duplication resulting
from hologram misalignment. As the reconstruction continues, the
correcting transformation matrices of the second and third hologram are
refined further, and feature duplication keeps getting reduced (the
first hologram was used as the reference upon which the other two were
aligned).  The refinement process also adjusted the
phase-absorption correlation coefficient $\kappa$ from
0.01 to 0.1, with convergence achieved before epoch 40; the
supplied propagation distances on the other hand were already
sufficiently accurate and the refinement did not significantly alter
their values.

\subsection{2D fly-scan ptychography with probe position refinement}
\label{sec:2idd}

While multidistance holography does not require any beam scanning,
ptychography has emerged as a powerful imaging technique for
extended objects with no optics-imposed limits on spatial resolution
and no approximations required on specimen material homogeneity
\cite{faulkner_prl_2004,rodenburg_prl_2007}.  However, practical
experiments often encounter two complications.  The first is that the
probe positions (the scanned positions of the spatially-limited
coherent illumination spot) might be imperfectly controlled or known;
this error can be corrected for using either correlative
alignment-based
\cite{maiden_ultramic_2012,zhang_optexp_2013,tripathi_optexp_2014} or
optimization-based
\cite{guizar_optexp_2008,odstrcil_optexp_2018,dwivedi_ultramic_2018}
approaches.  The second is that the probe is often in motion during
the recording of a diffraction pattern in a so-called ``fly scan'' or
continuous motion scanning approach; one can used mixed-state
pychographic reconstruction methods \cite{thibault_nature_2013} to
compensate for probe motion in ptychography
\cite{clark_optlett_2014,pelz_apl_2014,deng_optexp_2015,huang_scirep_2015}.
One can also correct for both complications together in one
optimization-based approach \cite{odstrcil_optexp_2018}.  We show here
that \adorym{} can handle both issues without explicitly implementing
the closed-form gradient.

\begin{figure}[t!]
  \centering \includegraphics[width=0.7\textwidth]{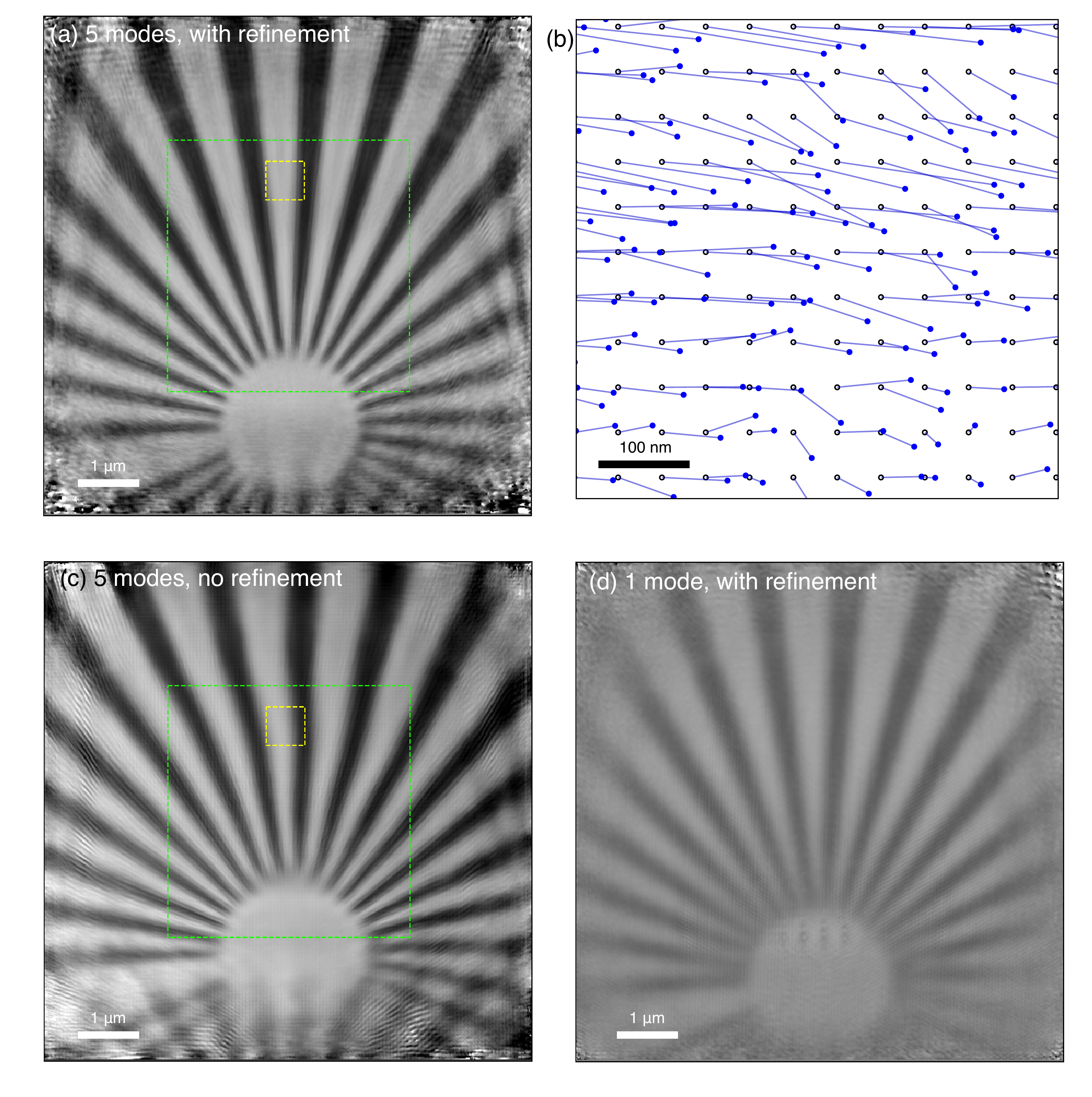}
  \caption{Ptychography reconstruction results of the Siemens star
    sample with deliberately-included errors in actual probe
    position. (a) Reconstruction with 5 probe modes, and probe
    position refinement. The area covered by the scan grid is shown in
    the green dashed boxes; the larger reconstructed image area arises
    due to the large finite size of the illumination probe. In (b), we
    show the probe position refinement results within the region
    bounded by the smaller yellow box of (a); here the assumed
    positions are represented by open dots on a square grid, with the
    refined positions shown by blue dots. Also shown is the
    reconstruction with probe modes included but no position
    refinement (c), or with position refinement included but no probe
    modes (d).  All reconstructed images are shown over the same
    dynamic range. Comparing the yellow boxes in (a) and (c), one can
    see that the bounded segment of the Siemens star spoke in (c) is
    shifted slightly to the left. Thus, the refined probe positions
    are mostly to the right of the original positions within that
    window, which agrees with (b).}
  \label{fig:2idd_posopt}
\end{figure}

A Siemens star test pattern of 500 nm thick gold was imaged using a
scanning x-ray microscope at the 2-ID-D beamline of the Advanced Photon Source
(APS).  A double-crystal monochromator was used to deliver a 8.8 keV
x-ray beam with a bandwidth of
$\Delta E / E \approx 1.4 \times 10^{-4}$ and an estimated flux in the
focus of $1\times 10^{9}$ photons/s. The beam was focused on the
sample using a gold zone plate with 160 \micron~diameter, 70 nm
outermost zone width, and 700--750 nm thickness; a 30
\micron~diameter pinhole combined with a 60 \micron~beam stop is placed
downstream serving as an order-selecting
aperture to isolate the first diffraction order focus. This produces a
modified Airy pattern at the focus, and a ring-shaped illumination
pattern at far upstream and downstream planes. The detector
was placed 1.75 m downstream the sample, which resulted in a
sample-plane pixel size of 13.2 nm. A $70\times 70$ scan grid with a
step size about 48 nm was used to collect 4900 diffraction patterns in
fly-scan mode, each with an exposure time of 50 ms.  A Dectris Eiger
500K detector was used to collect the diffraction patterns, and all
images were cropped to $256\times 256$ pixels before reconstruction.
This dataset included deliberate errors in illumination probe
positions for a demonstration of position refinement using an
established approach \cite{deng_spie_2020}.

We used \adorym{} on the workstation ``Godzilla''
(Sec.~\ref{sec:platforms}) to reconstruct this dataset to yield an
object of $618 \times 606 \times 1$ pixels, adding 50 pixels on each
side to accommodate specimen regions that were illuminated by the
tails of the probe. The probe was initialized using an
aperture-defocusing approach: a $256\times 256$ annular wavefront with
an outer radius of 10 pixels and an inner radius of 5 pixels was first
created to mimic the ring-shaped aperture used in experiment. It was
then defocused by 70 \micron~using Fresnel propagation, which
constituted the initial guess for the first probe mode; the other modes
were seeded with duplicates of the first mode with Gaussian noise
added to create perturbations. To accommodate the fly-scan scheme, we
used 5 probe modes, and we used the Fourier shift
method to optimize each probe position as described in
Sec.~\ref{sec:forward_model}. To prevent drifting of
the entire image, we subtracted the mean value of all probe positions
from the set after they were updated at the end of each iteration.
With both multi-mode reconstruction and probe position refinement
turned on, we obtained the reconstructed phase map shown in
Fig.~\ref{fig:2idd_posopt}(a), which has had the surrounding empty
regions cropped out. The reconstruction appears sharp, and the spokes
of the Siemens star stay straight as expected for this sample
fabricated using a precision electron beam lithography system.  The
image quality remains high even beyond the rectangular region of the
main scan (indicated by the green dashed box) due to the tails of the
beam, with decreasing quality only seen at the outer edges due to
reduced fluence.  To illustrate probe position correction, we show in
Fig.~\ref{fig:2idd_posopt}(b) the initial guesses of the probe
positions (which were known to be in disagreement with the actual
positions \cite{deng_spie_2020}) as hollow black circles, with lines
connecting to solid blue circles representing the refined positions.
If instead we
turn off probe position refinement but still use 5 probe modes, the
spokes of the Siemens star in the reconstructed image appear
distorted, as shown in Fig.~\ref{fig:2idd_posopt}(c).  For the whole
image, the mean distance by which the refined probe positions are
moved is 6.24 pixels or about 80 nm, which can be seen by the shift of the
yellow-boxed region between Fig.~\ref{fig:2idd_posopt}(a) and (c).
Alternatively, if we keep probe position refinement but use only 1
probe mode, we end up with the result of Fig.~\ref{fig:2idd_posopt}(d)
yielding straight spokes as in Fig.~\ref{fig:2idd_posopt}(c), but with
lower image resolution due to the probe motion being incorporated into
the image.  In addition, the central part
of the test pattern exhibits lower
contrast, indicating a increased reconstruction error at high
spatial frequencies.

\subsection{2D multislice ptychography}

\begin{figure}[tbp]
  \centering \includegraphics[width=0.7\textwidth]{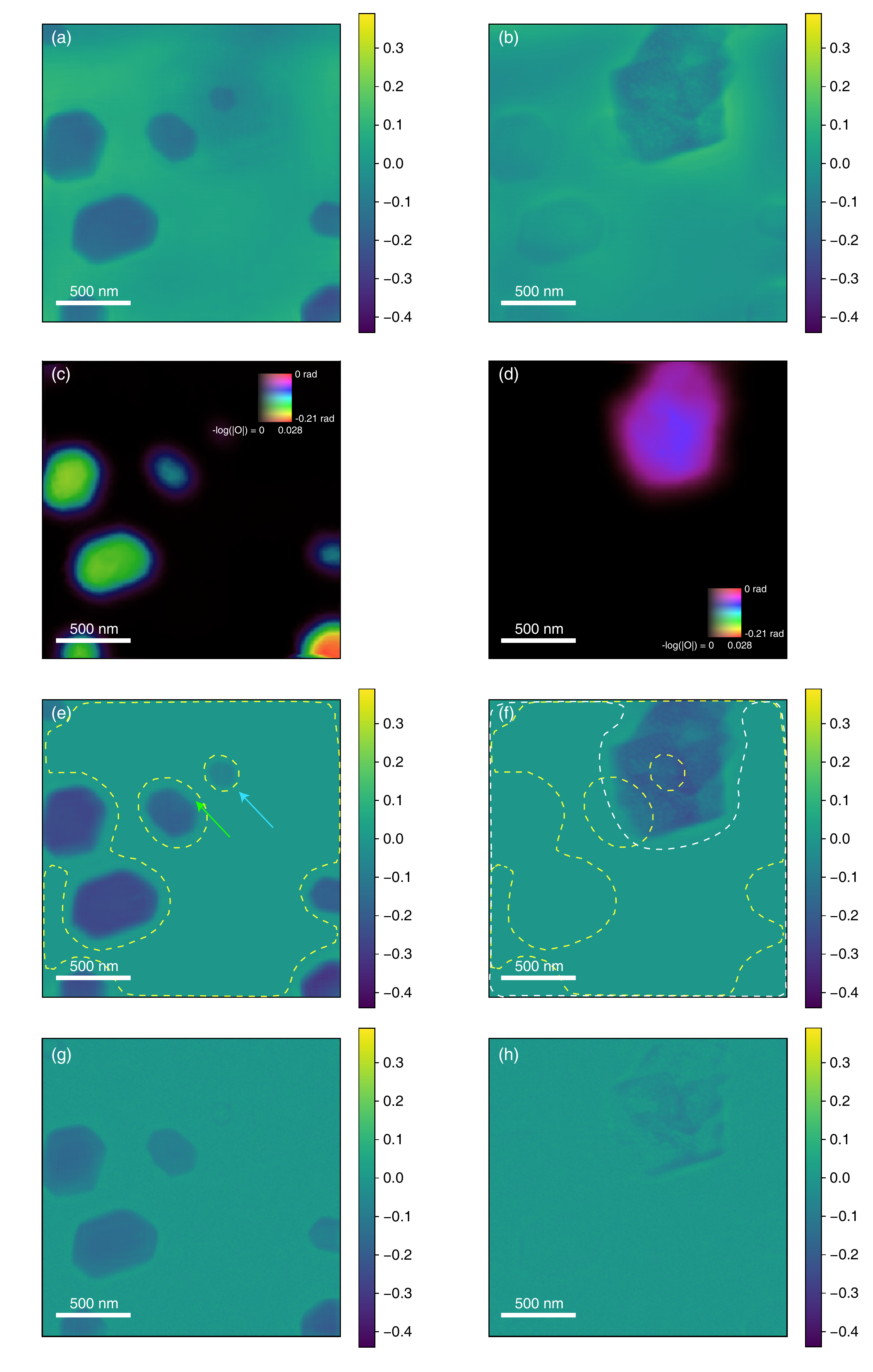}
  \caption{Reconstruction results of a multislice ptychography dataset
    \cite{ozturk_optica_2018} consisting of gold (upstream sample
    surface; left panels) and nickel (downstream sample surface; right
    panels) nanoflakes on either side of a 10 \micron{} thick silicon
    window. The initial retrieved phase images shown in (a) and (b)
    are from slice 1 and 2, respectively, where features appear to be
    sharp and clearly defined, yet with crosstalk at low spatial
    frequencies due to separation by only a small multiple of the 3.9
    \micron{} depth of focus of the illuminating optic. To improve the
    reconstruction, simultaneously-obtained x-ray fluorescence data was
    used to provide an initial estimate at lower resolution of the
    magnitude (whose minus-logarithm is coded by brightness)
    and phase (coded by hue) of each
    slice, as shown in (c) and (d).  With this improved starting
    guess, as well as a finite support constraint derived from the
    x-ray fluorescence maps, we obtained improved reconstructions as
    shown in (e) and (f) with the finite support constraint boundaries
    shown as dashed lines (yellow for Au, and white for Ni).  An
    alternative approach without finite support is to impose sparsity by incorporating a reweighted
    $l_1$-norm regularizer leading to images (g) and (h); in these
    images, crosstalk is suppressed at the cost of degrading the
    contrast and sharpness of the reconstructed images.}
  \label{fig:multislice_ptycho}
\end{figure}

\adorym{} can also handle a mixture of near-field and far-field
propagation in the forward model.  While we have shown that automatic
differentiation can be used for multislice ptychotomography
\cite{du_sciadv_2020}, we show here the simpler case of multislice
ptychography of reconstructing several planes at different distances
from the probe focus using data collected from a single viewing angle
\cite{maiden_josaa_2012,tsai_optexp_2016b}. We term it ``sparse multislice ptychography''
to distinguish it from the multislice ptychotomography in \cite{du_sciadv_2020}
which reconstructs a dense and continuous object. For
validation of our approach, we used a dataset that has already been
analyzed using SMP
\cite{ozturk_optica_2018}.  The specimen is of Au and Ni nanoflakes on
two sides of a 10 \micron{} thick silicon window (the silicon window
was largely transparent at the 12 keV photon energy used).  The
ptychography dataset consisted of 1414 $128\times 128$ diffraction
patterns acquired using a 12 nm FWHM beam focus from a multilayer Laue
lens (MLL), with the specimen located 20 \micron{} downstream of the
focus (the depth of focus for this optic of 3.9 \micron{}). The sample
plane pixel size is 7.3 nm. The
object shape in \adorym{} was $501\times 500\times 2$, where the last
dimension represents the number of slices.  In previous work, the
probe function reconstructed from a ptychographic dataset with only a
gold particle in the field of view was used as the initial probe
function \cite{ozturk_optica_2018}; with \adorym{}, we used the
inverse Fourier transform of the average magnitude from all
diffraction patterns to initialize the probe function.  While slice
spacing refinement was enabled in this reconstruction, the spacing was
known sufficiently well from both the previous SMP reconstruction, and
from electron microscopy, that no significant adjustment was made from
the 10 \micron{} separation specified at reconstruction
initiation. The \adorym{} reconstruction was run for 50 epochs with the
probe function fixed, after which it was run for an additional 1950
epochs with probe optimization, all using the Adam optimizer. This led to a reconstructed phase of both slices
shown in Fig.~\ref{fig:multislice_ptycho}(a) and (b), which closely
match the SMP results reported previously \cite{ozturk_optica_2018}.

Both this reconstruction and the previous reconstruction
\cite{ozturk_optica_2018} show low-spatial-frequency crosstalk between
the two slices, which is a well-known issue \cite{tsai_optexp_2016b}
in the case where the separation between slices is not much larger
than the depth-of-focus.  For this particular sample with known
differences in material between the two planes, one can use x-ray
fluorescence maps acquired simultaneously with the ptychography data
to greatly reduce this crosstalk \cite{huang_actacrysta_2019}.  We
employed an approach in \adorym{} of creating finite support
constraints based on the x-ray fluorescence images.  We first
convolved the x-ray fluorescence images (Au $L_\alpha$ for slice 1,
and Ni $K_\alpha$ for slice 2) with a Gaussian with
$\sigma=5$ pixels or about 36.5 nm,
and then creating a binary mask by thresholding the XRF maps using a
value determined through k-means clustering \cite{dhanachandra_proccs_2015} implemented
in the \scipy{} package \cite{scipy}. We also used the XRF maps to
generate the initial guess for the object function. For this,
the XRF images were normalized to their maxima, and
multiplied with an approximate phase shifting value estimated from the
refractive indices of Au and Ni. The magnitude and phase guesses for
both slices are shown in Fig.~\ref{fig:multislice_ptycho}(c) and
(d). These guessed slices roughly match the shape of the reconstructed
nanoflakes in Fig.~\ref{fig:multislice_ptycho}(a) and (b), and do not
contain any crosstalk from the other slice, but they also do not show
detail beyond that enabled by the focusing optic, unlike the case of a
ptychographic reconstruction.  Therefore, a subsequent ptychographic
reconstruction of 2000 epochs was performed using \adorym{} with the
addition of finite support constraint masks at each epoch, yielding
the crosstalk-free images of Fig.~\ref{fig:multislice_ptycho}(e) and
(f) with the boundaries of the respective finite support constraint
masks indicated by dashed lines. Additionally, the ``gap'' region
of the finite support in Fig.~\ref{fig:multislice_ptycho}(e), indicated
by the green arrow, is uniform and clean. This is not achievable if one
simply multiplies the support mask with (a).
This reconstruction also reveals a
small Au flake on the upstream plane which was previously obscured by
larger, higher contrast Ni flake on the downstream plane. Using the
XRF-derived initial image guesses alone did not lead to much
improvement over the images (a) and (b); it was mainly the finite
support constraint that led to the improved results of images (e) and
(f).

The loss-function based approach of \adorym{} allows one to test other
reconstruction approaches with ease.  We therefore tested out the use
of a reweighted $l_{1}$-norm regularizer \cite{candes_jfaa_2008} as a
Tikhonov regularizer \cite{tikhonov_dansssr_1963} in the
reconstruction, without using the finite support constraint.  The reweighted $l_1$-norm seeks to suppress weak pixel
values, and can therefore take advantage of the initial guess of the
object.  The reconstruction results using $\alpha_1 = \alpha_2 = 0.01$
as regularizer weights for planes 1 and 2 respectively are shown in
Fig.~\ref{fig:multislice_ptycho}(g) and (h). For slice 1, the
regularizer is indeed effective in suppressing the crosstalk coming
from the large Ni flake on slice 2. However, there are disadvantages:
the small Au flake pointed to with a the blue arrow in
Fig.~\ref{fig:multislice_ptycho}(e) becomes hardly visible in (g)
because its values were overly penalized. On slice 2, the Ni flake was
correctly reconstructed in terms of its overall shape, but at lower
contrast and a loss of internal detail so that it had a ``hollow''
appearance.  The $l_{1}$-norm reconstruction reduces the presence of
ghost images when comparing Figs.~\ref{fig:multislice_ptycho}(b) and
(h), and the images show increased high-spatial-frequency ``salt and
pepper'' noise. This type of noise comes from the XRF-derived
initial guess, which contains a slight scattering-based signal even in
empty regions, and the noise level of these pixels are slightly
magnified by the reweighted $l_1$-norm during reconstruction since
low-voxel pixels were subject to heavier suppression while high-value
pixels could grow in value.

The ability of \adorym{} to explore a wide range of constraints in an
automatic-differentiation-based image reconstruction approach is
illustrated by the above examples.  One can try other regularizers
beyond what is demonstrated here, change the forward models, and alter
the workflow in a flexible manner.

\subsection{Joint ptychotomography}

\begin{figure}[t!]
  \centering \includegraphics[height=0.75\textheight]{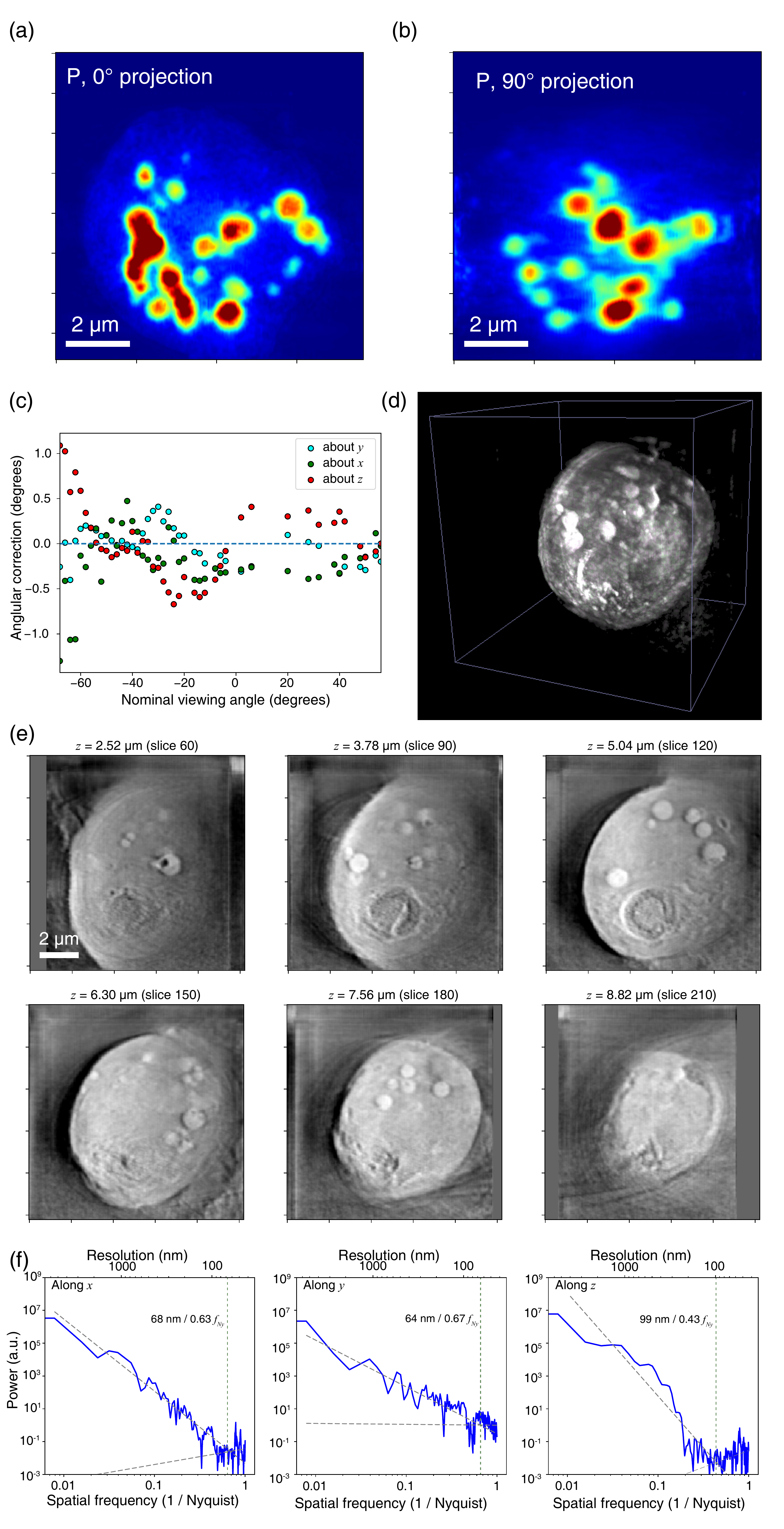}
  \caption{\adorym{} reconstruction of a tomographic x-ray
    fluorescence and ptychography dataset of a frozen hydrated alga cell
    \emph{Chlamydomonas reinhardtii} acquired using 5.5 keV X rays
    \cite{deng_sciadv_2018}.  The phosphorus $K$ fluorescence
    projection (a) at 0$\degree{}$ through the 3D volume highlights
    polyphosphate bodies, which are also seen at somewhat lower
    resolution in the 90$\degree{}$ projection through the
    reconstruction due to the limited tomographic tilt range of
    -68$\degree{}$ to +56$\degree{}$.  As in a previously-reported
    reconstruction of this dataset \cite{deng_sciadv_2018}, this
    phosphorus reconstruction used refinement of the tomographic tilt
    angles, leading to angular corrections about all three axes as
    shown in (c).  These refined angles were then employed in a
    subsequent ptychotomography reconstruction, yielding a 3D volume
    rendered in (d). Sub-images in (e) show optical sections from the
    reconstructed volume cut at indicated $z$-positions
    after it is rotated about the $y$-axis by
    30$\degree$. In (f), the power spectra of the reconstructions in
    the $x$-, $y$-, and $z$-direction are shown. For each case, two lines
    with the same color are fitted respectively using datapoints with
    spatial frequency in the range of 0.008--0.31$\fny$ and
    0.47--1.0$\fny$. As the latter fits the noise plateau, the
    intersection between both lines (marked by dotted vertical lines)
    provides a measure of spatial resolution. }
\label{fig:algae}
\end{figure}

Ptychotomography was originally done in a two-step fashion, where
ptychography was done to obtain high-resolution 2D projection images
at each viewing angle, and those projections were then used in a
conventional tomographic reconstruction algorithm to obtain a 3D
volume \cite{dierolf_nature_2010}. More recently, a joint
ptychotomography approach has been developed where the 3D object is
directly reconstructed from the set of diffraction patterns acquired
over scanned positions and rotations \cite{gursoy_optlett_2017}.  This
approach can relax the probe position overlap requirements in 2D
ptychography at one rotation angle, since there is sufficient
information overlap in position and angle to obtain a 3D
reconstruction as demonstrated in simulations
\cite{gursoy_optlett_2017} and in experiments
\cite{kahnt_optica_2019}.

We show here a joint ptychotomography reconstruction obtained using
\adorym{} to analyze previously-published data \cite{deng_sciadv_2018}
from the frozen-hydrated single-cell algae \textit{Chlamydomonas
  reinhardtii}. This data was acquired using the Bionanoprobe at the
Advanced Photon Source at Argonne \cite{chen_jsr_2014} at 5.5 keV
photon energy and a Fresnel zone plate with an approximate Rayleigh
resolution of 90 nm.  2D scans were acquired using 50 nm spacing in
continuous scan mode, with typically 38,000 diffraction patterns acquired to give a
field of view of nearly (10 \micron{})$^{2}$.  An Hitachi Vortex
ME-4 detector mounted at 90$\degree{}$ was used to record the x-ray
fluorescence signal from intrinsic elements, and a Dectris Pilatus
100K hybrid pixel array detector recorded the ptychographic
diffraction data with the center $256^{2}$ pixels used in the
subsequent reconstruction with voxel size of 10.2 nm. 2D images were
recorded at a total of 63 angles of a tilt range of -68$\degree{}$ to
+56$\degree{}$. The original
reconstruction in \cite{deng_sciadv_2018} was done following the
two-step method, using PtychoLib \cite{nashed_optexp_2014} for 2D
ptychographic reconstruction, and GENFIRE \cite{pryor_scirep_2017}, a
Fourier iterative reconstruction algorithm, for tomographic
reconstruction. The projection images used by GENFIRE were twice
downsampled so that the pixel size was about 21 nm, and the 3D
reconstruction was done using projections in a range of
124$\degree{}$.

Since x-ray fluorescence (XRF) data are available along with the
ptychographic diffraction patterns, we can use them to do a
preliminary cross-angle alignment before conducting
reconstruction. Similar to \cite{deng_sciadv_2018}, we estimated the
positional offsets at different viewing angles using the P-channel XRF
maps which have the best signal-to-noise ratio. Before alignment, we
performed a series of quick (3 epochs) 2D ptychography reconstructions
for all viewing angles using downsampled ptychographic data, and
selected 43 viewing angles from -68$\degree{}$ to 56$\degree{}$ whose
2D reconstructions exhibited acceptable quality; the data quality for
the rest is worse due to beam profile fluctuation.  Correspondingly, P
maps from these 43 angles were used in our own processing.  The
pre-alignment was performed using the iterative reprojection method
\cite{gursoy_scirep_2017} implemented in TomoPy
\cite{gursoy_jsr_2014}.  After acquiring the alignment data, we first
did a tomography reconstruction on the aligned P-channel XRF maps
using \adorym{} in order to examine the consistency with the images
shown in the original report \cite{deng_sciadv_2018}. The
reconstruction was run for 50 epochs using the CG optimizer.  The
reconstructed P map projected through the volume is shown at
0$\degree{}$ (a) and 90$\degree{}$ (b) in Fig.~\ref{fig:algae}(a) (the
corresponding images obtained previously are Figs.~2A and 4G in
\cite{deng_sciadv_2018}).  During reconstruction, we allowed \adorym{}
to refine both the cross-angle positional offsets, and the tilt angles
of the object along all 3 axes at each nominal specimen tilt about
$y$.  As the pre-alignment provided by iterative reprojection was
already highly accurate, the alignment refinement returned only
marginal corrections. On the other hand, the tilt refinement yielded
interesting results: as shown in Fig.~\ref{fig:algae}(c), the change
of $x$- and $z$-axis tilt corrections are relatively continuous with
the nominal viewing angle, instead of showing random fluctuations
around 0$\degree$. This is the expected outcome when the actual
rotation axis is not strictly vertical, or when it precesses
continuously during data acquisition.

We then applied the results of tilt refinement and the
reprojection-based pre-alignment to the joint ptychotomography
reconstruction.  Since joint ptychotomography reconstruction is more
computationally intensive than tomography, we downsampled the size of
diffraction patterns and the object function by 4 times, where all
diffraction patterns were cropped to $64\times 64$ (while the original
2D ptychography reconstruction reported in \cite{deng_sciadv_2018}
cropped the diffraction patterns to $256\times 256$ but tomography reconstruction
in that work downsampled the phase maps by twice), and the object
size was set to $256 \times 256 \times 256$ with a voxel size of 42 nm.
We used 5 probe modes to
account for fly-scan data acquisition; this probe function is commonly
used and jointly updated
by all specimen tilt angles.  Although this may not properly account
for beam fluctuation during acquisition, holding onto the same probe
further improves the utilization of information coupling among
different angles, which is favorable under the condition of angular
undersampling \cite{kahnt_optica_2019}.  The reconstruction was run
for 3 epochs using the Adam optimizer on the ``Godzilla'' workstation,
using the \torch{} backend with GPU acceleration. Using the Adam
optimizer, the learning rate was set to $1\times 10^{-6}$.  The probe
function was initialized to be the one retrieved from an individual 2D
ptychographic reconstruction, but was constantly optimized during
reconstruction to account for the fluctuation of beam profile
throughout acquisition. The probe and all refined parameters were held
fixed for the first 100 minibatches, and then optimized until the
reconstruction finished.

The ptychotomography reconstruction results are shown in
Fig.~\ref{fig:algae}(d) and (e).  To generate (d), the reconstructed
object array was thresholded with voxels below the threshold assigned
with zero opacity, so that the rendered 3D volume reveals a sharp
spherical outline of the cell, with the internal organelles clearly
visible. Next, we rotated the 3D array about the $y$-axis by
30$\degree{}$ (so that its orientation matches what is written as a
60$\degree{}$ orientation in Fig.~2D of \cite{deng_sciadv_2018}). We
then show 6 $x$--$y$ plane optical slices at indicated $z$ positions
in this new orientation in Fig.~\ref{fig:algae}(e). In order to
estimate the reconstructed spatial resolution, we also plot its power
spectra along all the 3 spatial axes in Fig.~\ref{fig:algae}(f).  Each
power spectrum is expected to be composed of 2 segments: a power-law
decline in from low to middle spatial frequencies of 0.008$\fny$ to
0.31$\fny$, and a more level section at higher frequencies of
0.47$\fny$ to 1$\fny$ corresponding to noise uncorrelated from one
pixel to the next (where $\fny$ is spatial frequency corresponding to
the Nyquist sampling limit).  The intersection between the two fitting
lines represents the point where high-frequency noise starts to
dominate, and can be considered as an approximate measure of the resolution.
With this, we found the average $xy$-resolution
is approximately 70 nm, and the $z$-resolution is approximately 100 nm.
The previous reconstruction \cite{deng_sciadv_2018} reports the resolutions
to be 45 and 55 nm, respectively; however, Given
that we downsampled the data twice more in our case, the joint
ptychotomography capability of \adorym{} is able to provide reasonably
good reconstruction quality compared to the more computationally
intensive iterative method of image reconstruction steps followed by
reprojection alignment steps.

\subsection{Conventional tomography on HPC}
\label{sec:tomography}

\adorym{}'s optimization framework can also work with conventional
tomography. When using the LSQ loss function and GD optimizer, \adorym{}'s
tomography reconstruction is equivalent to the basic algebraic reconstruction
technique (ART) \cite{kak_1988}, but more advanced optimizers like
Adam can be used to accelerate the reconstruction.

The tomographic data to be reconstructed was an activated charcoal
dataset \cite{vescovi_jsr_2018}. The original data were collected at
the APS 32-ID beamline using a beam energy of 25 keV, filtered using a
crystal monochromator, and the images were recorded by a CMOS camera
($1920 \times 1200$ GS3-U3-23S6M-C) after the x-ray beam was converted
into visible light by a LuAG:Ce scintillator.  This resulted in a
sample-plane pixel size of 600 nm.  Since the size of the pellet
(about 4 mm in diameter) was larger than the detector's field of view,
full rotation sets were taken with the beam located at different
offsets from the center of rotation in a \textit{Tomosaic} approach
\cite{vescovi_jsr_2018}. Each assembled projection image was
$4204 \times 6612$ pixels in size. For this demonstration, we used 900
projections evenly distributed between 0 and 180$\degree{}$,
downsampled the projections by 4 times in both $x$ and $y$, and then
selected the first 800 lines after downsampling for each projection
angle, so that the size of each projection image is $800 \times 1653$
pixels with an angular sampling that is coarser than required by the
Crowther criterion \cite{crowther_prsa_1970}. Before reconstruction,
all projection images were divided into tiles that are $25\times 52$
in size, so each viewing angle contained a tile grid of $32\times 32$,
or 1024 tiles in total.

Tomographic reconstruction was performed on the supercomputer
``Theta'' (Sec.~\ref{sec:platforms}).  We used 256 nodes with 4 MPI
ranks per node, so that the total number of ranks exactly matched the
number of tiles per angle. We used \adorym{}'s distributed object (DO)
scheme to parallelized the reconstruction; in this case, each of first
800 MPI ranks saved one slice of the object, and the corresponding
gradient, with a horizontal size $1653\times 1653$. The remaining 224
ranks did not serve as containers for object or gradient slabs, but
were still responsible for calculating the gradients of their assigned
tiles.  Since each tile contained 25 pixels in the $y$ direction,
assembling an object chunk on each rank required the incoming transfer
of 25 ranks (including itself); this rank, at the same time, also sent
parts of its own object slab to 32 ranks, which is the number of tiles
horizontally across the object.  Therefore the array size that a rank
sent to another rank is $1 \times 52 \times 1653 \times 2$, where the
last dimension is for both $\delta(\vecr)$ and $\beta(\vecr)$
(although only $\beta(\vecr)$ is reconstructed in conventional
absorption tomography). With this configuration, the \texttt{AlltoAll}
MPI communication for object chunk assembly took 0.7 s on average. The
gradient synchronization, which is essentially the reverse process and
requires each rank to send a $25 \times 52 \times 1653 \times 2$
gradient array to 25 ranks, took an average of 0.5 s.  Using a LSQ
loss function and the Adam optimizer, gradient calculation and object
update took about 0.7 s and 0.3 s on average. Along with rotation (1 s
for both object rotation and gradient rotation) and other overheads,
each batch (which involved 1024 tiles or exactly one viewing angle)
took roughly 5.8 s to process, excluding the time spent for
occasionally saving checkpoints for the object and other parameters.

\begin{figure}
  \centerline{\includegraphics[width=0.98\textwidth]{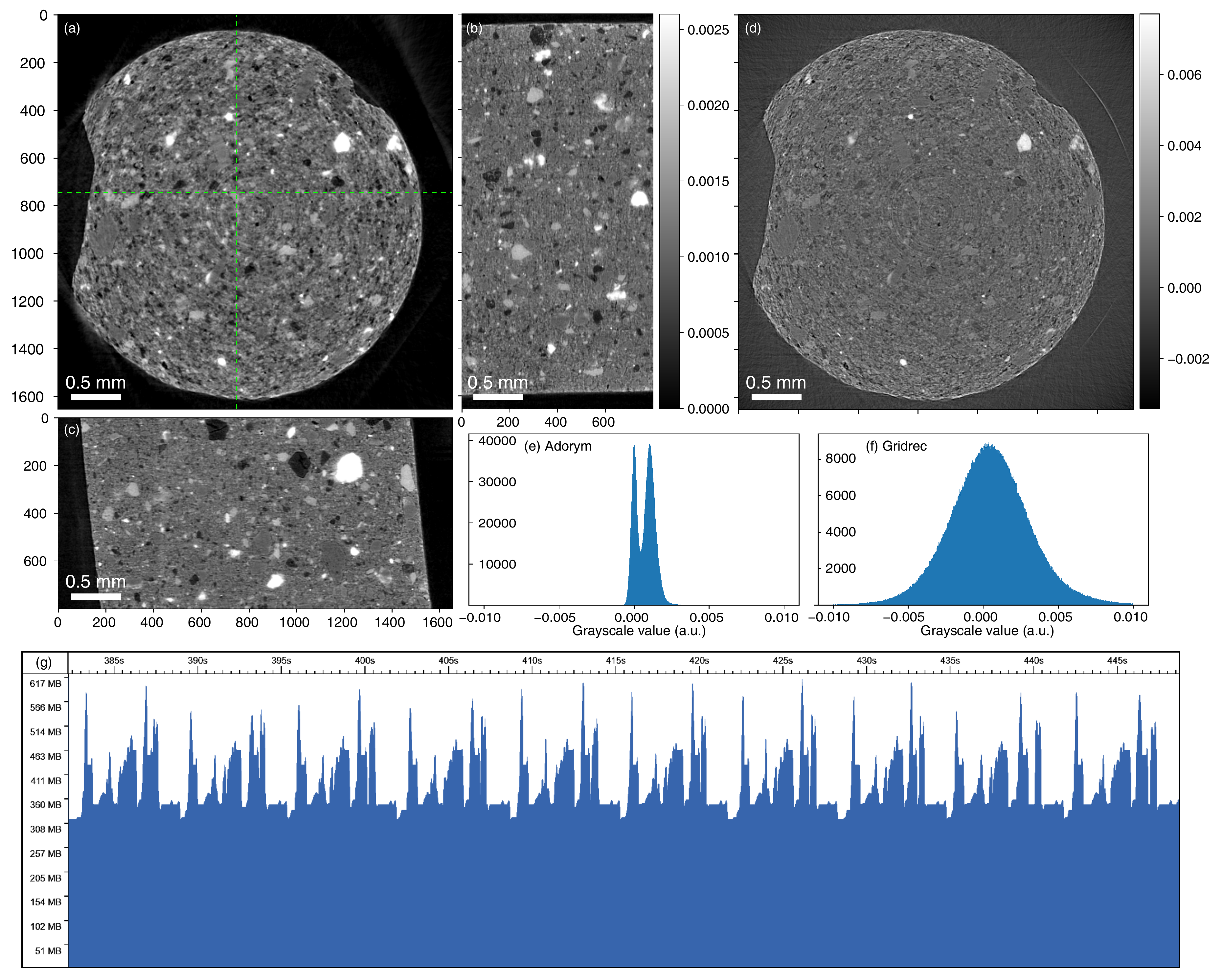}}
  \caption{Reconstruction of conventional projection tomography data
    from an activated charcoal pellet \cite{vescovi_jsr_2018}.  In
    (a-c), we show slices from the 3D volume reconstructed by
    \adorym{}: (a) shows the horizontal cross section cut at slice
    400, and (b, c) show the cross sections cut from locations
    indicated by the green dashed lines. In (d), we show the same
    slice as in (a) but reconstructed using conventional filtered
    backprojection (FBP), exhibiting much lower contrast.  The
    histogram of the \adorym{} reconstruction slice (a), and the
    filtered backprojection slice (d), are shown in (e) and (f)
    respectively; they indicate that \adorym{} better represents the
    density differences between different features in the sample.  In
    (g), we show the memory profile during 10 minibatches of \adorym{}
    reconstruction.}
  \label{fig:charcoal}
\end{figure}

After 3 epochs of reconstruction, we obtained the 3D volume whose
orthogonal slices are shown in Fig.~\ref{fig:charcoal}, where
sub-image (a) displays the $xz$-cross section cut from the vertical
center of the volume (slice 400 out of the 800
slices). Figures~\ref{fig:charcoal}(b) and (c) respectively show the
$xy$- and $yz$-cross sections cut at the positions indicated by the
dashed lines in (a). Since Crowther criterion \cite{crowther_prsa_1970}
requires $1653\pi=5193$ projection angles for complete coverage in
reciprocal space, while we used only 900 angles for this
reconstruction, it is to be expected that our ART-like approach would
outperform filtered backprojection methods such as \textit{Gridrec}
\cite{osullivan_ieeetmi_1985,dowd_spie3772,marone_jsr_2012}, with a
\textit{Gridrec} reconstructed middle slice shown in
Fig.~\ref{fig:charcoal}(d).  Comparing Fig.~\ref{fig:charcoal}(a) and
(d), we observed that the result of \adorym{} shows a much better
contrast than \textit{Gridrec}. The difference is made clearer when we
compare the histograms of the reconstructed slice. The histogram of
optical densities in the \adorym{}'s reconstruction shown in
Fig.~\ref{fig:charcoal}(e) exhibits 2 distinct peaks, which correspond
to the vacuum background surrounding the charcoal pellet and the
pellet itself. The background peak on the left is much sharper than
the charcoal peak, indicating that the background is uniformly
reconstructed and numerically well separated from the non-vacuum
features. The histogram of the \textit{Gridrec} result shown in
Fig.~\ref{fig:charcoal}(f), on the other hand, exhibits only one broad
Gaussian peak, meaning that the background and the sample features are
somewhat intermingled.

We utilized the Intel VTune$^{\text{TM}}$ profiler available on ``Theta'' to
obtain the memory profile of \adorym{} during
runtime. Fig.~\ref{fig:charcoal}(g) shows the memory consumption of
rank 0 through a 66-second window of the reconstruction process, where
10 minibatches were computed sequentially (the average per-batch
walltime was slightly longer than 5.8 s due to the added overhead of
VTune itself). The profile shows obvious periodicity, with each period
corresponding to a minibatch. The two highest peaks of around 650 MB
in each minibatch are caused by rotating the object slice and the
gradient slice, respectively, as both rotation operations involve
considerable intermediate memory usage (contributed by loading the
transformation coordinates, performing bilinear interpolation, and so
on).  The other peaks result from gradient calculations and the update
step of Adam, which involves several intermediate variables that have
the same size as the object array. Beyond that, the baseline memory is
jointly comprised of the stored object array, gradient array, Adam
parameter arrays, and auxiliary data including the tile allocation
over all ranks. The overall memory usage is steady, and no memory
leakage is found. If data parallelism was used, each rank would then
require at least 70 GB of memory to store the whole object function
and the associated gradient and optimizer parameters, not to mention
the additional usage caused by intermediate and auxiliary variables.
The reconstruction times are included in Table~\ref{tab:perf}.

\begin{table}
  \centering
  \scriptsize
  \begin{tabular}{lllllll}
    \hline
    \textbf{Case}                  & \textbf{Backend} & \textbf{Platform} & \textbf{Tile size}        & \makecell[l]{\textbf{Number of} \\ \textbf{tiles per batch}} & \makecell[l]{\textbf{Mean walltime} \\ \textbf{per batch (s)}} & \textbf{Data size (GB)}\\
    \hline
    MDH (simulation)               & \torch{}          & Godzilla     & $512\times 512$                             & \makecell[tl]{4 \\ (1 set of \\ 4 distances)} & 0.08 & 0.004  \\
    MDH (experimental)             & \torch{}          & Godzilla      & $2160\times 2560$                           & \makecell[tl]{3 \\ (1 set of \\ 3 distances)} & 1.4 & 166 (all angles)  \\
    2D ptychography w/ pos. refin. & \torch{}          & Godzilla      & \makecell[tl]{$256\times 256$ \\ (5 modes)} & 35                                            & 0.4 & 2.6 \\
    Sparse MS ptychography         & \torch{}          & Godzilla      & $128\times 128$                             & 101                                           & 0.3 & 0.19 \\
    Ptychotomography               & \torch{}          & Godzilla      & \makecell[tl]{$64\times 64$ \\ (5 modes)}   & 50                                            & \makecell[tl]{5.2/2.2 \\ (w/wo.~tilt refin.)} & 29.5 \\
    Tomography with DO             & \autograd{}       & Theta            & $25\times 52$                               & 1                                             & 5.8 & 4.8 \\
    \hline
  \end{tabular}
  \caption{Performance data of all test cases shown in
    Sec.~\ref{sec:results}, using the compute platforms described in
    Sec.~\ref{sec:platforms}.  Walltimes shown in seconds do not
    include the time spent for saving intermediate results, or for
    providing diagnostic checkpoint results after each epoch.}
  \label{tab:perf}
\end{table}

\section{Discussion}
\label{sec:discussion}

We have shown that \adorym{} provides a reconstruction framework for
several imaging methods, including multi-distance near-field
holography (MDH), sparse multislice ptychography (SMP) and 2D
ptychography, and projection as well as ptychographic tomography.
This framework allows for the refinement of imperfectly-known
experimental parameters, resulting in substantial improvements in
reconstruction quality. The forward model class \modfm{} allows for
the treatment of ptychotomography with multiple viewing angles and
multiple images per angle; a simpler variant allows for simple
projection tomography with a per-angle tile number of 1 and a single
illumination mode.

This flexibility is made possible through the use of automatic
differentiation (AD) for gradient derivation.  While AD drastically
reduces the development cost, does it mean that the runtime speed has
to be sacrificed as a trade-off? There are several reasons to think
this might be the case:
\begin{itemize}

\item Some AD libraries (like \torch{} and \autograd{}) create
  dedicated data types for node variables, so that each of them knows
  the operations that connect it with other nodes, as well as the list
  of nodes that it is connected to. This genre of AD is known as
  ``operator overloading'' \cite{nashed_procedia_2017}, and is
  favorable for its flexibility and ease of implementation, but
  logging these interconnections at runtime means that each operation
  causes a slight additional overhead \cite{merrienboer_arxiv_2018}.

\item In manual differentiation, the expression of the gradient can be
  often simplified to reduced the number of numerical evaluations
  needed. In contrast, it is difficult for AD to fully take this
  advantage because each step in backpropagation is local to a
  primitive function.  An illustrative example is a scalar that
  contains $x\log(x)$: the derivative with regards to $x$ is
  $x(1/x) + \log(x)$, so that in a manual differentiation one would
  simplify the first term to 1 while AD would mindlessly follow the
  backpropagation procedure and evaluate both $x$ and $1/x$.
  Nevertheless, the influence of this kind of situation is usually
  minor, and one can define custom functions in many AD libraries to
  explicitly inform AD of the derivatives of these special composite
  functions.

\end{itemize}
However, the runtime penalties are not necessarily unacceptably high.
In the case of multislice x-ray ptychotomography of a $256^{3}$ voxel
simulated object that extends beyond the depth-of-focus limit, a C
code with manually derived gradients took 6.48 core hours per
illumination angle to arrive at a solution \cite{gilles_optica_2018}
using the HPC system ``Blues'' (Sec.~\ref{sec:platforms}),
while a Python code using \tensorflow{} for the AD implementation
\cite{du_sciadv_2020} took 8.25 core hours for a similar
reconstruction step using the CPUs only on the HPC system ``Cooley''
(Sec.~\ref{sec:platforms}).  This may be in part
due to AD's ``cheap derivative principle'' \cite{griewank_2008}, where
the cost of evaluating the gradient with reverse-mode AD is generally
bound to small multiple of the cost needed to evaluate the forward
model.  Moreover, well-developed AD libraries can incorporate
low-level optimizations, which can help them relative to basic manual
differentiation implementations.  For example, when working with GPUs,
\torch{} uses the CUDA stream mechanism to queue on-device
instructions to the GPU's command buffer, so that tensor operations on
GPU can be concurrent with Python interpretation on CPU
\cite{pytorch}.  The most effective way to accelerate image
reconstruction computations is to increase the degree of
parallelization and improve interprocess communication -- a principle
that applies to both AD and manually coded implementations.  Finally,
it is known that interpreted languages like Python and MATLAB might
have inferior execution speeds compared to compiled languages, but the
advent of just-in-time (JIT) compilers \cite{lam_llvm_2015} is likely
to change the situation.  We are aware of a few AD libraries which are
already using JIT, such as JAX \cite{schoenholz_arxiv_2019};
meanwhile, efforts on incorporating JIT support into \torch{} are also
ongoing \cite{merrienboer_arxiv_2018}, which \adorym{} can gain from
when it becomes available.

The memory consumption of AD is another concern, though this can be
offset in part by the use of well-developed AD libraries.  In some
cases, one might have similar storage requirements for a manual
differentiation implementation. For example, in multislice imaging
where the wavefield at the $j$-th slice, $\psi_j(\vecrxy)$, is
modulated by this object slice $O_j(\vecrxy)$, the modulated wavefield
is given by $\psi_{j+1}(\vecrxy) = \psi_j(\vecrxy)O_j(\vecrxy)$.  The
derivative of $\psi_{j+1}(\vecrxy)$ with regards to $O_j(\vecrxy)$ is
given by $\psi_j^*(\vecrxy)$. That is, to update the object slice
$O_j(\vecrxy)$, one needs to know $\psi_j(\vecrxy)$ which is
calculated in the forward pass. Both AD and manual differentiation may
inevitably choose to store $\psi_j(\vecrxy)$ during the forward pass
and reuse it for gradient calculation, especially when recalculating
$\psi_j(\vecrxy)$ in the differentiation stage is too costly. Most AD
libraries do not blindly store all intermediate variables, but only
those needed to be reused for differentiating a non-linear operation,
or a linear operation with a variable operator. 
The advanced memory management strategies in well-developed AD libraries
also help.
For example, \torch{}
features a reference counting scheme which deallocates the memory of
variables immediately when their usage counts down to zero, preventing
the extra memory consumption in the traditional garbage collection
scheme \cite{pytorch}. Additionally, while in manual differentiation
one can freely adjust the workflow of the program, it is also possible
in AD to not store some intermediate variables, but recalculate them
in the backward pass. This strategy, known as checkpointing
\cite{margossian_arxiv_2018}, leads to a tradeoff between memory
consumption and computation time.  Manually scheduled checkpointing is
available with \torch{} \cite{pytorch_docs} and can be conveniently
added into the forward model if desirable; on the other hand,
automatic scheduling of checkpointing has also been proposed and
implemented \cite{griewank_revolve}, which could potentially reach a
better balance between storage and recomputation.


AD has a profound value when developing new methods, as making
improvements on existing forward models or devising new forward models
are made much less laborious without needing to hand-derive
gradients. Beyond the parameter refinement capabilities demonstrated
in Section~\ref{sec:results}, one could also consider adding the following
example capabilities:
\begin{itemize}

\item Optimizing structured illumination function for ptychographic or
  holographic imaging.  Structured illumination is often used as a way
  to improve the resolution and reconstruction robustness of these
  techniques by enhancing the contrast transfer function of the
  imaging system, and choosing the optimal illumination function for a
  certain imaging technique or sample condition is usually of interest
  \cite{stockmar_scirep_2013,odstrcil_optexp_2019a,ching_arxiv_2020}.
  With AD, one possible way to get an optimal illumination function
  would be to add the structured modulator into the forward model as
  an optimizable variable, and set the loss function to measure the
  disparity between the predicted intensity and the desired far-field
  intensity distribution.

\item Speckle tracking. For imaging techniques like point-projection
  microscopy, where x-ray beams are focused using optics such as
  multi-layer Laue lenses, it is often important to accurately
  retrieve the illuminating wavefield in order to reconstruct the
  phase images with good quality. When the reconstruction algorithm is
  based on the generalized Fresnel-scaling theorem
  \cite{morgan_arxiv_2020}, the forward model involves resampling of
  the measured intensity following a function without a closed-form
  expression.  While an optimization-based speckle tracking algorithm
  applicable in this case has been proposed \cite{morgan_arxiv_2020},
  AD can potentially address the problem using a similar strategy, but
  without the need of hand-deriving and implementing the
  differentiation through the resampling operation.  One could instead
  use the \texttt{grid\_sample} function in \torch{} for a
  well-optimized implementation.

\item In-situ 3D imaging. In many cases, it is of interest to track
  the structural evolution over time inside a 3D sample. This means
  that conventional tomography reconstruction algorithms must be
  modified to account for the continuous deformation of the sample
  during acquisition. One way to address this issue is to represent
  the time-space dependence of the object function as the linear
  superposition of a series of spatial basis functions, where the
  superposition coefficients depends only on time
  \cite{nikitin_ieeetrans_2018}. With AD, this strategy can be
  conveniently implemented by adding the time-dependent coefficient as
  a refinable variable. Alternatively, AD may also solve the
  time-dependency by refining a vector field deformer that dictates
  the distortion of the object function
  \cite{odstrcil_natcomm_2019}. This approach involves grid-based
  resampling, but as mentioned above, AD libraries like \torch{}
  provide easy-to-use interfaces for this operation. This application
  is very closely related to the optical flow problem in computer
  vision, where AD has been demonstrated to provide good performance
  \cite{shen_arxiv_2019}.

\end{itemize}
In \adorym{}, these additional capabilities can be achieved by modifying
or creating new forward models.

In addition to extending towards more forward models, another field of
future improvement for \adorym{} or any optimization-based
reconstruction methods is the underlying optimization algorithm.
First-order methods like Adam and CG are fast and robust in many
cases, but they are not advantageous compared to second-order methods
when the loss is a well structured convex function.  Second order
algorithms based on Gauss-Newton approximation provide a relatively
easy and low-cost way to utilize the curvature of the loss function,
with the Curveball algorithm \cite{henriques_arxiv_2018} being a
particularly light-weight variant. As Gauss-Newton methods approximate
the Hessian matrix using the ``internal'' Jacobian of the forward
model (\textit{i.e.}, the derivatives of the prediction function), the
power of AD can be again exploited in this case to calculate the
desired vector-Jacobian products and Jacobian-vector products without
manual derivation. Finally, since the capability of AD is often part
of common machine learning tools like \torch{}, it is also easy to
seamlessly combine a physical model and a neural network into one
pipeline and optimize the parameters in both parts. This has already
been demonstrated in full-field coherent diffraction imaging
\cite{chan_arxiv_2020}.

AD toolkits are often built and optimized by operators of
supercomputer systems, allowing an \adorym{} user to piggyback on
those efforts. Thus the DO parallelization scheme described in
Sec.~\ref{sec:do_mode} can provide a well-supported approach to
reconstructing gigavoxel datasets on HPCs.

\section{Conclusion}

We have developed \adorym{} as a generic image reconstruction
framework that utilizes the power of automatic differentiation. It
works with a variety of imaging techniques, ranging from near-field 2D
holography to far-field 3D ptychographic tomography.  It is able to
refine a series of experimental parameters along reconstructing the
object.  It has a modular architecture so that it can be extended to
cover new forward models or refinable parameters without lengthy
manual derivations of gradients.  It can be operated on platforms
ranging from workstations to HPCs, and comes with several
parallelization schemes that provide different balances between
computational overhead and memory usage. Future efforts will be aimed
at improving speed and memory consumption.  Adding JIT support, and
incorporating ``smarter'' checkpointing strategies, are both potential
approaches towards this objective.

\section*{Acknowledgement}

The authors would like to thank Taylor Childers
and Corey J. Adams for their support in the scaling and benchmarking
of \adorym{} on ALCF Theta. The authors would also like to thank
Sajid Ali for helpful discussions on parallel array operations on
Theta and on the distribution of the software package.

\section*{Funding}

This research used resources of the Advanced Photon Source (APS) and
the Argonne Leadership Computing Facility (ALCF), which are
U.S. Department of Energy (DOE) Office of Science User Facilities
operated for the DOE Office of Science by Argonne National Laboratory
under Contract No. DE-AC02-06CH11357.
It also used 3ID of the National Synchrotron Light Source II,
a U.S. Department of Energy (DOE) Office of Science User Facility
operated for the DOE Office of Science by Brookhaven National
Laboratory under Contract No. DE-SC0012704.
We also thank the Laboratory
Computing Resource Center (LCRC) at Argonne for additional
computational access. We thank the Argonne Laboratory Directed Research
and Development for support under grant 2019-0441.
We also thank the National Institute of Mental
Health, National Institutes of Health, for support under grant R01
MH115265.

\section*{Disclosure}

The authors declare no conflicts of interest.

\bibliographystyle{unsrt}
\bibliography{mybib}

\end{document}


\maketitle

\begin{abstract}
We describe and demonstrate an optimization-based x-ray image
reconstruction framework called \adorym{}. Our framework provides a generic
forward model, allowing one code framework to be used for a wide range
of imaging methods ranging from near-field holography to and fly-scan
ptychographic tomography.  By using automatic differentiation for
optimization, \adorym{} has the flexibility to refine experimental
parameters including probe positions, multiple hologram alignment, and
object tilts.  It is written with strong support for parallel
processing, allowing large datasets to be processed on high-performance
computing systems.  We demonstrate its use on several experimental
datasets to show improved image quality through parameter refinement.
\end{abstract}

\tableofcontents

\section{Modules and components}

\begin{figure}
  \centerline{\includegraphics[width=0.98\textwidth]{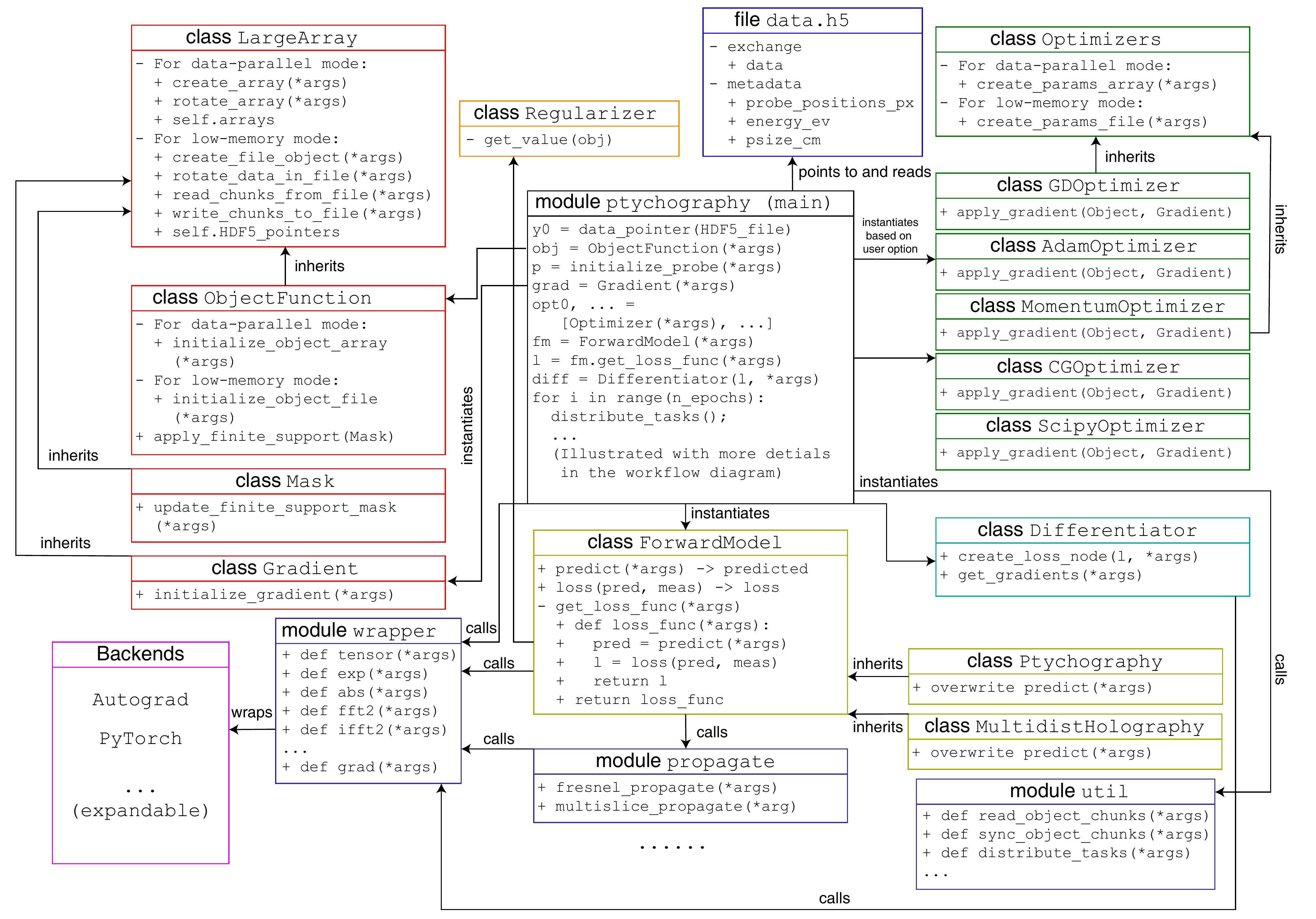}}
  \caption{Architecture of Adorym, listing the interrelation of
    all modules, classes, and child classes.  This is the same figure
    as Fig.~1 in the main manuscript.}
  \label{fig:block_chart}
\end{figure}

As noted in Section 2 of the main manuscript, \adorym{} is written in
a modular and object-oriented programming style, consisting primarily
of the following classes:
\begin{itemize}

\item The \modfm{} class discussed in Sec.~\ref{ssec:forward_model},
  which contains the forward model corresponding to a particular
  imaging method.

\item The \modopt{} class discussed in Sec.~\ref{ssec:class_optimizer},
  which provides options on what optimization method to use for
  minimizing the loss function $L$.

\item The \modla{} class discussed in Sec.~\ref{ssec:class_largearray},
  which manages the large array of the object, or $\vecxx$.

\item The \modprop{} class discussed in Sec.~\ref{ssec:propagate},
  which handles wavefield propagation for those forward models that
  require it.

\item The \modwpr{} module that provides a common interface for
  functions in the two AD engines provided: \autograd{}, and \torch{}.

\end{itemize}
Modules like \modfm{} and \modopt{} each contain several child
classes, providing support for different types of imaging techniques
or setups, numerical optimization algorithms, distributed computation
methods, and automatic differentiation backends (and computation
devices, namely CPU and GPU).  We have therefore attempted to make it
easy to combine varying detailed methods within the above categories.
The image reconstruction task is streamlined by the \modmain{} module,
which is named in such a way because it supports the 4D-type (rotation
angle, scan position, and 2D diffraction patterns) data processing
that is common in ptychography. However, this module is also capable
of handling a wide range imaging techniques other than ptychography.
A full picture illustrating the architecture design is shown in Fig.~2
of the main manuscript, which is repeated here as
Fig.~\ref{fig:block_chart}.

\subsection{Data format}
\label{ssec:data_format}

\begin{figure}
  \centerline{\includegraphics[width=0.6\textwidth]{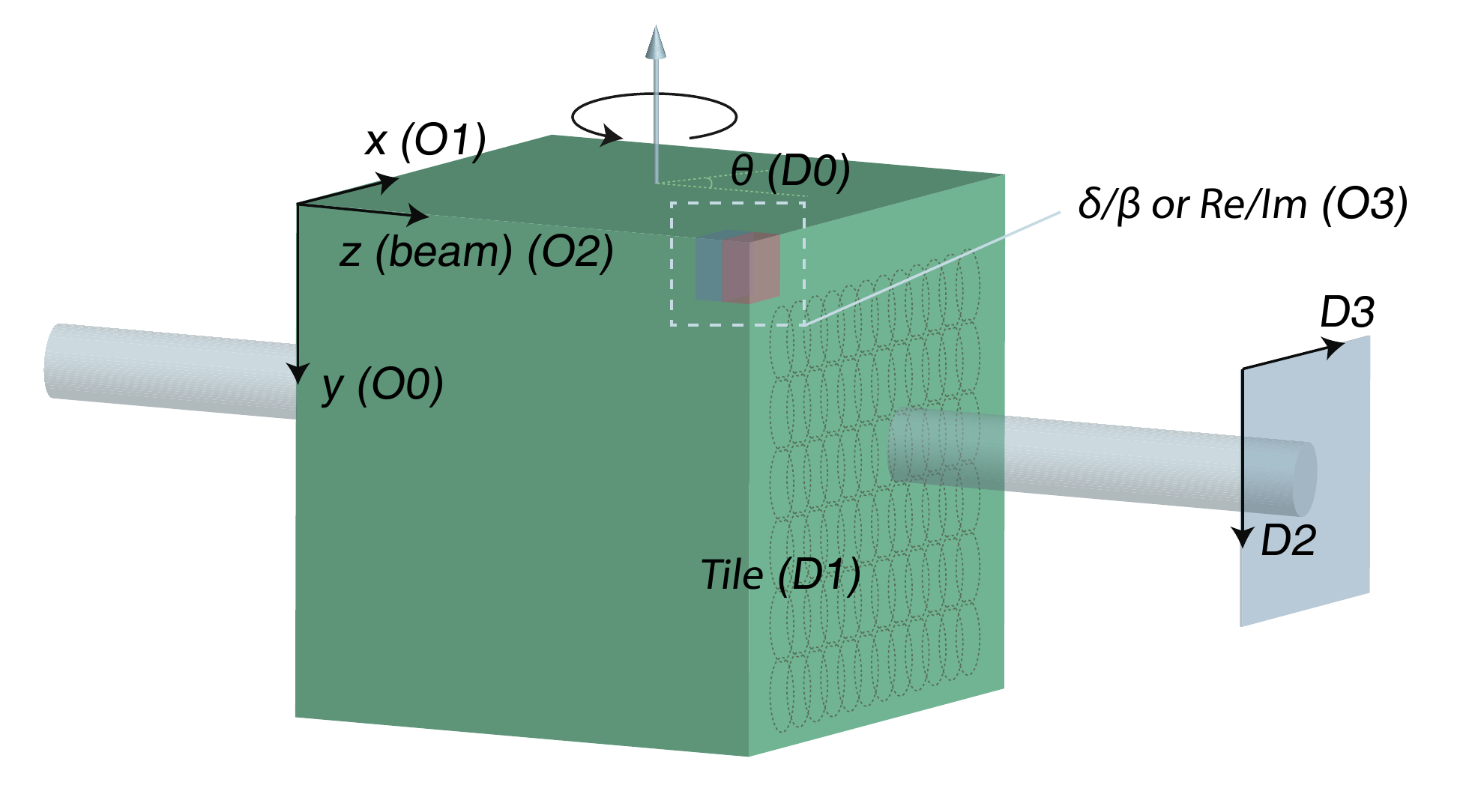}}
  \caption{Representation of experimental coordinates in \adorym{}'s
  readable dataset ($D$) and object function array ($O$). Directions
  and quantities are labeled with the index of dimension in the corresponding
  array; for example, $O2$ means that the associated object axis is
  stored as the 2nd dimension of the object array. }
  \label{fig:geo}
\end{figure}

Because of its ability to contain tagged scalar and array data, as
well as its support for parallel access, \adorym{} uses HDF5 files
\cite{hdf5} for input and output as noted in Sec.~2.1 of the main
manuscript.  Acquired images are saved as a 4D dataset \texttt{data}
under the \texttt{exchange} group \cite{decarlo_jsr_2014}, and
metadata such as beam energy, probe position, and propagation
distances are stored under the \texttt{metadata} group. In compliance
with the \modmain{} module of \adorym{}, \texttt{data} should have a
shape of \texttt{[num\_angles, num\_tiles, len\_detector\_y,
  len\_detector\_x]}, where \texttt{num\_tiles} can be the (maximum)
number of diffraction spots per angle in the case of ptychography or
ptychotomography, or the number of subdivided image ``tiles'' per
angle in the case of full-field imaging; the latter will be explained
in more details later in this section (Fig.~\ref{fig:geo}). This is a
general data format, and is compatible towards more specific imaging
types: for example, 2D ptychography has \texttt{num\_angles = 1} and
\texttt{num\_tiles > 1}, while full-field tomography has
\texttt{num\_tiles = 1} and \texttt{num\_angles > 1}. In the case of
multi-distance holography (MDH) where in-line holograms of the object
at difference sample-to-detector distances are acquired to provide
additional information for phase retrieval \cite{cloetens_apl_1999},
holograms collected from the same viewing angle but different defocus
distances are treated as different tiles --
\textit{i.e.}, they are contiguous along the second
(\texttt{num\_tiles}) dimension.

When dealing with full-field data from large detector arrays,
\adorym{} provides a script to divide each image into several
subblocks called ``tiles.''  It follows that the divided image data
can be treated in a way just like ptychography, where only a small
number of these tiles are processed each time.  This can be an
effective way of reducing per-node memory usage especially when using
GPU acceleration, and is also desirable when a large number of
processes are run in parallel, a typical scenario when using \adorym{}
on a multi-node supercomputer. For the user-given number of tiles of
$N_{\text{tile},y}$ and $N_{\text{tile},x}$, the original image of
size $L_y\times L_x$ is divided into interconnected tiles with size
given by $L_{\text{tile},y} = \ceil{L_y / N_{\text{tile},y}}$ and
$L_{\text{tile},x} = \ceil{L_x / N_{\text{tile},x}}$, where $\ceil{x}$
is the ceiling function that returns the smallest integer larger or
equal to $x$. If MDH
data are being tiled, all the tiles of images acquired at different
distances are collectively regarded as ``scan spots'', and are saved
contiguous in the second dimension of the dataset following a
``tile-then-distance'' order.

It follows from our introduction above that a ``tile'' is the
fundamental level of data organization scheme used in \adorym{} for
all range of imaging techniques.  In far-field ptychography, a tile
refers to a diffraction pattern; in full-field imaging methods, a tile
may refer to the image acquired at a certain viewing angle if the raw
data are undivided, or a subblock of the image if they are divided.
We shall thus use the word ``tile'' as a general term when referring
to these data elements.

\subsection{Module LargeArray}
\label{ssec:class_largearray}

The \modla{} class for large arrays is used to hold the object (when
it is kept or partially kept in memory; more about this will be
introduced in subsection \ref{ssec:distribution_modes}), or arrays that
are comparable in size as the object (such as the object gradient and
the finite support mask).  The parent class of \modla{} contains
common methods such as array rotation (important for processing
tomographic data), HDF5 read/write (used in HDF5-mediated distribution
mode), and MPI-based data synchronization (used in distributed object
mode). The latter two methods mentioned above are specifically
relevant to distributed reconstruction modes where the object function
is either scattered over processes or nodes, or stored in a HDF5 file
on the hard drive; distribution schemes for these cases are discussed
in Section \ref{ssec:distribution_modes}.  Three child classes, namely
\texttt{ObjectFunction}, \texttt{Mask}, and \texttt{Gradient}, inherit
properties from the parent class, and possesses additional methods
developed for their specific roles and purposes. For example, the
\texttt{ObjectFunction} child class contains methods to initialize the
object function with Gaussian random values or a user-supplied initial
guess, and the \texttt{Mask} class includes a
\texttt{update\_finite\_support} function, which revises the mask by
setting voxels associated with low object function values to zero -- a
technique known as shrink-wrapping in coherent diffraction imaging
\cite{marchesini_prb_2003}.

The unknown object to be solved is represented by an instance of the
\texttt{ObjectFunction} class: a 4D array where the first 3
dimensions correspond to the $y$ (vertical), $x$ (horizontal), and $z$
(beam axial) coordinates of the object (also see Fig.~\ref{fig:geo}
for a illustration of the geometry). The last dimension depends on
the mathematical representation of the object: \adorym{} allows the
object function to be reconstructed as either a complex modulation
function, where a wavefield $\psi(\vecr)$ is modulated by the object
$O(\vecr)$ as
\begin{equation}
  M[\psi(\vecrxy)] = \psi(\vecrxy)\cdot \left(\Re[O(\vecrxy)] + i\Im[O(\vecrxy)]\right),
  \label{eqn:complex_modulation}
\end{equation}
where $\Re(\cdot)$ and $\Im(\cdot)$ respectively take the real
and imaginary part of the argument, or a distribution of refractive indices
$n(\vecr) = 1 - \delta(\vecr) - i\beta(\vecr)$
\cite{henke_adndt_1993}, where the modulation to a input x-ray
wavefield is given as
\begin{equation}
  M[\psi(\vecrxy)] = \psi(\vecrxy)\exp[-k\beta(\vecrxy)\Delta z]\exp[ik\delta(\vecrxy)\Delta z]
  \label{eqn:ri_modulation}
\end{equation}
where $k = 2\pi / \lambda$ is the wavenumber, and $\Delta z$ is the
thickness of the modulating object. Since the object is assumed to be
a 3D function while the wavefield is 2D, we use $\vecrxy$ to represent
the lateral coordinates of a plane in the object.  In the case
of $\Re[O(\vecrxy)] + i\Im[O(\vecrxy)]$, the last dimension of the object
array stores the real and imaginary part of the complex modulation
function $O(\vecr)$, while in the case of
$\delta(\vecr)+i\beta(\vecr)$, the last dimension
stores $\delta(\vecr)$ and $\beta(\vecr)$, respectively. The complex
multiplicative modulation function representation used in
Eq.~\ref{eqn:complex_modulation} is more common in phase retrieval
techniques such as ptychography \cite{rodenburg_aiep_2008}, and we
have found it sometimes shows better numerical robustness in
optimization-based reconstruction because it avoids the
ill-conditioned exponential function.  However, since the phase angle
$\phi(\vecr)$ can only be recovered from the reconstructed complex
object function through $\mbox{Arg}[O(\vecr)]$, phase wrapping is
present in high phase contrast objects.  In that case, post-processing
is required to unwrap the phase \cite{goldstein_radiosci_1988}, but
this can be complicated when noise or strongly varying phases are
present \cite{xia_optexp_2016}. The situation is easier in the
refractive index representation of the object
(Eq.~\ref{eqn:ri_modulation}).  When the refractive indices are solved
without regularization, the phase part $\delta(\vecr)$ may still
suffer from the $2\pi$ ambiguity since it is contained in the periodic
function $\exp[ik\delta(\vecr)\Delta z]$ as in
Eq.~\ref{eqn:ri_modulation}.  However, $\delta(\vecr)$ can be
regularized to promote spatial smoothness.  That is, if the values of
$\exp[ik\delta(\vecr)\Delta z]$ at two adjacent object pixels are
$-1 + \epsilon_1$ and $-1 - \epsilon_2$ (where $\epsilon_1$ and
$\epsilon_2$ are complex numbers conjugate about the real axis), then
the corresponding value of $\delta(\vecr)$ for the former is given by
\begin{align}
  \begin{split}
    \exp[ik\delta(\vecr_1)\Delta z] &= -1 + \epsilon_1 \\
    \exp(i\pi) + i\exp(i\pi)[ik\delta(\vecr_1)\Delta z - \pi] &\approx -1 + \epsilon_1 \\
    k\delta(\vecr_1)\Delta z &= i\epsilon_1 + \pi + 2\pi m
  \end{split}
\end{align}
and similarly,
$k\delta(\vecr_2)\Delta z = i\epsilon_2 + \pi + 2\pi n$, where $m$ and
$n$ can be any arbitrary integers. It follows that
$|\delta(\vecr_2) - \delta(\vecr_1)| = \frac{1}{k\Delta z}[2\pi(n - m)
+ i(\epsilon_2 - \epsilon_1)]$ is unrestricted.  Yet, if one applies
TV regularization on $\delta(\vecr)$, then a small
$|\delta(\vecr_2) - \delta(\vecr_1)|$ is favored. While
$i(\epsilon_2 - \epsilon_1)$ is preserved by the data fidelity term of
the loss function, the TV term favors $|m - n| = 0$, which leads to a
solution that is spatially continuous without the $2\pi$ jumps in
phase wrapping.  More importantly, since the range of $\delta(\vecr)$
itself is continuous and unbounded, the solved values of
$\delta(\vecr)$ can be ones that push the phase angle
$k\delta(\vecr)\Delta z$ out of the bound of $(-\pi, \pi]$, which
would be impossible when one solves for the real and imaginary parts
of $O(\vecr)$ instead: in the latter case, even if one regularizes
$\mbox{Arg}[O(\vecr)]$, the computed phase is still bounded between
$(-\pi, \pi]$, and TV only puts a $\pi$-ceiling or $-\pi$-floor for
slightly out-of-bound pixels aside from smoothing the reconstructed
image.  By solving for $\delta(\vecr)$, one may directly obtain a
phase-unwrapped solution without any post-processing.
Furthermore, since $\delta(\vecr)$ of natural samples is
positive except near an absorption edge, this physical property can be
used to further regularize $\delta(\vecr)$ through a non-negativity constraint.
The
$\delta(\vecr)$-representation is common in direct phase retrieval
methods, and one example is the algorithm based on the contrast
transfer function (CTF) \cite{turner_optexp_2004, cloetens_apl_1999},
which derives a linear relation between the phase and the Fourier
transform of the detected intensity from Eq.~\ref{eqn:ri_modulation},
based on the assumption that the object is homogeneous and weak in
phase. Moreover, solving the refractive indices of the object is also
proven to work with iterative reconstruction algorithms in
\cite{fabian_arxiv_2020} and by us in \cite{du_sciadv_2020}.

\subsection{Module Propagate}
\label{ssec:propagate}

The \modprop{} module is essentially a toolbox for simulating
wave propagation, the central component in building the forward
model. For a wavefield propagating in vacuum (or air, without any
substantial loss of accuracy), two scenarios are considered: when the
Fresnel number of propagation is small as in the case of far-field
imaging techniques, Fraunhofer diffraction can be used to calculate
the wavefield at the destination plane, which is simply done by
Fourier transforming the source wavefield:
\begin{align}
  \funcpfar[\psi(\vecr)] &= \int{\psi(\vecr)\exp\left[i2\pi(\vecr\cdot\vecnu)\right]d\vecr} \\
                         &= \mathcal{F}[\psi(\vecr)].
                           \label{eqn:fraunhofer}
\end{align}
One thing to note here is that there exist two sign conventions in
formulating wave propagation equations.  Most equations in this paper
are written assuming a ``negative phase convention'' -- that is, the
phase factor of a wavefield evolves with propagation distance as
$\exp(-ikz)$, and accordingly, the refractive index is expressed as
$n(\vecr) = 1 - \delta(\vecr) - i\beta(\vecr)$
\cite{jacobsen_2020}. In another often-used sign convention, phase
evolves as $\exp(ikz)$ and refractive index takes the form of
$n(\vecr) = 1 - \delta(\vecr) + i\beta(\vecr)$
\cite{goodman_fourier_2017}.  Eq.~\ref{eqn:fraunhofer} is consistent
with the negative phase convention, which results in a positive
argument in the exponential phase factor of the Fourier transform's
integrand. Using the positive phase convention, on the other hand, is
associated with a negative sign in the Fourier phase factor's
argument. Some scientific computation packages used in \adorym{}
including \autograd~and \torch~implement Fourier transform
(\emph{i.e.}, the \texttt{fft} function) with a negative exponential
argument for forward Fourier transform, and a positive exponential
argument for inverse Fourier transform. Therefore, Fraunhofer
diffraction in the negative phase convention should in fact be
implemented using the \texttt{ifft} functions of these packages, and
vice versa for the positive phase convention.  While both conventions
should lead to the same intensity prediction, it is important stay
consistent with the sign convention throughout the forward model.  In
\adorym{}, most optical propagation and modulation functions come with
a \texttt{sign\_convention} argument, so that users who attempt to
create new forward models can switch the convention according to their
preference.

When not satisfying the requirements of the Fraunhofer approximation
(such as in holography, or cases where the multislice method needs to
be applied), Fresnel propagation is used.  For propagation over a
short distance $d$ where the Fresnel number is large
\cite{voelz_fourier_optics_matlab}, Fresnel propagation can be
expressed as
\begin{equation}
  \funcpfr[\psi(\vecr)] = \mathcal{F}^{-1}\{\mathcal{F}[\psi(\vecr)] H(\vecnu, d)\}
  \label{eqn:fresnel_prop}
\end{equation}
where the Fresnel propagator kernel $H$ in Fourier space can be either
from the paraxial approximation of
\begin{equation}
  H_1(\vecnu, d) = \exp(i\pi\lambda d |\vecnu|^2)
\label{eqn:paraxial_kernel}
\end{equation}
or the Sommerfeld-Rayleigh formulation
\cite{goodman_fourier_2017} of
\begin{align}
  H_2(\vecnu, d) =
    \begin{cases}
      \exp\left[-i kd \sqrt{1 - \lambda^2|\vecnu|^2}\right], & 1 - \lambda^2|\vecnu|^2 > 0 \\
      0, & 1 - \lambda^2|\vecnu|^2 \leq 0
    \end{cases}
           \label{eqn:sommerfeld_kernel}
\end{align}
where the frequency components with
  $1 - \lambda^2|\vecnu|^2 \leq 0$ result in evanescent waves that die
  down quickly.

The process through which the sample interacts with the wavefield is
modeled using the \texttt{multislice\_propagate\_batch} function. The
multislice method was first used in electron microscopy
\cite{cowley_actacryst_1957a,ishizuka_actacryst_1977}, and is also
known as the beam propagation method in optics \cite{vanroey_josa_1981}.
Multislice propagation treats a 3D object
as a stack of thin slices along the beam axis, assuming the refractive
index at each lateral ($x$-$y$) point in a slice is axially constant.
At each slice, the wavefield is modulated using either
Eq.~\ref{eqn:complex_modulation} or \ref{eqn:ri_modulation} depending
on how the object is represented, and then propagated to the next
slice using Fresnel propagation (Eq.~\ref{eqn:fresnel_prop}). The
steps above are repeated until the wavefield reaches the last slice of
the object.  The entire process can be compactly represented in
matrix form as
\begin{equation}
    \vecpsi{\text{exit}} = \prod_j^{L_z}\left(\matpdz\boldsymbol{M}_{j,\Delta z}\right)\vecpsi{}
\end{equation}
where $\matm$ extracts the $j$-th slice of the object with thickness
(or spacing) $\Delta z$, and uses it to modulate wavefield $\vecpsi{}$;
$\matpzf$ is the linear operator that implements Fresnel propagation
over slice spacing $\Delta z$.

The implementation of multislice propagation in \adorym{} is again
generalizable: a 2D object is interpreted to contain just 1 slice, and
the wavefield is modulated by the only slice before it is propagated
to the detector plane.  On the other hand, when the object to be
reconstructed is within the depth of focus, the
\texttt{pure\_projection} option can be turned on: in this case, the
modulation of the wavefield by the entire object is computed as
\begin{equation}
  M_{\text{obj}}[\psi(\vecrxy)] = \psi(\vecrxy)\prod_j^{L_z}O(\vecrxy, j)
  \label{eqn:obj_complex_modulation}
\end{equation}
if the object is represented as complex modulation function, or
\begin{equation}
  M_{\text{obj}}[\psi(\vecrxy)] = \psi(\vecrxy)\exp\left[-k\sum_j^{L_z}\beta(\vecrxy)\Delta z\right] \exp\left[ik\sum_j^{L_z}\delta(\vecrxy)\Delta z\right]
  \label{eqn:obj_ri_modulation}
\end{equation}
if it is represented as refractive indices.

The \texttt{multislice\_propagate\_batch} function is intended for
reconstructing a continuous 3D object with isotropic voxel size --
that is, it assumes that the spacing $\Delta z$ between slices is
equal to the lateral pixel size $\Delta x$. By binning the slices, the
function can work with $\Delta z$ being multiples of $\Delta x$, but
the slice spacings are assumed to be constant. This setting works well
with common 2D ptychography (which does not need slice spacing at all)
and beyond-depth-of-focus ptychotomography
\cite{du_sciadv_2020}.  On the other hand, in multislice ptychography
\cite{maiden_josaa_2012,suzuki_prl_2014,tsai_optexp_2016b},
multiple slices of the object are reconstructed using ptychography
data from one or just a few tilt angles. The slices in these cases are
generally much sparser, and the slice spacing can be variable. Thus,
we also implemented a sparse-and-variable version of multislice
propagation in \adorym, where the positions of each slice are saved as
an array. The array is optimizable in the AD framework if slice
positions need to be refined.

\subsection{Module ForwardModel}
\label{ssec:forward_model}

The \modfm{} class and its child classes define the prediction models
corresponding to various types of imaging techniques, which take as
inputs the object function to be solved, the probe function that may
be known or unknown, and a collection of experimental parameters.
They predict (generally the magnitude of) the wavefields at the
detector plane according to given experimental setups and optical
theories. Although \adorym{} can be easily adapted to work with
complex-valued measurements (for example, obtained using
interferometry \cite{kamilov_optica_2015, kamilov_ieeetci_2016}),
intensity-only measurements are more common and easier to setup. If
multiple probe modes are used, the intensity will be the incoherent
sum of all probe modes:
\begin{equation}
	\ipred = \sum_i^{n_{\text{modes}}}|\psi_{i,\text{detector}}|^2
	\label{eqn:multimode}
\end{equation}
and magnitude is calculated as the square root of the summed
intensity.

\begin{algorithm}
	\SetAlgoLined\DontPrintSemicolon
	\SetKwFunction{TakeObjectChunk}{TakeObjectChunk}
	\SetKwFunction{MultislicePropagate}{MultislicePropagate}
	\SetKwFunction{FourierShift}{FourierShift}
	\SetKwFunction{FarFieldPropagate}{FarFieldPropagate}
	\SetKwFunction{FourierShift}{FourierShift}
	\SetKwFunction{Rotate}{Rotate}
	\SetKwFunction{append}{append}
	\SetKwInput{Input}{Input}
	\SetKwBlock{Initialization}{Initialization}{end}
	\SetKwBlock{Main}{Procedure}{end}
	\SetKwInput{Output}{Output}
	\Input{
	\par
	detector size [$l_y$, $l_x$], object size [$L_y$, $L_x$, $L_z$], full object function $O$, a list of probe modes $\boldsymbol{\Psi}$, probe positions $\boldsymbol{R}$, probe position corrections $\boldsymbol{\Delta R}$, wavelength $\lambda$, pixel size $\delta$, current rotation angle $\theta$
    }
    \Initialization{
    \par
	$\boldsymbol{o} \leftarrow []$ \tcp*{List of object chunks each in shape of $[l_y, l_x, L_z]$}
	$\boldsymbol{\Psi_{\text{detector}}} \leftarrow []$ \tcp*{List of detector-plane waves}
	$\boldsymbol{\Psi_{\text{shifted}}} \leftarrow []$ \tcp*{Shifted probe functions}
    }
    \Main{
    $O' \leftarrow \Rotate{O}$ \;

	\For {$\vecr$ \textup{in} $\boldsymbol{R}$}{
		\tcc{Get local object chunks corresponding to diffraction patterns (tiles).}
		$\boldsymbol{o}$.\append(\TakeObjectChunk{$O'$, $\vecr$})\;
		\tcc{Shift probes according to the list of position corrections.}
		$\boldsymbol{\Psi_{\text{shifted}}}$.\append(\FourierShift{$\boldsymbol{\Psi}$, $\boldsymbol{\Delta R}$})
		}
	
	\tcc{Do propagation for all probe modes.}
	\For {$\psi_i$ \textup{in} $\boldsymbol{\Psi_{\text{shifted}}}$}{
	$\psi_i \leftarrow$ \MultislicePropagate{$\psi_i$, $\boldsymbol{o}$, $\lambda$, $\delta$} \;
  	$\psi_i \leftarrow$ \FarFieldPropagate{$\psi_i$} \;
  	$\boldsymbol{\Psi_{\text{detector}}}$.\append{$\psi_i$}
     }
    }
	\Output{
	$\boldsymbol{\Psi_{\text{detector}}}$
    }
	
    \caption{An example \texttt{predict} function in the forward model
      of far-field ptychotomography.  This is for data parallelism
      mode, with probe position refinement and multiple probe modes. }
	\label{algo:example_ptycho}
\end{algorithm}

Any experimental variables that are to be refined, such as probe
positions, geometric transformation of projection images, and
propagation distances, should also be incorporated as a part of the
model. Collectively, these procedures, or the prediction function, are
given in the \texttt{predict} method inside each child class. An
example of the \texttt{predict} method used for far-field
ptychotomography with probe position correction is shown in Algorithm
\ref{algo:example_ptycho}. The example assumes simple data
parallelism, where each rank possesses its own copy of the whole
object function. In this case, the 3D object is rotated to the current
tilt angle $\theta$, and sub-chunks of the object corresponding to the
beam path of each diffraction pattern are extracted. If probe position
correction is enabled, the prediction function takes in an optimizable
array of probe position corrections, $\boldsymbol{\Delta R}$, which
has a length equal to the number of probe positions. Each element
in the list is a 2D vector $\Delta \vecr$ which represents the
correction that needs to be made to the user-given position $\vecr$,
so that the corrected position is
$\vecr^{\prime} = \vecr + \Delta \vecr$;
additionally, when scan positions are not integer
  pixel, the array $\boldsymbol{\Delta R}$ can also be used to hold
  floating point residuals of probe positions that should be added to
  the integer pixel positions regardless whether probe position
  refinement is enabled or not.  The correction shift $\Delta \vecr$
is then applied to the probe function using the Fourier shift theorem:
\begin{equation}
    \psi_{\text{shifted}}(\vecr) = \mathcal{F}^{-1}\left\{ \mathcal{F}[\psi(\vecr)]\exp[ -i2\pi(\Delta \vecr\cdot\vecnu) ] \right\}.
    \label{eqn:fourier_shift}
\end{equation}
Now this is already enough for the refinement: the gradient of the
loss function with regards to the position correction vector
$\Delta \vecr$ is left for the AD engine to calculate behind the
scene, and no explicit derivation or implementation is needed. More
information about parameter refinement will be introduced in Section
\ref{ssec:refinement}.

When the size of the raw data is large, \adorym{} can be set to
process only a subset, or a minibatch, of the measured data containing
\texttt{batch\_size} tiles at a time. Thus, we hereby define the terms
we will use for describing the workload of reconstruction: an
``iteration'' is used to refer to the process of finishing a minibatch
of data by all parallelized processes (the inner loop), and an
``epoch'' refers to the work done to cover the entire dataset (all
tiles on all viewing angles; the outer loop).  Propagation through the
object is done using multislice propagation which is capable of
simulating multiple scattering within the object volume and is useful
when the object thickness is beyond the depth of focus
\cite{cowley_actacryst_1957a, du_sciadv_2020}. On the other hand, when
projection approximation is valid, the multislice simulation can be
reduced to a line-integration of the object function. The prediction
function is concluded by propagating the wavefield onto the detector
plane, and the final complex wavefields are returned.

The above illustrated routine can be easily generalized beyond the
settings of far-field ptychography.  Full-field CDI data, for
instance, is interpreted by \adorym{} as a special type of
ptychography with 1 tile per angle, and with detector size equal to
the $y$- and $x$-dimensions of the object array.  Furthermore,
propagation from the object's exiting plane to the detector plane is
modeled using far-field diffraction in the example of Algorithm
\ref{algo:example_ptycho}, but when working with near-field
ptychography or tiled holography data (described in Section
\ref{ssec:data_format}), it can be replaced with a Fresnel propagation
function. In addition, when processing tiled data with bright
illumination (in contrast to far-field ptychography, where the
illumination at each tile or diffraction spot is localized and has
most of its energy concentrated to an area much smaller than the
detector size), Fresnel propagation in object (when multislice
modeling is enabled) and in free space of a wavefield tile may cause
diffraction fringes to wrap around, due to the non-zero edges in the
modulated wavefield. In this case, \adorym{} can select a larger area
for each object chunk and wavefield tile, leaving a ``buffer zone''
around the wavefield for wrapping-around fringes to spread, and
discard the buffer zones after the wavefield reaches the detector
plane \cite{blinder_optexp_2019, ali_arxiv_2020}.

When the illumination is only partially coherent
\cite{thibault_nature_2013} or when fly-scan is used in data
acquisition \cite{pelz_apl_2014,deng_optexp_2015,huang_scirep_2015},
multiple mutually incoherent probe modes can be used for image
reconstruction to account for the limited coherence of illumination
wavefield.  In this case, the prediction function gives the outputs of
all probe modes individually, assuming each of them does not interact
with others before arriving at the detector plane. As a general
convention, the prediction functions in virtually all forward models
return the detector-plane wavefields as two separate arrays that
respectively stores the real and imaginary parts of the complex
magnitude. Each of these arrays has a shape of \texttt{[batch\_size,
  n\_probe\_modes, len\_detector\_y, len\_detector\_x]}. In this way,
multi-mode reconstruction can be realized for various types of imaging
experiments in addition to far-field ptychography.

\subsubsection{Loss function}

Each child class of \modfm{} contains also a \texttt{loss} method that
takes in the predicted detector-plane magnitude $\sqrt{\ipred}$ and
the associated experimental measurement $\sqrt{\imeas}$, and calculates
the loss from them (\emph{i.e.}, the last-layer loss function). Combining \texttt{predict} and \texttt{loss},
another method \texttt{get\_loss\_function}, constructs an end-to-end
loss function mapping the input variables to the final loss,
optionally with added regularization terms. The data mismatch
term in the last-layer loss function can take different forms, each of them
has specific advantages in terms of numerical robustness and noise resistance.
The expressions
of these data mismatch terms are defined in the parent \modfm{} class
which are inherited by all its children classes
the \texttt{loss} method, so users can easily select from existing types of
the data mismatch term, modify them, or plug in new types, and then apply
the change to all forward model classes. \adorym{} has two built-in types
of data mismatch term, namely the least-square type (LSQ) and the Poisson
maximum likelihood type.

The LSQ data mismatch term
measures the Euclidian distance between the prediction and the
measurement. As pointed out by Godard
\emph{et al.}  \cite{godard_optexp_2012}, when the measured intensity
$I_{\text{meas}}$ follows Poisson statistics whose standard deviation
is the square root of its mean, then the variance of
$\sqrt{I_{\text{meas}}}$ is insensitive to the expectation of
$I_{\text{meas}}$. That is, using the square root of intensities,
instead of intensities themselves in the LSQ loss function, may lead to
better numerical robustness. As such, \adorym{} uses an LSQ-type loss
function expressed as
\begin{equation}
	\mathcal{D}_{\text{LSQ}} = \frac{1}{N_d}\sum_{m}^{N_d}\norm{\sqrt{\ipredm} - \sqrt{\imeasm}}^2
	\label{eqn:lsq_loss}
\end{equation}
where $N_d$ is the number of detector (or tile) pixels.
If $I_{\text{meas}}$ and $I_{\text{pred}}$ are
  spatially sparse, the intensity terms $I_k$ in
  Eq.~\ref{eqn:lsq_loss} can be implemented as $I_k + \epsilon$, where
  $\epsilon is a small positive number$, to improve numerical
  robustness, since $I_k$ appears in the denominator of the square root
  function's derivative.

Alternatively, when the measured data are noisy, one may also
explicitly introduce Poisson statistics to account for the noise in
$\imeas$, which leads to the Poisson maximum likelihood mismatch
\cite{godard_optexp_2012}:
\begin{equation}
	\mathcal{D}_{\text{Poisson}} = \frac{1}{N_d}\sum_{m}^{N_d}\ipredm - \imeasm\log(\ipredm).
	\label{eqn:poisson_loss}
\end{equation}
In a previous publication \cite{du_jac_2020}, we have shown with
numerical simulations that the Poisson loss function of
Eq.~\ref{eqn:poisson_loss} can provide better resolution and contrast
under low-dose conditions compared to the LSQ loss function of
Eq.~\ref{eqn:lsq_loss}, but it takes longer to converge, and may
result in indeterministic artifacts for less noisy
images. \adorym{} allows users to conveniently add new loss function
  types (for example, mixed Gaussian-Poisson) into the parent class of
  \modfm{}.

When processing data subject to photon noise or information
deficiency, it is a common practice to supply the reconstruction
algorithm with additional prior knowledge about the sample.  The
finite support constraint, which is commonly used in full-field
coherent diffraction imaging \cite{marchesini_prb_2003} and is also
made available in \adorym, is one type of such prior knowledge used to
ensure the algorithm converges to a unique solution. Alternatively,
some other categories involve assumptions on one or more quantitative
indicators of the object function, such as its smoothness or
sparsity. The $l_1$-norm, for example, is known to enhance the
sparsity of the solution, encouraging more voxels to have near-zero
absolute values \cite{tibshirani_jrssb_1996,candes_jfaa_2008}. If such a quantity
$\mathcal{R}[O(\vecr)]$ constrained to be below a certain value, the
constrained optimization can be equivalently solved by adding
$\mathcal{R}[O(\vecr)]$ to the original loss function of
Eq.~\ref{eqn:lsq_loss} or Eq.~\ref{eqn:poisson_loss} as a Tikhonov
regularizer \cite{mccann_arxiv_2019}. The full form of the loss
function to be minimized is then
\begin{equation}
  \mathcal{L} = \mathcal{D}(\ipred, \imeas) + \sum_i\alpha_i\mathcal{R}_i[O(\vecr)]
  \label{eqn:tikhonov}
\end{equation}
where $\alpha_i$ is the ``weighting coefficient'' applied to
regularizer $i$. The regularizers already available in \adorym{}
include $l_1$-norm \cite{tibshirani_jrssb_1996,candes_jfaa_2008}, reweighted $l_1$-norm, and
total variation.  The expressions, purposes, and example scenarios of
application, are shown in Table \ref{tab:regularizers}.
Moreover, these regularizer functions are defined as
individual \modreg{} classes, and users can conveniently add their
  own regularizers.

\begin{table}
	\centering
	\begin{tabular}{p{2cm}p{6cm}p{3cm}p{4cm}}
		\hline
		\textbf{Name} & \textbf{Expression of $\mathcal{R}$} & \textbf{Purpose} & \textbf{Application} \\
		\hline
		
		$l_1$-norm &
		{\vspace{-0.2cm}$$\frac{1}{N_o}\sum_{i_r}\alpha_1|\delta_{i_r}| + \alpha_2|\beta_{i_r}|$$ \vspace{-0.3cm} \newline
		if object is represented by refractive indices, or \newline
		\vspace{-0.2cm}$$\frac{1}{N_o}\sum_{i_r}\alpha_1\left||O_{i_r}| - \overline{|O|}\right| + \alpha_2|\mbox{arg}(O_{i_r})|$$ \vspace{-0.3cm} \newline
	    if object is represented by complex modulation function. }
		& Promotes sparsity. & Reduce artifacts in objects with finite extent (smaller than the object array being solved) or porous objects. \\
		\hline
		
				Reweighted $l_1$-norm \cite{candes_jfaa_2008} &
		{\vspace{-0.2cm}$$\frac{1}{N_o}\sum_{i_r}\alpha_1 W_1|\delta_{i_r}| + \alpha_2 W_2|\beta_{i_r}|$$ \vspace{-0.3cm} \newline
		if object is represented by refractive indices, or \newline
		\vspace{-0.2cm}$$\frac{1}{N_o}\sum_{i_r}\alpha_1W_1\left||O_{i_r}| - \overline{|O|}\right| + \alpha_2W_2|\mbox{arg}(O_{i_r})|$$ \vspace{-0.3cm} \newline
		if object is represented by complex modulation function. In both equations, the quantity $W_i$
	    as in $W_i|C|$ is the adaptive weight defined as $W_i = \max(|C|) / (|C| + \epsilon)$. Values of $W_i$
        are calculated outside the loss function using ``snapshots'' of the object function, and are treated
        as no-gradient constants by AD. Their values are updated once upon a fixed number of minibatches are done.}
		& Promotes sparsity adaptively, so that near-zero voxels are penalized more. & Similar to $l_1$-norm,
		and more desirable if a good initial guess is available.  \\
		\hline
		
		Total variation (TV) \cite{grasmair_amo_2010} &
		{\vspace{-0.2cm}$$\frac{1}{N_o}\gamma\sum_{i_r}\norm{\nabla\delta_{i_r}}_1 + \norm{\nabla\beta_{i_r}}_1$$ \vspace{-0.3cm} \newline
			if object is represented by refractive indices, or \newline
			\vspace{-0.2cm}$$\frac{1}{N_o}\gamma\sum_{i_r}\norm{|O_{i_r}| - \overline{|O|}}_1 + \norm{\mbox{arg}(O_{i_r})}_1$$ \vspace{-0.3cm} \newline
			if object is represented by complex modulation function. $\norm{\nabla C}_1$ returns
		the $l_1$-norm of the spatial gradient of $C$, given as $|\nabla_x C| + |\nabla_y C| + |\nabla_z C|$.
        }
		& Promotes smoothness. & Reconstructing the object from noisy data. \\
		\hline
		
		\hline
	\end{tabular}
    \caption{A list of regularizers available in \adorym. In all equations, $N_o = L_xL_yL_z$ is the total
    number of voxels in the object function, and $i_r$ indexes the object voxels;
    $\overline{O}$ is the mean of object modulus $\mbox{arg}(\cdot)$ represents
    the complex argument function which returns the phase of the complex input.  }
    \label{tab:regularizers}
\end{table}

\subsubsection{Parameter refinement}
\label{ssec:refinement}

In Section \ref{ssec:forward_model}, we have shown the incorporation
of probe position correction in the forward model, which is done using
Fourier shift theorem. In addition to that, \adorym{} also supports
the refinement of a range of experimental variables, and the list can
be readily expanded by users upon needed. At present, the following
parameters and variables can be refined:
\begin{itemize}

\item{Probe defocus}: in imaging techniques like ptychography, it is
  often the case where the probe function retrieved from a previous
  experiment is used to reconstruct a new dataset.  However, as the
  axial position of the sample may differ, the previously recovered
  probe might be deviated from the incident plane of the new
  sample. In this case, \adorym{} can Fresnel propagate
  (Eq.~\ref{eqn:fresnel_prop}) the probe function according to a list
  of defocusing distances, $\boldsymbol{z}_{\text{defoc}}$, which
  stores the defocusing that should be applied for each tilt
  angle. The gradient of $\boldsymbol{z}_{\text{defoc}}$ can be
  automatically calculated by AD for optimization.

\item{Probe position offset}: Section \ref{ssec:forward_model}
  presented an example where all probe positions are refined
  individually. In the case of ptychotomography, it is often the case
  that the relative positions of probe positions on a viewing angle
  are accurate, but there exist offsets among positions on different
  angles in the sample frame, such as in the case of a
  not-well-centered rotation axis
  \cite{holler_scirep_2014}. Individually refining all probe positions
  from all angles might still be feasible, but would introduce too
  many degrees of freedom, making the reconstruction problem badly
  underconstrained. \adorym{} can instead refine a list of position
  offsets across different tilt angles, where each correction vector
  as an element of the list applies to all probe positions in a
  certain angle.

\item{Slice position}: in sparse multislice imaging, slice positions
  enter the forward model through Fresnel propagation from a slice to
  the next, and the gradient of slice positions is thus obtainable
  through AD.

\item{Free propagation distance}: the distance between the exiting
  plane of the sample to the detector in the case of near-field
  imaging can be refined as well. In the case of multi-distance
  holography, the distances from the sample to all projection images
  can be included in the framework.

\item{Object orientation}: the sample stage in nanotomography may
  suffer from random wobbling, which causes the angular orientation of the
  sample to fluctuate in all three axes at different viewing angles.
  Refinement of orientations has already been demonstrated in combined
  x-ray ptychography and fluorescence microscopy
  \cite{deng_sciadv_2018}, so orientation refinement has been
  implemented in \adorym{} as well. When enabled, the object is
  rotated in its three axes according to an optimizable orientation
  correction array of shape \texttt{[3, num\_angles]}. For each axis,
  the pixel coordinates on each plane of the rotated object
  perpendicular to the axis are mapped back to the original object
  function at $0\degree{}$, using the linear transform
  \begin{equation}
    \begin{pmatrix}
      x_0 \\
      y_0
    \end{pmatrix} =
    \begin{pmatrix}
      \cos\theta & -\sin\theta \\
      \sin\theta & \cos\theta
    \end{pmatrix}
    \begin{pmatrix}
      x_\theta \\
      y_\theta
    \end{pmatrix}.
    \label{eqn:rotation_transform}
  \end{equation}
  Subsequently, the rotated object is computed by taking voxel values
  according to the mapped coordinates from the $0\degree{}$ object
  array using bilinear interpolation. A way to understand how the loss
  function can be differentiable with regards to rotation angle
  $\theta$ through this interpolation operation is to regard the
  interpolation as an $(L_p L_q) \times (L_p L_q)$ matrix
  $\boldsymbol{R_\theta}$ parameterized by $\theta$, where $L_p$ and
  $L_q$ are the numbers of pixels along an object plane perpendicular
  to the rotation axis. It follows that the interpolation is in fact a
  linear operation of evaluating
  $\boldsymbol{\overline{o_\theta}} = \boldsymbol{R_\theta}
  \boldsymbol{\overline{o_0}}$, where $\boldsymbol{\bar{o}}$ is a
  vectorized ``$pq$-plane'' of the object, formed by concatenating
  each row of it. The matrix $\boldsymbol{R_\theta}$ is sparse,
  containing at most 4 non-zero elements on each row in the case of
  bilinear interpolation; therefore, calculating the derivative of
  $\boldsymbol{R_\theta}$ to $\theta$ with a gradient vector can be
  done efficiently by exploiting this sparsity
  \cite{jiang_arxiv_2019}. In \adorym{}, we implement differentiable
  rotation by using the \texttt{grid\_sample} function of
  \torch~\cite{pytorch,pytorch_docs}.

\item{Affine transform of projection images}: this feature is
  particularly useful for near-field imaging and especially
  multi-distance holography. The misalignment of detector or sample
  stage may cause the acquired projection image to suffer from one or
  more of three types of distortions, namely translation, tilting, and
  scaling. These distortions are collectively refined to as the affine
  transformation.  We denote the translation, scaling, and shearing in
  $x$ and $y$ as $\delta x$, $\delta y$, $S_x$, $S_y$, $c_x$, $c_y$,
  respectively, and also represent the tilt angle as $\phi$.  This
  leads to an affine transformation matrix $\boldsymbol{A}$ which is
  the composite of 4 matrices responsible for translation, scaling,
  shearing, and tilt:
  \begin{equation}
    \boldsymbol{A} =
    \begin{pmatrix}
      \cos\phi & -\sin\phi & 0 \\
      \sin\phi & \cos\phi & 0 \\
      0 & 0 & 1
    \end{pmatrix}
    \begin{pmatrix}
      1   & c_x & 0 \\
      c_y & 1   & 0 \\
      0 & 0 & 1
    \end{pmatrix}
    \begin{pmatrix}
      S_x & 0 & 0 \\
      0 & S_y & 0 \\
      0 & 0 & 1
    \end{pmatrix}
    \begin{pmatrix}
      1 & 0 & \delta x \\
      0 & 1 & \delta y \\
      0 & 0 & 1
    \end{pmatrix}.
    \label{eqn:affine_transform}
  \end{equation}
  Eq.~\ref{eqn:affine_transform} is apparently an expanded form of the
  square matrix in Eq.~\ref{eqn:rotation_transform} for situations
  beyond rotation. \adorym{} can refine an affine transformation
  matrix for the projection image at each defocus distance to minimize
  the loss function, and the transformation is differentiable based on
  the same reason as in the case of sample rotation refinement.  The
  forward model of affine transformation and the associated bilinear
  interpolation is also implemented using \torch's
  \texttt{grid\_sample} function.

\item{Linear relation coefficient between phase and absorption.} The
  homogeneous assumption about the object, which assumes that the
  $\delta(\vecr)$ and $\beta(\vecr)$ terms in its refractive index is
  related through $\delta(\vecr) = (1 / \kappa)\beta(\vecr)$
  \cite{turner_optexp_2004}, is sometimes desirable when the measured
  data do not provide enough information to recover both the phase and
  magnitude of the object function. By enforcing a linear relation
  between the phase shifting part and the absorption part of the
  object, the number of unknowns is effectively reduced to half, which
  could prevent the reconstruction problem from being too loosely
  constrained. However, since most real samples have multiple
  materials present, it is hard to accurately calculate the value of
  $\kappa$ using prior knowledge about the object. \adorym{} on the
  other hand can include $\kappa$ into the automatic differentiation
  loop, so that its value can be adjusted constantly to minimize the
  loss function.

\end{itemize}

\subsection{Module Wrapper and backends}
\label{ssec:module_wrapper}

The AD engine is the cornerstone of \adorym{}: it provides the
functionality to calculate the derivative of the loss function with
regards to the object, the probe, and other parameters. Because the AD
engine needs to keep track of all variables (or ``nodes''
\cite{margossian_arxiv_2018}), and the relationship between them and
the loss function (this relationship is known as the ``graph''), most
AD libraries have their own data types and functions for the parent
and child nodes in the forward model. Many AD libraries are quite
similar: they come with a comprehensive look-up table that logs the
derivatives for many common mathematical and array-manipulation
operators (like stacking and reshaping), and some
  provide an easy interface for users to expand the package by
  defining derivatives for new functions \cite{autograd_docs}.  On
the other hand, packages may differ in performance depending on the
environment in which they run, in the ease of setting up the
environment they require, and in the size of their user
communities. \autograd{} builds its data type and functions on the
basis of the popular scientific computation package NumPy \cite{numpy}
and SciPy \cite{scipy}; these in turn can make use of hardware-tuned
libraries such as the Intel Math Kernel library
\cite{intel_mkl}. However, \autograd{} does not have built-in support
for graphical processing units (GPUs).
Another AD package which provides GPU support is
\torch{}, which has perhaps a larger user community.  \autograd{} and
\torch{} are dynamic graph tools which allow the computational graph
or the forward model to be altered at runtime \cite{looks_arxiv_2017}.
This
means the forward model structure can be modified on the fly using
Python's native workflow control clauses (such as an
\texttt{if}$\ldots$\texttt{else}$\ldots$ block), which is very helpful
for implementing a multi-stage optimization strategy.  In this way one
can use one type of loss function for the first several iterations,
and switch to another after that. Moreover, a dynamic graph tool means
the runtime values of any intermediate variables can be easily
retrieved by inserting a print function inside the forward model
code. With a static graph tool, this can only be done outside the
forward model, which makes debugging less intuitive and
straightforward. Therefore, \adorym{} is designed
to specifically work with dynamic graph tools.

As we aim to incorporate both \autograd{} and \torch{} (and more dynamic graph AD libraries in the future) in \adorym{}, we
have built a common front end for them in the \modwpr{} module.  This
module provides a unified set of APIs to create optimizable or
constant variables (arrays in \autograd{}, or tensors in \torch), call
functions, and compute gradients. In order to make sure all frontend
APIs know the right backend to use, we keep the setting parameter
\texttt{backend} in a specific module named
\texttt{global\_settings}. This common module is imported by reference
(instead of by value) in every relevant module of \adorym{}, and once
its value is changed in any module or script, it is changed for all
modules after that point. For example, the line \texttt{import
  adorym.global\_settings} is used in both \modmain{} and \modwpr{};
once it is changed in \modmain{} by
\texttt{adorym.global\_settings.backend = "pytorch"} upon user
setting, the value of \texttt{adorym.global\_settings.backend} is
changed in \modwpr~immediately as well, so all frontend functions in
\modwpr~referring to the backend setting will use functions from
\torch{} in this case.

\subsection{Module Optimizer}
\label{ssec:class_optimizer}

Optimizers define how gradients should be used to update the object
function or other optimizable quantities; these choices are provided
by the class \modopt{}. \adorym{} currently provides
  4 built-in optimizers: gradient descent (GD), momentum gradient
  descent, adaptive momentum estimation (Adam), and conjugate gradient
  (CG). Moreover, \adorym{} also has a \texttt{ScipyOptimizer} class
  that wraps the \texttt{optimize.minimize} module of the Scipy
  library \cite{scipy}, so that users may access more optimization
  algorithms coming with Scipy when using the \autograd{} backend on
  CPU.  The simplest GD algorithm updates an optimizable vector
$\vecx$ using the gradient of
the loss function with regards to it, or
\begin{equation}
  \vecxx \leftarrow \vecxx - \rho\nabla_{\vecxx}\mathcal{L}
  \label{eqn:gd}
\end{equation}
where $\rho$ is the step size. GD is very low in overhead in that it
does not require any additional parameters to be stored and updated.
However, when the loss function is ill-conditioned (so that the loss
is steep at the vicinity of its minimum) and convex, update iterations
can cause it to overshoot the global minimum. As a consequence, the
solution may oscillate around the minimum point, taking many
iterations to finally converge. One viable workaround is to make
$\rho$ variable. If $\rho$ is reduced to $\rho/2$ each time one
reduces by half the suboptimality gap
$|\mathcal{L}_j - \mathcal{L}^{*}|$ between the current loss
$\mathcal{L}_j$, and the theoretically asymptotic loss value
$\mathcal{L}^{*}$, then GD will monotonically converge to the minimum
\cite{ruder_arxiv_2016}.  In practice, we found the actual halving of
the suboptimality gap can be tricky to decide as the loss curve
sometimes exhibits a periodically oscillating pattern even if the
unknowns are still far away from the global minimum. Nevertheless, if
$\rho$ is reduced according to the above mentioned strategy, then the
number of iterations needed for the suboptimality gap to drop from
$|\mathcal{L}_j - \mathcal{L}^{*}|$ to
$|\mathcal{L}_j - \mathcal{L}^{*}| / 2$ is twice the number needed for
it to drop from $2|\mathcal{L}_{j} - \mathcal{L}^{*}|$ to
$|\mathcal{L}_{j} - \mathcal{L}^{*}|$.  In other words, the number of
iterations before which $\rho$ should be halved doubles each time that
halving occurs.  Therefore, in our implementation of GD in the
\texttt{GDOptimizer} child class, users are allowed to define a ``base
iteration number'' $N_\text{bi}$, so that $\rho$ is halved each time
the iteration index hits
$\{ \sum_{s=0}^S 2^S N_\text{bi} | S = 0, 1, \cdots \}$.

Another known problem of GD is that it can also
  oscillate excessively when it travels through a ravine in the
  solution space. As the gradient vector can sometimes point to the
  ``side walls'' of the ravine, so that GD may dangle between the side
  walls instead of going straight down along the ravine. The momentum
  algorithm \cite{qian_nn_1999} augments GD by adding the history of past
  gradients into the update vector, so that Eq.~\ref{eqn:gd} becomes
\begin{align}
  \begin{split}
    \boldsymbol{v} &\leftarrow \gamma\boldsymbol{v} + \rho\nabla_{\vecxx}\mathcal{L} \\
    \vecxx &\leftarrow \vecxx - \boldsymbol{v}.
    \label{eqn:mgd}
  \end{split}
\end{align}
This algorithm is implemented as the \texttt{MomentumOptimizer} class
of \adorym{}.

The Adam algorithm \cite{kingma_arxiv_2014} makes use of the history
of gradients in past iterations, and is among the most popular
optimization algorithms in the machine learning community
\cite{ruder_arxiv_2016}.  What distinguishes Adam from many other
algorithms in the same category is that it stores both the first-order
moment (the mean) and the second-order moment (the variance) of past
gradients, and that it involves a correction step for countering the
bias of the momenta towards zero.  These features provide an advantage
in comparison to similar momentum-based algorithms
\cite{ruder_arxiv_2016}.  Another merit of Adam in our specific
application is that the magnitude of its update vector is relatively
insensitive to the scale of gradient \cite{kingma_arxiv_2014}, so it
can work well with raw data that have various numerical scalings.  As
an example, full-field tomography and holography data are usually
normalized using ``white field'' images, while far-field ptychography
data generally have a high dynamic range and are often used without
normalization.  Because the update rate in Adam can be imperfectly
tuned even by several orders of magnitude, Adam can also work with
different types of loss functions which may have very different
values.  This aligns well with the generalizability concept of
\adorym, so we have implemented the algorithm in the
\texttt{AdamOptimizer} class.

Our implementations of GD, Momentum, and Adam are designed
for stochastic minibatch optimization -- in the literature, these are
the algorithms of choice for when we have large data sets \cite{ruder_arxiv_2016}.
For small or medium-sized problems, when the raw data and computational graph
fit into the memory, more sophisticated (and hyperparameter-free) conjugate gradient
algorithms or higher order algorithms are typically the methods of choice.
In \adorym{}, we implement the non-linear conjugate gradient (CG) method
\cite{hestenes_cg_1952,fletcher_nlcg_1964}, which is implemented as
the \texttt{CGOptimizer} in \adorym{}.  Our implementation uses the
Polak–Ribi\`{e}re method \cite{polak_mmean_1969,polyak_ucmmp_1969}
to find the conjugate gradient vectors, followed by an adaptive line
search \cite{nocedal_2006} to determine the update step size.
We also implement a \texttt{ScipyOptimizer} wrapper that can be used to access a
variety of sophisticated algorithms (such as BFGS, Newton-CG, and so on) from \scipy{}
within \adorym{}. One distinction is that while our GD, Momentum, Adam, and CG
implementations can be used in both CPUs and GPUs, the Scipy optimizers can
only be used in the CPU. However, for small-sized convex problems, this can often
be balanced by the reduced number of iterations that some advanced algorithms
require \cite{bottou_arxiv_2016}.

\subsection{Parallelization modes}
\label{ssec:distribution_modes}

\begin{figure}[H]
  \centering \includegraphics[width=0.98\textwidth]{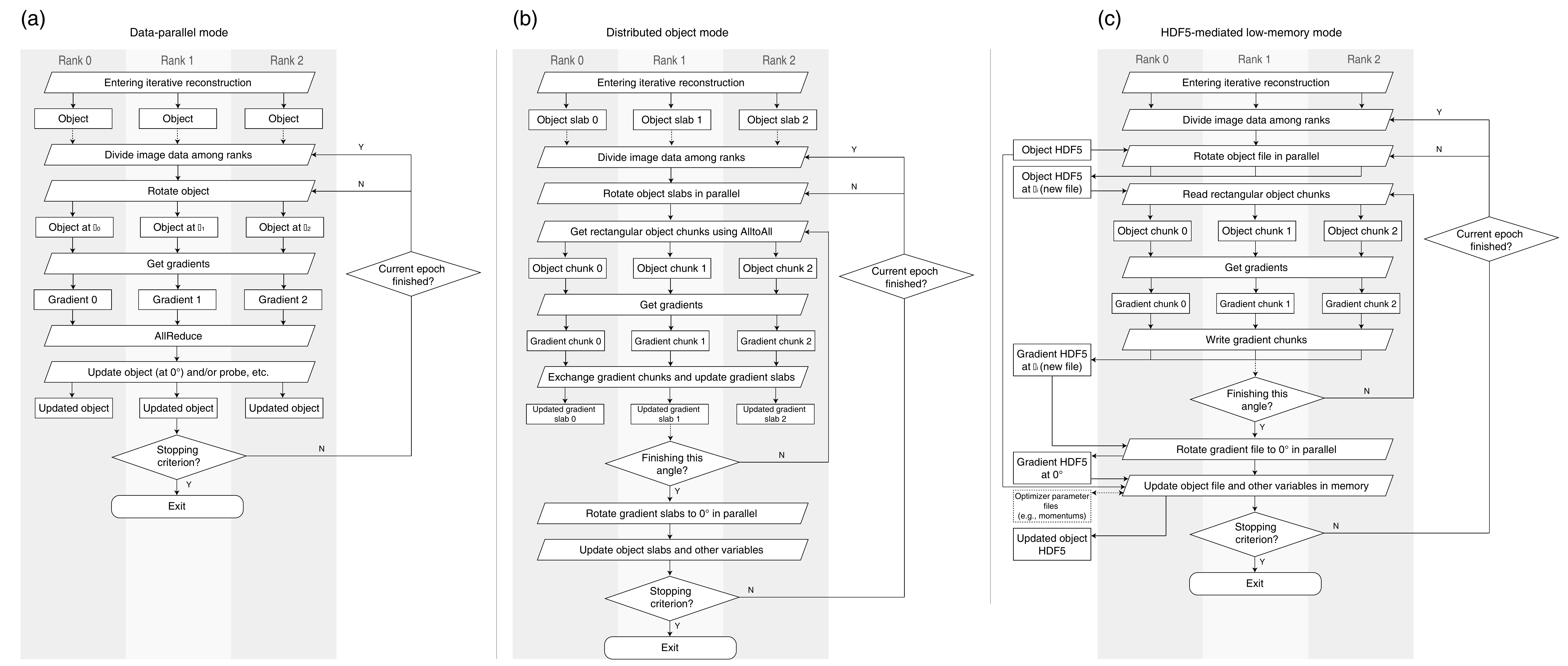}
  \caption{Workflow diagram of \adorym{} in DP mode, DO mode, and H5
    mode.}
  \label{fig:workflow}
\end{figure}

\adorym{} supports parallelized processing based on the message
passing interface (MPI) \cite{mpi_31_standard}, a multiprocessing
programming model that is compatible with most modern computation
platforms ranging from laptops to HPCs. Specifically, in \adorym{} we
use MPI4py Python API \cite{dalcin_jpdc_2005}. The MPI interface
allows one to conveniently communicate data or synchronize program
workflow among multiple processes (or ranks) on CPU cores.  Because
each spawned MPI rank is independent from others on the Python level,
it can individually send data back and forth between RAM and a
specified GPU, and run GPU operations without interfering with other
ranks as long as the GPU memory does not overflow. Therefore, our
MPI-based parallelization is also inherently capable of working with
an arbitrary number of GPUs.  When it is necessary to communicate
between all ranks, each rank needs to retrieve the data back from the
GPU to the CPU, and use collective MPI operations such as
Allreduce. This is inefficient compared to direct GPU-to-GPU
communications using tools such as NVLink \cite{foley_ieeemicro_2017},
but it works on a wider range of devices.

Parallel processing based on MPI can be implemented in several
different ways, each of which differs from others in terms of
computational overhead and memory consumption. Here we introduce the
three types of parallelization modes used in \adorym{}: data
parallelism (DP) mode, distributed object (DO) mode, and the
HDF5-file-mediated low-memory (H5) mode.

\subsubsection{Data parallelism (DP) mode}

In machine learning, ``data parallelism'' refers to distributed neural
network training where each rank has a copy of the full model stored
in memory, and process a subset of training data that is distinct from
that on other ranks \cite{xing_engineering_2016}.  The model
parameters are periodically synchronized by taking their averages over
all ranks. In \adorym{}, the equivalent of a neural network is the
forward optical model, where the parameters are the object function
and all optimizable variables. When working in DP mode, each rank is
allocated with a batch of \texttt{batch\_size} tiles, which is used to
calculate the gradient of the loss function with regards to the entire
object volume and other parameters (probes, probe positions,
\etc{}). Once all ranks have finished their gradient computation,
their gradient arrays are synchronized to the same average values
through an MPI Allreduce operation, after which they are used to
update the optimizable variables using a chosen optimization algorithm. Since
gradient averaging is the only major inter-rank communication, the DP
mode is low in overhead, but it has high memory consumption due to the
need of keeping a copy of the whole object function for every rank.
This in turn limits the ability to reconstruct objects with limited
resources.  A illustrative workflow diagram for the DP mode with 3 MPI
ranks is shown in Fig.~\ref{fig:workflow}(a).

\subsubsection{Distributed object (DO) mode}

\begin{figure}[H]
  \centering \includegraphics[width=0.9\textwidth]{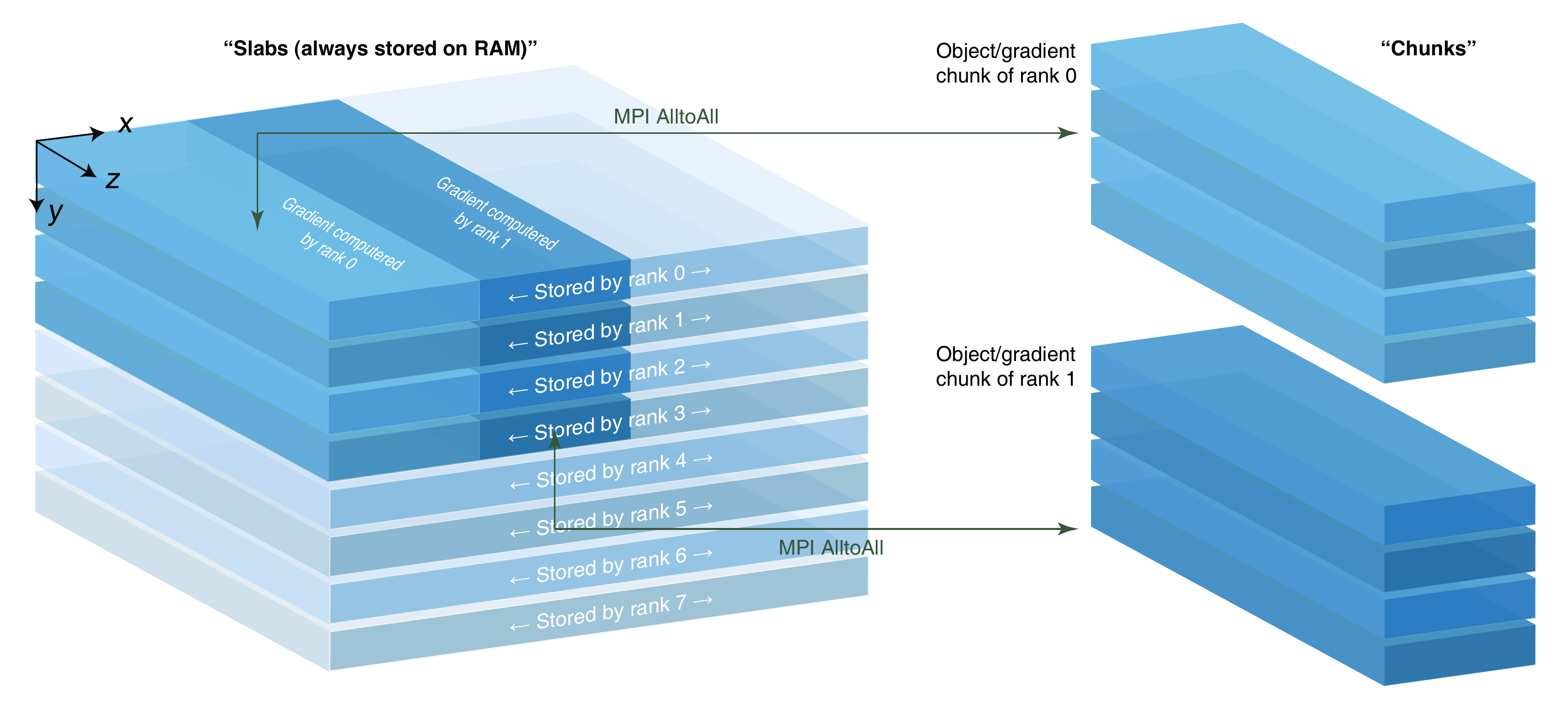}
  \caption{Illustration of the distributed object (DO) scheme. With
    multiple MPI ranks, the object is divided into a vertical stack of
    slabs, which are kept separately by all ranks. When a rank
    processes a diffraction image, it gets data from ranks that
    possess the parts of the object function that it needs, after
    which it assembles these partial slabs into the object chunk
    corresponding to the beam path of the diffraction image that it is
    processing.  After gradient of the chunk is calculated, it is
    scattered back into the same positions of the gradient slabs kept
    by relevant ranks. Object update is done by each rank individually
    using the gathered gradient array.}
  \label{fig:distributed_object}
\end{figure}

In spite of its simplicity and low overhead, keeping a copy of the
object function for every rank as in the DP mode can be impractical
with limited memory per rank. We have therefore implemented the
distributed object (DO) mode as an alternative option. In DO mode,
only 1 object function is jointly kept by all ranks. This is done by
severing the object along the vertical axis into several slabs, and
letting each rank store one of them in its available RAM as 32-bit
floating point numbers.  In this way, standard tomographic rotation
can be done independently by all ranks in parallel. The stored object
slab does not enter the device memory as a whole when GPU acceleration
is enabled; instead, object chunks are extracted from the object slabs
in the CPU, and only these partial chunks are sent to GPU memory for
gradient computation. Since the ``thickness'' of a slab is usually
different from the vertical size of an object chunk (whose $x$-$y$ cross
sections should match the size of a tile), we use MPI's AlltoAll
communication for a rank to gather the voxels it needs to assemble an
object chunk from other ranks.

This is illustrated in
Fig.~\ref{fig:distributed_object}, where we assume the MPI command
spawns 8 ranks, and the object function is evenly distributed among
these ranks. For simplicity, we assume a batch size of 1, meaning a
rank processes only 1 tile at a time. Before entering the
differentiation loop, the object slab kept by each rank is duplicated
as a new array, which is then rotated to the currently processed
viewing angle. In our demonstrative scenario, rank 0 processes the top
left tile in the first iteration, and the vertical size of the tile
spans four object slabs. As ranks 0--3 have knowledge about rank 0's
job assignment, they extract the part needed by rank 0 from their own
slabs, and send them to rank 0 through AlltoAll operation. Rank 0 then
assembles the object chunk from the partial slabs by concatenating
them along $y$ in order. Besides, since rank 0 itself contains the object slab
needed by other ranks, it also needs to send information to them.
The collective send/receive can be done using MPI's AlltoAll communication
in a single step.
In reality, the vertical size of an object
chunk might not be divisible by the number of slices in each slab, so
certain ranks may send a partial object that is ``thinner'' than the
slab it keeps. The ranks
then send the assembled object chunks to their assigned GPUs for
gradient computation, yielding gradient chunks on each rank that have
exactly the same shape as the object chunks. These gradient chunks are
scattered back to relevant ranks, but this time to
their ``gradient slab'' arrays that identically match the object slabs
in position and shape. After all tiles on this certain
viewing angle are processed, the gradient slabs kept by all ranks are
rotated back to 0$\degree{}$. Following that, the object slab is
updated by the optimizer independently in each rank. When
reconstruction finishes, the object slabs stored by all ranks are
dumped to a RAM-buffered hard drive, so that they can be stacked to
form the full object. The optimization workflow of the DO mode is
shown in Fig.~\ref{fig:workflow}(b).

While it has the advantage of
requiring less memory per rank, a limitation of the DO mode is that
tilt refinement about the $x$- and $z$-axis is hard to implement;
tilting about these axes requires a rank to get voxel values from
other slabs, which not only induces excessive MPI communication, but
also demands AD to differentiate through MPI operations. The latter is
not impossible, but existing AD packages may need to be modified in
order to add that feature. Another possible approach among the slabs
kept by different MPI ranks is to introduce overlapping regions along
$y$; in this case, tilting about $x$ and $z$ can be done individually
by each rank, though the degree of tilt is limited by the length of
overlap.  This is not yet implemented in \adorym{}, but could be added
in the future.

\subsubsection{HDF5-file-mediated low-memory mode (H5)}

When running on an HPC with hundreds or thousands of computational nodes,
the DO mode can significantly reduce the memory needed by each node to store the object
function if one uses many nodes (and only a few ranks per node)
to distribute the object. On a single workstation or laptop, however, the total volume
of RAM is fixed, and very large-scale problems are still difficult to solve
even if one uses DO.

For such cases, \adorym{} comes with an alternative
option of storing the full object function in a parallel HDF5 file
\cite{hdf5}, which is referred to as the H5 mode. The HDF5 library works with the MPI-IO driver to allow a
file to be read or written by several MPI ranks, where any
modifications to the file's metadata are done collectively by all
ranks to avoid potential conflicts. This feature provides a viable
scheme of further reducing the RAM usage of \adorym{}. At the
beginning of the reconstruction job, the object function is
initialized and saved as an HDF5 file. Before entering the
differentiation loop, the object function is duplicated as a new
dataset in the same HDF5 file, and rotated to the viewing angle being
processed in a similar fashion as the DO mode: the rotation is done in
parallel by all ranks, with each rank handling several
$y$-slices. Following that, object chunks are read from the rotated
object by each rank, and sent to GPU if GPU acceleration is
enabled. The calculated gradient chunks are written by each rank into
a separate gradient dataset, which, after all tiles on the current
angle are processed, is rotated back to 0$\degree{}$ collectively. To
update the object function, each rank reads in a slice of the object
and the corresponding gradient at a time, performs update using the selected optimizer,
then writes the updated object slice back to the HDF5 file. The
next slice is then processed, until the entire object is updated
jointly by all ranks. The workflow [Fig.~\ref{fig:workflow}(c)] in
fact resembles that of the DO mode, except the distribution of object
chunks and synchronization of gradient chunks is done through HDF5
file reading and writing instead of using MPI AlltoAll.

While the H5 mode
allows the reconstruction of large objects on limited memory machines,
I/O with a hard drive (even with a solid state drive) is slower than
memory access.  Additionally, writing into an HDF5 with multiple ranks
is subject to contention, which may be mitigated if a parallel file
system with multiple object storage targets (OSTs) is available
\cite{behzad_2013}.  This is more likely to be available at an HPC
facility than on a smaller, locally managed system, and precise
adjustment of striping size and HDF5 chunking are needed to optimize
OST performance \cite{howison_hdf}.
Despite these challenges, the HDF5 mode is valuable because it
enables reconstructions of large objects and/or complex forward models
on limited memory machines. It also provides future-proofing because,
as fourth-generation synchrotron facilities deliver higher brightness and
enable one to image very thick samples \cite{eriksson_jsr_2014}, we may eventually encounter
extra-large objects which might be
difficult to reconstruct even in existing HPC machines.

With all the above descriptions, we summarize in
Fig.~\ref{fig:block_chart} the inter-relations among different
modules as parts of \adorym{}'s software architecture.

\subsubsection{User interface}

If a reconstruction job can be done using the provided \modfm{}
classes, users generally just need to call the \texttt{reconstruct\_ptychography}
function inside the \modmain{}. Optimizers for different variables can either
be explicitly declared using the child classes of \modopt{}, or be specified by providing
the optimizer name (for object function) and step size while using default values for other
parameters. Below is a code example that can be used to generate the fly-scan ptychography
results shown in the main text:
\begin{framed}
\small
\begin{verbatim}
import adorym
from adorym.ptychography import reconstruct_ptychography

output_folder = "recon"
distribution_mode = None
optimizer_obj = adorym.AdamOptimizer("obj", output_folder=output_folder,
                                     distribution_mode=distribution_mode,
                                     options_dict={"step_size": 1e-3})
optimizer_probe = adorym.AdamOptimizer("probe", output_folder=output_folder,
                                       distribution_mode=distribution_mode,
                                       options_dict={"step_size": 1e-3, "eps": 1e-7})
optimizer_all_probe_pos = adorym.AdamOptimizer("probe_pos_correction",
                                               output_folder=output_folder,
                                               distribution_mode=distribution_mode,
                                               options_dict={"step_size": 1e-2})

params_ptych = {"fname": "data.h5",
                "theta_st": 0,
                "theta_end": 0,
                "n_epochs": 1000,
                "obj_size": (618, 606, 1),
                "two_d_mode": True,
                "energy_ev": 8801.121930115722,
                "psize_cm": 1.32789376566526e-06,
                "minibatch_size": 35,
                "output_folder": output_folder,
                "cpu_only": False,
                "save_path": ".",
                "initial_guess": None,
                "random_guess_means_sigmas": (1., 0., 0.001, 0.002),
                "probe_type": "aperture_defocus",
                "forward_model": adorym.PtychographyModel,
                "n_probe_modes": 5,
                "aperture_radius": 10,
                "beamstop_radius": 5,
                "probe_defocus_cm": 0.0069,
                "rescale_probe_intensity": True,
                "free_prop_cm": "inf",
                "backend": "pytorch",
                "raw_data_type": "intensity",
                "optimizer": optimizer_obj,
                "optimize_probe": True,
                "optimizer_probe": optimizer_probe,
                "optimize_all_probe_pos": True,
                "optimizer_all_probe_pos": optimizer_all_probe_pos,
                "save_history": True,
                "unknown_type": "real_imag",
                "loss_function_type": "lsq",
                }

reconstruct_ptychography(**params_ptych)
\end{verbatim}
\end{framed}
\normalsize
In the code example, we passed \texttt{adorym.PtychographyModel} to \texttt{"forward\_model"}.
This argument can be replaced by a user-defined \modfm{} class whenever desirable. Instructions
on creating new forward models and defining new refinable parameters can be found in the documentation
of \adorym{}, which is currently hosted on the GitHub repository (\url{https://github.com/mdw771/adorym/}).

\section{Definition of structural similarity index (SSIM)}

The SSIM \cite{wang_ieeetip_2004} is
a metric that measures the ``degree of match'' between two images,
and, unlike pixel-to-pixel error metrics such as the mean squared
error, it takes into account the interdependence among pixels lying in
their local neighborhoods.
For reconstructed image $I_a$ and reference
image $I_r$, the SSIM is computed as
\begin{equation}
  \mbox{SSIM}(I_a, I_r) = l(I_a, I_r) \cdot c(I_a, I_r) \cdot s(I_a, I_r).
\end{equation}
where $l$, $c$, and $s$ respectively gauge the similarity of the
images in terms of luminance, contrast, and structure. These factors
are defined as
\begin{eqnarray}
  l(I_a, I_r) &=& \frac{2\mu_a\mu_r + c_1}{\mu_a^2 + \mu_r^2 + c_1} \label{eqn:ssim_l} \\
  c(I_a, I_r) &=& \frac{2\sigma_a\sigma_r + c_2}{\sigma_a^2 + \sigma_r^2 + c_2} \label{eqn:ssim_c} \\
  s(I_a, I_r) &=& \frac{\sigma_{a,r} + c_3}{\sigma_a\sigma_r + c_3} \label{eqn:ssim_s}
\end{eqnarray}
where $\mu_a$ and $\mu_r$ are the mean value of the reconstructed
image and the reference image, $\sigma_a$ and $\sigma_r$ are their
standard deviations, and $\sigma_{a,r}$ is their correlation
coefficient. The parameters appearing in the equations above are given
by
\begin{eqnarray}
  c_1 &=& (k_1 L)^2 \\
  c_2 &=& (k_2 L)^2 \\
  c_3 &=& c_2 / 2
\end{eqnarray}
where $k_1$ and $k_2$ are set to 0.01 and 0.03 in our case, and $L$ is
the dynamic range of the grayscale images.

\bibliographystyle{unsrt}
\bibliography{mybib}